\documentclass[aps,twocolumn,
prb,
amsmath,
amssymb,
superscriptaddress,
]{revtex4-2}

\usepackage[
total={6.5in,8.75in}, top=1.2in, left=0.9in, includefoot,
]{geometry}

\usepackage{dsfont}
\usepackage{epstopdf}
\usepackage{dcolumn}
\usepackage{color}
\usepackage{amsmath,amssymb}
\usepackage[english]{babel}
\usepackage{bm}
\usepackage{resizegather}
\usepackage[utf8]{inputenc}
\usepackage{graphicx}
\usepackage{float}
\usepackage{caption}
\usepackage{subcaption}
\usepackage{geometry}
\usepackage{braket}
\usepackage{xcolor}
\usepackage{mathtools}
\usepackage{amssymb} 
\usepackage{amsmath}
\usepackage{hyperref}
\usepackage{cleveref}
\usepackage[export]{adjustbox}

\usepackage{url}
\usepackage{notes2bib}
\usepackage{mathbbol}
\usepackage{ulem}
\usepackage{tikz}
\usepackage{comment}
\usepackage{amsfonts}
\usepackage{makecell}

\usepackage{graphicx}

\DeclareUnicodeCharacter{2212}{\textendash}
\newcommand{\up}{\uparrow}
\newcommand{\down}{\downarrow}

\DeclareUnicodeCharacter{2212}{\textendash}

\begin{document}

\title{Bands renormalization and superconductivity in the strongly correlated Hubbard model using composite operators method}
\author{L. Haurie}
\affiliation{Institut de Physique Th\'eorique, Universit\'e Paris Saclay, CEA
CNRS, Orme des Merisiers, 91190 Gif-sur-Yvette Cedex, France \\email \href{louis.haurie@ipht.fr}{louis.haurie@ipht.fr}}
\author{M. Grandadam}%
\affiliation{%
Department of Physics and Physical Oceanography, Memorial University of Newfoundland, St. John's, Newfoundland \& Labrador, Canada A1B 3X7
}%
\author{E. Pangburn}
\affiliation{Institut de Physique Th\'eorique, Universit\'e Paris Saclay, CEA
CNRS, Orme des Merisiers, 91190 Gif-sur-Yvette Cedex, France}

\author{A. Banerjee}
\affiliation{Department of Physics, Ben-Gurion University of the Negev, Beer-Sheva 84105, Israel}

\author{S. Burdin}
\affiliation{Université de Bordeaux, CNRS, LOMA, UMR 5798, F-33400 Talence, France}

\author{C. Pépin}
\affiliation{Institut de Physique Th\'eorique, Universit\'e Paris Saclay, CEA
CNRS, Orme des Merisiers, 91190 Gif-sur-Yvette Cedex, France}

\date{\today}

\begin{abstract}
We use the composite operator method (COM) to analyze the strongly correlated repulsive Hubbard model, investigating the effect of nearest-neighbor hoppings up to fourth order on a square lattice. We consider two sets of self-consistent equations, one enforcing the Pauli principle and the other imposing charge-charge, spin-spin, and pair-pair correlations using a decoupling scheme developed by L. Roth [Phys. Rev. \textbf{184}, 451–459, (1969)]. We extract three distinct solutions from these equations: COM1 and COM2 by imposing the Pauli principle and one from Roth decoupling. An overview of the method studying the validity of particle-hole symmetry and the Luttinger theorem for each solution is presented. Additionally, we extend the initial basis to study superconductivity, concluding that it is induced by the Van Hove singularity. Finally, we include higher-order hoppings using realistic estimates for tight binding parameters and compare our results with ARPES measurements on cuprates.
\end{abstract}
\maketitle

\section{\label{sec:intro} Introduction}
Exotic behavior and multiple phases exhibited by strongly correlated materials have attracted significant interest in the quantum condensed matter community\cite{morosan_strongly_2012,sachdev_order_2003}. One prominent example is cuprates, where superconductivity survives up to very high critical temperatures compared to conventional superconductors described by the BCS theory \cite{bardeen_theory_1957}. Since their discovery in 1987 by G. Bednorz and K.A. Müller \cite{Bednorz1988}, extensive research has focused on modeling and understanding the pairing mechanism in these materials to obtain a microscopic theory of cuprate superconductors. The complexity of their phase diagram and the numerous unexplained quantum phases \cite{norman_electronic_2003} have resulted in various theoretical models \cite{georges_dynamical_1996}.

The repulsive Hubbard model remains the paradigmatic tool for studying strongly correlated quantum matter even after sixty years of its proposal. It offers a minimal description of such systems with just two ingredients - hopping of electrons between lattice sites $t$ and on-site electron repulsion $U$ creating an energy cost for double occupancy on the same lattice site. It is given by
\begin{equation}
H= -\sum \limits_{ i, j \sigma} t_{ij} c^\dagger_{i \sigma} c_{j \sigma} + U \sum \limits_i n_{i \up} n_{i \down} - \mu \sum \limits_{i \sigma} n_{i \sigma},
\label{Hubbard}
\end{equation}
where the chemical potential $\mu$ fixes the total number of electrons. In Eq.~(\ref{Hubbard}) $t_{ij}$ is such that $t_{ij}=t$ if $i$ and $j$ are nearest neighbouring sites, and zero otherwise. One can extend the model to longer ranged hoppings. $c_{i \sigma} / c_{i\sigma}^\dagger$ destroys/creates an electron on site $i$ with spin $\sigma\in \{ \up, \down \}$ on the lattice and $n_{i \sigma}= c_{i \sigma}^\dagger c_{i \sigma}$ is the number operator for site $i$ and spin $\sigma$. However, this seemingly straightforward model has eluded unbiased theoretical solutions.

The Hubbard Hamiltonian is known to be solvable only in one dimension (1d) and infinite dimensions \cite{lieb_one-dimensional_2003}. In the intermediate dimensions, the equations of motion of the electronic Green's function involve higher order Green's function and cannot be solved exactly~\cite{Hubbard_electron_1963}. In the past, approximations such as , II and III \cite{Hubbard_electron_1963} have been made to truncate the higher-order terms and close the equations of motion.

We revisit a perturbative scheme called "Composite Operators Method" (COM) \cite{avella_hubbard_1998} \cite{beenen_superconductivity_1995}. This method originates from the Hubbard II approximation, where the electronic Green's function is decomposed into two poles with respective self-energies of 0 and $U$. Unlike Hubbard I and III approximations, Hubbard II approximation satisfies particle-hole symmetry \cite{gebhard_mott_1997}. In contrast to other methods, such as Kotliar-Ruckenstein approximation of slave bosons or Gutzwiller's wavefunction approach \cite{gebhard_mott_1997}, COM is exact in the atomic limit, establishing an ideal framework for studying strongly correlated regimes. However, Hubbard II approximation requires self-consistency to establish the effect of electron hopping, treated as a perturbation.
Originally developed by L. Roth \cite{roth_electron_1969}, the composite operators (or Hubbard operators \cite{phillips_advanced_2012}) exactly diagonalize the Hubbard Hamiltonian in the atomic limit. They are a good choice to treat the hopping term $t$ as a perturbation. The composite operator method starts from Hubbard II approximation and set-up a self-consistent scheme to compute the effect of the hopping term.

We consider two sets of such self-consistency equations. The first enforces charge-charge, spin-spin, and pair-pair correlations using L. Roth's decoupling scheme, and the second imposes the Pauli principle instead of correlations. We perform a numerical minimization of these two sets of equations. The self-consistency equations enforcing the Pauli principle yield two distinct solutions named COM1 and COM2. The other sets of equations using Roth decoupling only exhibit one solution, which is referred to as the Roth solution.  The spectral and magnetic properties of the Roth solution have been studied in Ref. \cite{mehlig_single-particle_1995} and Ref. \cite{nolting_band_1989}. Despite violating the Pauli principle \citep{avella_hubbard_1998}, the Roth solution shows a band structure in agreement with quantum Monte-Carlo simulations \citep{beenen_superconductivity_1995}. In contrast, COM1 and COM2, which have been widely studied in Refs. \cite{mancini_hubbard_2004} and \cite{avella_hubbard_1998}, exhibit a Fermi surface consistent with cuprate ARPES experiments only under some approximations \cite{avella_anomalous_2008}. Furthermore, some extensions of COM method to the $t-J$ models have been made in Ref. \cite{eskandari-asl_local_2022}.

Although the self-consistency equations enforcing the Pauli principle have been previously minimized with next-nearest neighbor hoppings \cite{avella_t_t'_2001}, this is not the case for the Roth decoupling. The latter has been studied with next-nearest neighbor hoppings for a three-band Hubbard model in cuprates \cite{calegari_pseudogap_2015}, but never for the one-orbital Hubbard model.

Superconductivity has also been studied using the COM\cite{beenen_superconductivity_1995}. Past studies indicate that the proximity to the Van Hove singularity enhances SC within Roth minimization scheme \cite{calegari_superconductivity_2005}. The enhancement of the density of the state associated with this singularity allows electrons to form more pairs readily. However, a detailed study comparing and contrasting different self-consistent solutions and including realistic hopping parameters to model the cuprates is still lacking.

We aim to compare and contrast the solutions obtained in previous studies using the COM, in order to benchmark the results for a later study where we will break translational invariance. We evaluate their physical consistency with experiments in cuprates by testing the particle-hole symmetry and the validity of the Luttinger theorem. This theorem states that the enclosed volume by the Fermi Surface is proportional to the electron density \cite{a._a._abrikosov_methods_1963}. It is important to note that there is no consensus on when this theorem is expected to be violated, although its violation is routinely observed \cite{heath_necessary_2020} \cite{dzyaloshinskii_consequences_2003} \cite{skolimowski_luttingers_2022}. Interestingly, it is always violated for this method. Our study of Fermi Surfaces reveal that only Roth solution is close to the non-interacting Fermi surface, but is enlarged because of the violation of Luttinger theorem. Finally, we include the superconductivity with longer-ranged hoppings. The plan for the rest of the paper is as follows: Section \ref{sec:I} details the formalism and self-consistency we use in the manuscript. In section \ref{sec:II}, we restrict to the nearest neighbor hoppings and compare the solutions obtained by the different minimizations \cite{beenen_superconductivity_1995} \cite{avella_composite_2012} while discussing their physical implications, particularly with regard to particle-hole symmetry and the Luttinger theorem. Finally, we allow for the superconductivity in the model, and in the last section \ref{sec:III}, we study the impact of the longer-ranged hopping orders.

\section{\label{sec:I}Composite operator formalism}
Composite operators are introduced to solve the equations of motion of composite Green's functions (Green's functions are composed of two composite operators) exactly in some limits. Since we are interested in a strongly correlated regime of the Hubbard model, we introduce composite operators solving exactly the Hubbard model in the atomic limit (no hopping term). This is a good starting point to treat the hopping term $t$ with a perturbation expansion.

\subsection{Presentation of the method}

We start by introducing the following composite operators
\begin{equation}
\begin{cases}
\xi_{i \sigma}&= c_{i \sigma} - c_{ i \sigma} n_{i \bar{\sigma}} \\
\eta_{i \sigma}&= c_{i \sigma} n_{i \bar{\sigma}}
\end{cases}
\label{spinor}
\end{equation}

Where $\bar{\sigma}$ means we take a spin $\up$ if $\sigma=\down$ and a spin $\up$ if $\sigma=\down$. These composite operators represent respectively the transition from an empty site i to a site filled with one electron of spin $\sigma$, and the transition from a site i filled with an electron of spin $\bar{\sigma}$ to a doubly occupied state \cite{mancini_hubbard_2004}. Indeed, applying $\xi_{i \sigma}$ to a state with an electron of spin $\sigma$ at site i will remove this electron, while applying $\eta_{i \sigma}$ on a doubly occupied state on site i will only let one electron of spin $\bar{\sigma}$ on site i. We introduce the following spinors
\begin{equation}
\begin{aligned}
     \psi_{i\sigma}=&\begin{pmatrix}  \xi_{i \sigma} \\  \eta_{i \sigma} \end{pmatrix} 
\end{aligned}
\end{equation}


Hereafter, $\psi^1_{i\sigma}=\xi_{i \sigma}$ and $\psi^2_{i \sigma}=\eta_{i \sigma}$ denotes the first and second component of the spinor at site i and with spin $\sigma$.\\

\subsubsection{Atomic limit}

To illustrate the interest of introducing these operators, we first consider the atomic limit of Eq. (\ref{Hubbard}) : we put the hopping term $t$ to zero. The Hamiltonian is now local (ie electrons are not hopping anymore and each site is independent) and is given by 
\begin{equation}
H_{loc}= U \sum \limits_i n_{i \up} n_{i \down} - \mu \sum \limits_{i \sigma} n_{i \sigma} 
\end{equation}

We introduce the $2 \times 2$ composite Green's function matrix at sites $i$ and $j$, with spins $\sigma$ and $\sigma'$ and defined for an imaginary time $\tau$ by

\begin{equation}
\textbf{S}_{ij\sigma\sigma'}^{loc}(\tau,\tau')= \langle \langle \psi_{i\sigma}( \tau ); \psi_{j\sigma'}^\dagger(\tau ') \rangle \rangle_{loc} 
\label{compositeGF}
\end{equation}

Where, for two operators X and Y,
\begin{equation}
\langle \langle X(\tau); Y(\tau') \rangle \rangle_{loc}= \theta_H(\tau-\tau') \langle \{X(\tau); Y(\tau') \} \rangle_{loc}
\end{equation}

Where $\theta_H(\tau-\tau')$ is one if $\tau > \tau'$ and zero otherwise (Heaviside function). $\langle ... \rangle_{loc}$ denotes the thermal average taken with the Hamiltonian $H_{loc}$ and $\{ X(\tau); Y(\tau') \}$ is the anticommutator of X and Y.  Since we are at equilibrium we have $\textbf{S}^{loc}_{ij\sigma \sigma'}(\tau,\tau')=\textbf{S}^{loc}_{ij\sigma\sigma'}(\tau-\tau')$. By differentiating with respect to time, we get the following equations of motion for the composite Green's function matrix 

\begin{equation}
\begin{aligned}
\frac{d}{d\tau} \textbf{S}^{loc}_{ij\sigma \sigma'}(\tau) = & \delta(\tau) \delta_{\sigma \sigma'} \langle \{ \psi_{i\sigma}(\tau); \psi_{j\sigma}^\dagger (0) \} \rangle_{loc} \\&+ \delta_{\sigma \sigma'} \langle \langle  [ \psi_{i\sigma}(\tau); H_{loc} ] ; \psi_{j\sigma}^\dagger (0)  \rangle \rangle_{loc}
\end{aligned}
\label{eqofmotion}
\end{equation}

Where $[A;B]$ is the usual commutator between two operators A and B. We enforce a paramagnetic solution by adding $\delta_{\sigma \sigma^\prime}$ prefactor. The currents in the atomic limit are given by

\begin{equation}
J^{loc}_{i\sigma}(\tau) = \frac{d}{d\tau} \psi_{i\sigma}(\tau) = [ \psi_{i\sigma}(\tau);H_{loc}] = \textbf{A} \psi_{i\sigma}(\tau)
\label{currents}
\end{equation}

With

\begin{equation}
\textbf{A} = \begin{pmatrix}
\mu & 0 \\
0 & U-\mu
\end{pmatrix}
\end{equation}

The equations of motion become

\begin{equation}
\begin{aligned}
\frac{d}{d\tau} & \textbf{S}^{loc}_{ij\sigma\sigma'}(\tau) = \\ &\delta(\tau) \delta_{\sigma\sigma'} \langle \{ \psi_{i\sigma}(\tau); \psi_{j\sigma}^\dagger (0) \} \rangle_{loc} - \textbf{A} \  \textbf{S}^{loc}_{ij\sigma}(\tau)  
\end{aligned}
\end{equation}

By time fourier transform we get 

\begin{equation}
    \textbf{S}^{loc}_{ij\sigma\sigma'}(\omega)=\delta_{\sigma\sigma'} (\omega - \textbf{A} +  i 0^+)^{-1} \textbf{I}_{i\sigma}^{loc} \delta_{ij}
    \label{GFloc}
\end{equation}

With $\textbf{I}_{i\sigma}^{loc}= \langle \{ \psi_{i\sigma}; \psi_{i\sigma}^\dagger \} \rangle_{loc} $ the normalization matrix and $0^+$ a small positive parameter used for analytic continuation since we are working with Matsubara time $\tau$. This matrix can be explicitly computed. A bit of algebra leads to
\begin{equation}
    \textbf{I}_{i\sigma}^{loc}= \begin{pmatrix}
    1-\langle n_{i \sigma}\rangle_{loc} & 0 \\
    0 & \langle n_{i\sigma} \rangle_{loc}
    \end{pmatrix}
    \label{Imatloc}
\end{equation}

We finally obtain
\begin{equation}
\textbf{S}_{ij\sigma\sigma}^{loc}(\omega)= \delta_{ij} \delta_{\sigma \sigma'}\begin{pmatrix} \frac{ 1-\langle n_{i \sigma} \rangle_{loc}}{\omega-\mu + i 0^+} &0 \\
0 & \frac{\langle n_{i \sigma} \rangle_{loc}}{\omega-U+\mu+ i 0^+}
\end{pmatrix}
\end{equation}

In the atomic limit the equations of motion can therefore be closed. The solution is given in Eq. (\ref{GFloc}), and by using the relation between composite and electronic operators $\xi_{i \sigma} + \eta_{i \sigma} = c_{i \sigma}$, we can deduce the electronic Green's function

\begin{equation}
\begin{aligned}
G_{ij\sigma\sigma'}^{loc}(\tau)=& \delta_{\sigma\sigma'} \langle \langle c_{i \sigma}(\tau); c_{j \sigma'}^\dagger \rangle \rangle_{loc} \\=&\delta_{\sigma\sigma'}(S^{11 \ loc}_{ij\sigma}(\tau) + S^{12 \ loc}_{ij\sigma}(\tau) \\&+ S^{21 \ loc}_{ij\sigma}(\tau) + S^{22 \ loc}_{ij\sigma}(\tau))
\end{aligned}
\label{elecGF}
\end{equation}

Where $S^{nm \ loc}_{ij\sigma}= \langle \langle  \psi^{n}_{i\sigma}(\tau) ; \psi_{j \sigma}^{m}(0) \rangle \rangle_{loc}$. Therefore we have showed that the composite operators we introduced solve the Hubbard model exactly at the Atomic limit. The electronic Green's function can be directly recovered, allowing to extract information such as the Fermi surface and the density of states.

\subsubsection{General case}

Let us consider the full Hamiltonian Eq. (\ref{Hubbard}) that includes both the local term $H_{loc}$ and the hopping term t. We consider the limit $U \gg t$ and we build an approximation from the atomic limit. From now on, $\langle ... \rangle$ are the thermal averages taken with the full Hamiltonian. We then introduce

\begin{equation}
\delta J_{i\sigma}= [\psi_{i\sigma}, H-H_{loc}] = J_{i\sigma} - J_{i\sigma}^{loc} 
\end{equation}

Where the current operator in the atomic limit is given by Eq. (\ref{currents}) and $J_{i\sigma}=[\psi_{i\sigma},H]$ is the current operator taken with the full Hubbard Hamiltonian Eq. (\ref{Hubbard}). Because of the tight-binding term, higher order Green's functions will appear in the equations of motion of the composite Green's function. We will not be able to solve the problem exactly as in the atomic limit. We thus need to do an approximation to be able to obtain the composite Green's function by truncating the equations of motion. 
Indeed, let us consider the composite Green's functions $2 \times 2$ matrix $\textbf{S}_{ij\sigma\sigma'}(\tau)=\delta_{\sigma \sigma'} \langle \langle \psi_{i \sigma}(\tau) ; \psi_{j \sigma'}(0) \rangle \rangle$ with thermal average on the full Hamiltonian. Its equations of motion can be written as

\begin{equation}
    \begin{aligned}
    \frac{d}{d\tau} \textbf{S}_{ij\sigma\sigma'}(\tau)=& \delta_{\sigma\sigma'}(\delta(\tau) \delta_{ij}  \textbf{I}_{i\sigma} + \theta_H(\tau) \textbf{M}_{ij\sigma}(\tau))
    \end{aligned}
    \label{EqMotion}
\end{equation}

Where we introduced the normalization matrix $\textbf{I}$ and the overlap matrix $\textbf{M}$ respectively as 

\begin{equation}
\begin{aligned}
\textbf{I}_{i\sigma}=& \langle \{ \psi_{i\sigma}, \psi_{i\sigma}^\dagger \} \rangle = \delta_{ij} \begin{pmatrix}
    1-\langle n_{i \sigma}\rangle & 0 \\
    0 & \langle n_{i\sigma} \rangle
    \end{pmatrix}
\\ 
\textbf{M}_{ij\sigma}=& \langle \{ J_{i\sigma}, \psi_{j\sigma}^\dagger \} \rangle =\begin{pmatrix}
m_{ij}^{11} & m_{ij}^{12} \\
m_{ij}^{12} & m_{ij}^{22}
\end{pmatrix}
\end{aligned}
\label{matrices}
\end{equation}

To solve the equations of motion for the composite Green's function matrix, we need to compute the $\textbf{I}$ and $\textbf{M}$ matrix. We directly computed the $\textbf{I}$ matrix. The current of the total Hamiltonian is given by
\begin{equation}
\begin{aligned}
    J_{i \sigma} =& \sum \limits_l \textbf{E}_{il\sigma} \psi_{l\sigma}+ \delta \phi_{i\sigma} \\
    \textbf{E}_{il\sigma}=& \textbf{A} \delta_{il} + \textbf{P}_{il\sigma}
\label{JandE}
\end{aligned}
\end{equation}


The $\textbf{E}$ matrix contains all the terms proportional to $\psi$, and $\delta \phi_{i\sigma}$ contains all terms which are not. The $\textbf{A}$ matrix appearing in $\textbf{E}$ is the contribution of the atomic limit terms. It is given by
\begin{equation}
\textbf{A}=\begin{pmatrix}
\mu &0   \\
0 & U-\mu
\end{pmatrix}
\end{equation}

The $\textbf{P}$ matrix appearing in $\textbf{E}$ is defined by
\begin{equation}
\textbf{P}_{ij\sigma}= \langle \{ \delta J_{i\sigma} ; \psi^\dagger_{j\sigma} \} \rangle \textbf{I}^{-1}_{j\sigma} 
\end{equation}

$\textbf{P}$ is the contribution of the terms proportional to $\psi$ in $\delta J_{i\sigma}$. 

With this rewriting, the $\textbf{M}$ matrix is now given by the following expression

\begin{equation}
\begin{aligned}
    \textbf{M}_{il\sigma}(\tau)=&  \sum \limits_j \textbf{E}_{ij\sigma} \langle \{ \psi_{j\sigma}(\tau); \psi^\dagger_{l\sigma} \} \rangle \\&+ \langle \{ \delta \phi_{i \sigma}(\tau); \psi_{l\sigma}^\dagger \}\rangle
\end{aligned}
\label{Mrel}
\end{equation}

The first term is proportional to $\textbf{S}_{jl\sigma \sigma'}$. However the second term is not and will introduce higher-order Green's function in the equations of motion

\begin{equation}
\begin{aligned}
\frac{d}{d\tau} \textbf{S}_{ij \sigma\sigma'}(\tau)&=\delta_{\sigma\sigma'}(\delta(\tau)\delta_{ij} \textbf{I}_{i\sigma} +  \sum \limits_l \textbf{E}_{il \sigma} \textbf{S}_{lj\sigma}(\tau) \\&+ \theta_H(\tau) \langle \{ \delta\phi_{i\sigma}(\tau); \psi_{l\sigma}^\dagger \} \rangle 
\label{EqofS}
\end{aligned}
\end{equation}

An approximation is needed: we will assume that $\delta \phi_{i\sigma}$ is negligible. Therefore, after a time Fourier transform, Eq. (\ref{EqofS}) becomes:

\begin{equation}
\sum \limits_l ((\omega+i0^+) \textbf{Id}_2 \delta_{il} - \delta_{\sigma \sigma'} \textbf{E}_{il\sigma}) \textbf{S}_{lj \sigma \sigma'} = \delta_{\sigma \sigma'} \delta_{ij} \textbf{I}_{i\sigma}
\label{EqonS2}
\end{equation}

One can perform a spatial Fourier transform and use translational invariance in order to have diagonal elements only in the momentum space. Therefore, in Fourier space inverting Eq. (\ref{EqonS2}) leads to:

\begin{subequations}
\begin{align}
\textbf{S}_{k\sigma\sigma'}(\omega)&\approx \delta_{\sigma \sigma'}   ((\omega+ i 0^+)\textbf{Id}_2 - \textbf{E}_{k\sigma})^{-1} \textbf{I}_{\sigma} \label{EMIS:S}  \\
J_{i\sigma} (\tau) & \approx \sum \limits_{l} \textbf{E}_{il\sigma} \psi_{l\sigma}(\tau) \label{EMIS:curr} \\
\textbf{E}_{k\sigma}&\approx \textbf{M}_{k\sigma}(0) \textbf{I}_{\sigma}^{-1}\label{EMIS:E} 
\end{align}
\end{subequations}

These three equations are a direct consequence of the COM approximation (neglecting $\delta \phi_i$ in the $U \gg t$ limit). The first one has been derived from the equation of motion. The second equation is the current from Eq. (\ref{JandE}) and the last equation is Eq. (\ref{Mrel}) at $\tau=0$(notice that $\textbf{S}_{ij\sigma}(\tau=0)=\delta_{ij} \textbf{I}_{i\sigma}$), after a spatial Fourier transform in order to have only diagonal elements in momentum space using translational invariance. 

In appendix A, this approximation is studied in depth and we detail the physical consequences of neglecting $\delta \phi$.
$\textbf{E}$ acts as an effective energy matrix. Note that Eq. (\ref{EMIS:curr}) is similar to a Schrödinger equation for the composite operators.
We can also include higher order terms in the basis to go further in the approximation. This has been done in Ref. \citep{avella_composite_2012}.

The equations of motion of the composite Green's function only depend on the $\textbf{E}$ and $\textbf{I}$ matrices under our approximation. In order to perform a self-consistent scheme, we introduce the $2 \times 2$ correlation function matrix
\begin{equation}
\textbf{C}_{ij\sigma }=\langle \psi_{i\sigma} \psi_{j\sigma}^\dagger \rangle
\end{equation} 

We want to find an expression of $\textbf{C}_{ij\sigma\sigma'}$ as a function of the eigenvalues of the $\textbf{E}$ matrix. We use the spectral representation to get

\begin{equation}
\textbf{C}_{ij\sigma}= \int d\omega d^2k \ e^{i k(r_i-r_j)} (1-f_D(\omega))  \left( -\frac{1}{\pi} \right) Im( \textbf{S}_{k\sigma}(\omega))
\label{CfctS}
\end{equation}
With $f_D=\frac{1}{Exp(\beta \omega)+1}$ the Fermi Dirac distribution. From there, we can use Eq. (\ref{EMIS:S}) to apply the residue theorem on $\textbf{S}$ and express it as a function of the eigenvalues of the $\textbf{E}$ matrix (proof in appendix B) 
\begin{equation}
\textbf{S}_{k\sigma}(\omega)= \sum \limits_{a=1}^2 \frac{\boldsymbol{\kappa}^a_{k\sigma} }{(\omega - \epsilon_k^a + i 0^+)}
\label{Sresidue}
\end{equation}

With
\begin{equation}
\boldsymbol{\kappa}^a_{k\sigma}=  \frac{(-1)^{a+1}  Com(\epsilon_{k\sigma}^a\mathbf{Id_2}-\textbf{E}_{k\sigma})^T \textbf{I}_{\sigma}}{(\epsilon_{k\sigma}^1 - \epsilon_{k\sigma}^2)}
\label{kappa}
\end{equation}
Where $Com(A)$ is the cofactor matrix of A, and $\textbf{A}'$ is the matrix that results from deleting row i and column j of $\textbf{A}$.
In these equations, $a\in \{ 1, 2\}$ and $\epsilon_{k}^1$ and $\epsilon_{k}^2$ are the two  eigenvalues of the energy matrix $\textbf{E}_{k\sigma}$. Finally, combining Eq. (\ref{CfctS}) and (\ref{Sresidue}) leads to an expression of the correlation function as a function of $\epsilon^a$

\begin{equation}
\begin{aligned}
\textbf{C}_{k\sigma}=& \frac{1}{2} \left( 1+\tanh{\frac{\beta \epsilon^1_{k\sigma}}{2}} \right) \boldsymbol{\kappa}^1_{k\sigma} \\& \frac{1}{2}\left( 1 + \tanh{\frac{\beta \epsilon^2_{k\sigma}}{2}} \right) \boldsymbol{\kappa}^2_{k\sigma} \end{aligned}
\label{CfctE}
\end{equation}

Where $\beta=\frac{1}{k_B T}$ comes from the Fermi Dirac distribution. To summarize, from the $\textbf{M}$ and $\textbf{I}$ matrices we can obtain the energy matrix $\textbf{E}$. We showed that its eigenvalues are directly related to the composite correlation function $\textbf{C}_{ij \sigma \sigma}$. To close the system, in the next section we explicitly compute the $\textbf{M}$ and $\textbf{I}$ matrices and express them as a function of the correlation functions.

\subsection{Self-consistent scheme}


We can compute the algebra associated with our composite operators defined in Eq. (\ref{spinor}). We assume spin rotational symmetric solutions,, allowing us to have $\delta_{\sigma\sigma'}$ in our Green's and correlation functions. This implies \begin{equation}
\langle n_{i \up} \rangle = \langle n_{i \down} \rangle= \frac{n_i}{2}
\end{equation} 

We compute explicitly the currents
\begin{equation}
\begin{aligned}
j_{i\sigma}^1=&  -\mu \xi_{i \sigma} - \sum \limits_l t_{il} \left(c_{l\sigma} - n_{i \bar{\sigma}} c_{l \sigma} + S_i^{sign(\bar{\sigma})} c_{l \bar{\sigma}} + sign(\bar{\sigma}) \Delta_{ i} c^\dagger_{l \bar{\sigma}} \right) \\
j_{i\sigma}^2=&  -( \mu - U) \eta_{i \sigma} + \sum \limits_l t_{il} \left( -n_{i \bar{\sigma} }c_{l \sigma } + S_i^{sign(\bar{\sigma})} c_{l \bar{\sigma} } +sign(\bar{\sigma}) \Delta_{i} c^\dagger_{l \bar{\sigma}} \right)
\end{aligned}
\label{Explicitcurrents}
\end{equation}

With $S_i^- = c^\dagger_{i \down} c_{i \up}$, $S_i^+ = c^\dagger_{i \up} c_{i \down}$ and $\Delta_i =  c_{i \uparrow } c_{i \down}$. We take the following convention: $ sign( \up )= 1 $ and $sign( \down )= - 1 $. 

The coefficients of the $\textbf{M}$ matrix are (some details of the algebra can be found in appendix C) :

\begin{equation}
\hspace{-0.1cm}\begin{aligned}
m_{ij}^{11}=& \hspace{-0.05cm}  -\mu \left(1-\frac{n_i}{2}\right)\delta_{ij} - \delta_{ij} \sum \limits_l t \alpha_{il}^1 e_{il}  - t \alpha_{ij}^1 (1-\frac{n_i+n_j}{2} + p_{ij}) \\
m_{ij}^{12}=&  \delta_{ij} \sum \limits_l t \alpha_{il}^1 e_{il} - t \alpha_{ij}^1 \left(\frac{n_j}{2}-p_{ij} \right) \\
m_{ij}^{22}=& -(\mu - U) \frac{n_i}{2} \delta_{ij} - \delta_{ij} \sum \limits_l t \alpha_{il}^1 e_{il} - t \alpha_{ij}^1 p_{ij}
\end{aligned}
\label{Mcoefs}
\end{equation}

With
\begin{equation}
\begin{aligned}
e_{ij}&= \langle \xi_{j \sigma} \xi^\dagger_{i \sigma} \rangle - \langle \eta_{j \sigma} \eta_{i \sigma}^\dagger \rangle + \langle \xi_{i \sigma} \eta^\dagger_{j \sigma} \rangle - \langle \xi_{j \sigma} \eta^\dagger_{i \sigma} \rangle \\
p_{ij}&=  \langle n_{i \sigma} n_{j \sigma} \rangle + \langle S_i^- S_j^+ \rangle - \langle \Delta_i \Delta_j^* \rangle
\end{aligned}
\label{paramdef}
\end{equation}
$\alpha_{il}^1=\frac{t_{il}}{t}$ is equal to 1 if $i$ and $l$ are nearest neighbors, and 0 otherwise.

\begin{figure*}[ht]
\centering
\includegraphics[width=1.\linewidth, height=9cm]{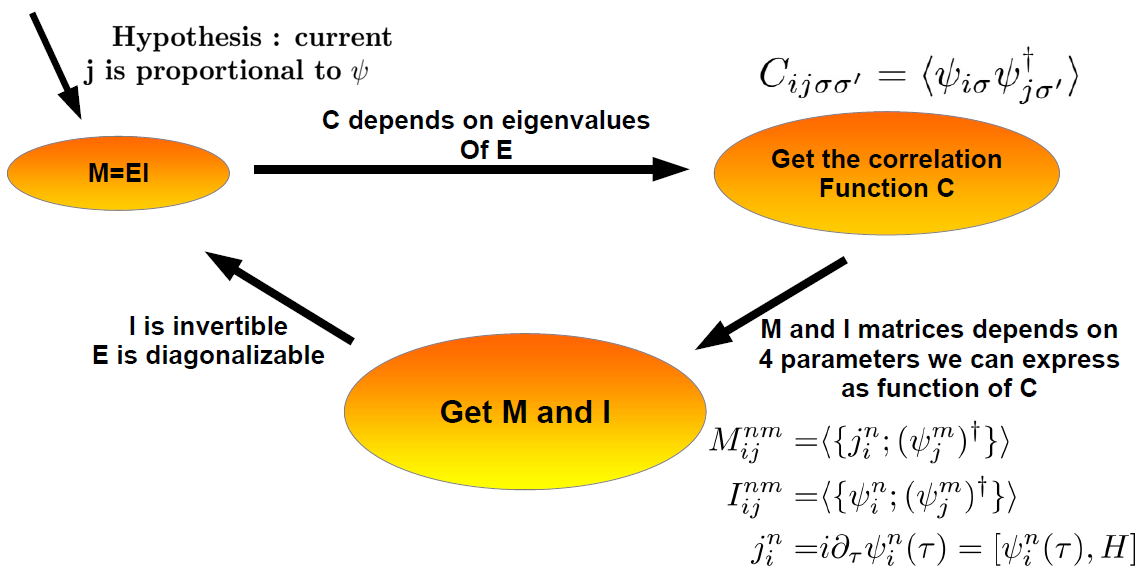}
\caption{The self consistency loop. Our hypothesis on the currents allows us to get a relation between the $\textbf{M}$ and $\textbf{I}$ matrices (Eq. (\ref{EMIS:E})). The energy matrix $\textbf{E}$ is diagonalizable. We can express correlation functions as a function of its eigenvalues with Eq. (\ref{CfctE}). Then, we can rewrite the $\textbf{M}$ and $\textbf{I}$ matrices in term of these correlations functions and do a self-consistency (Eq. (\ref{selfcon})).}
\label{Fig1:selfcon}
\end{figure*}

\begin{figure*}[ht]
\begin{subfigure}{\linewidth}
\begin{subfigure}{0.31\linewidth}
\centering
\includegraphics[width=5.5cm, height=6cm]{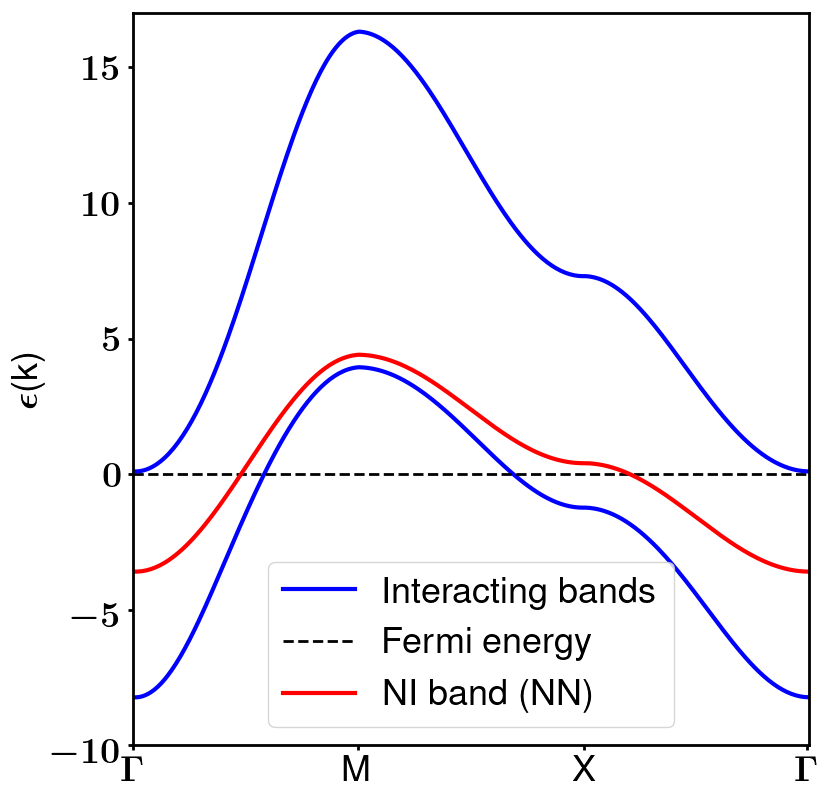}
\caption{First Pauli Solution (COM1)}
\end{subfigure}
\hspace{0.02\linewidth}
\begin{subfigure}{0.31\linewidth}
\centering
\includegraphics[width=5.5cm,height=6cm]{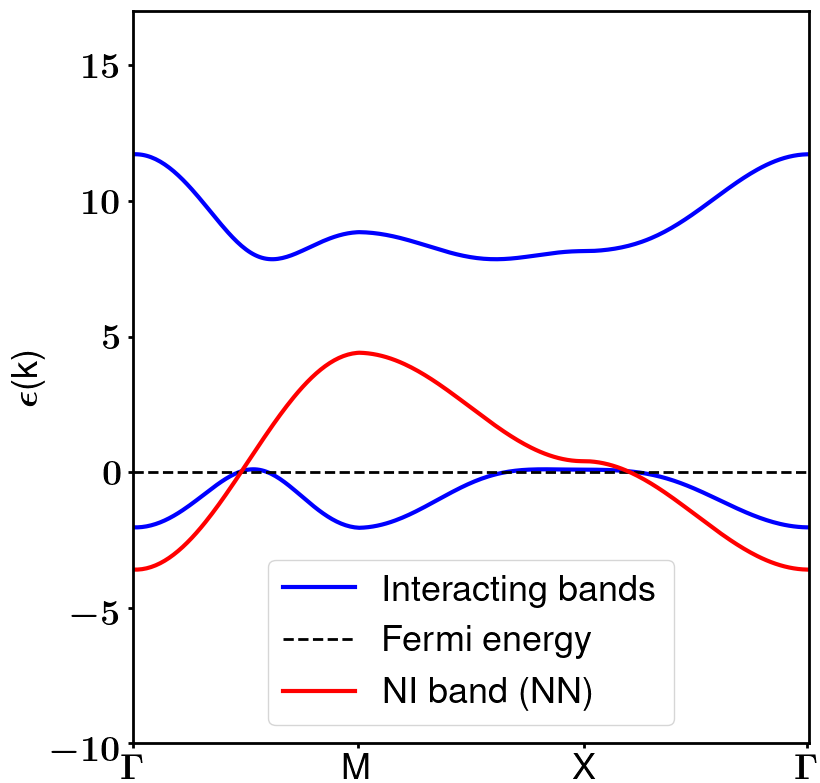}
\caption{Second Pauli solution (COM2)}
\end{subfigure}
\hspace{0.02\linewidth}
\begin{subfigure}{0.31\linewidth}
\centering
\includegraphics[width=5.5cm,height=6cm]{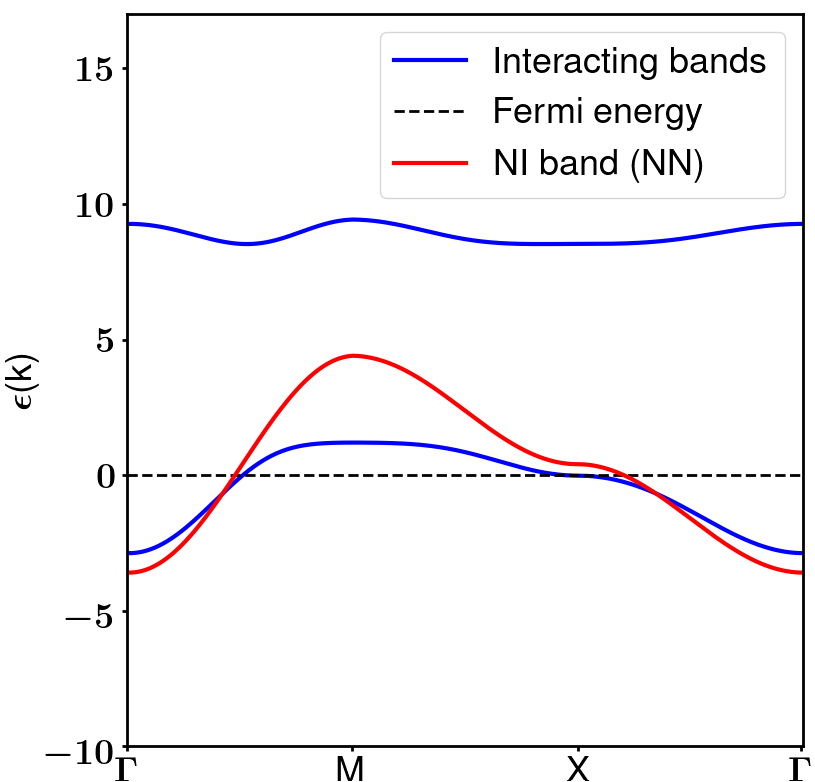}
\caption{Roth Minimization.}
\end{subfigure}
\end{subfigure}
\caption{Bands along high symmetry points with $t=1, U=8t, T=0, n=0.8$. The non interacting (NI) band (red line) is splitted into 2 Hubbard bands obtained for every solutions with the composite operator method (blue lines). Therefore at half filling the chemical potential lies in between the two Hubbard bands and we get a Mott insulator. (a) One of the two solutions obtained with Pauli minimization (using Eq. (\ref{PauliC012}) instead of the p parameter in Eq. (\ref{selfcon})). It has no renormalization from the interaction since the shape of the Hubbard bands is similar to the non-interacting band. (b) The second solution obtained with the Pauli minimization. It has a minimum at M=($\pi, \pi$) and exhibits two hole pockets. The two dispersions obtained from the Pauli minimization we obtain are analogous to Ref. \citep{avella_hubbard_1998} (c) The solution obtained with the Roth minimization. We observe a flattening of the bands around X=$(\pi,0)$.}
\label{Fig2:bandsNN}
\end{figure*}

\begin{figure}[h]
\begin{minipage}{0.2\textwidth}
\includegraphics[scale=0.41]{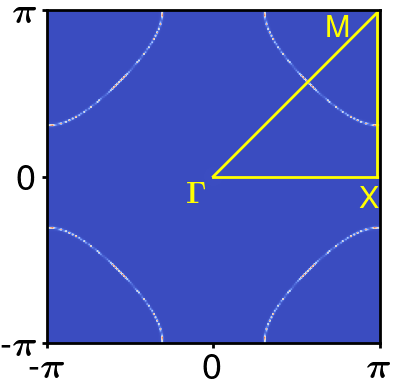}
\includegraphics[scale=0.41]{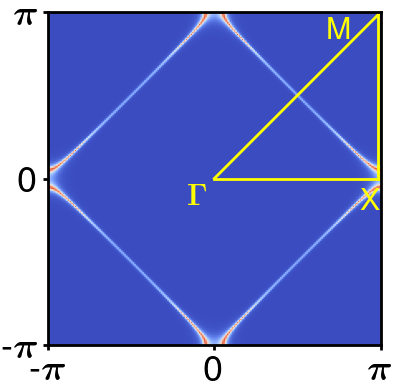}
\end{minipage}
\hspace{0.1mm}
\begin{minipage}{0.2\textwidth}
\includegraphics[scale=0.41]{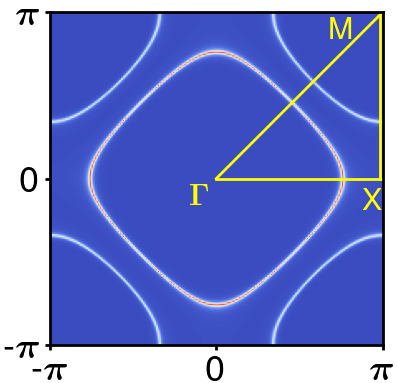}
\includegraphics[scale=0.41]{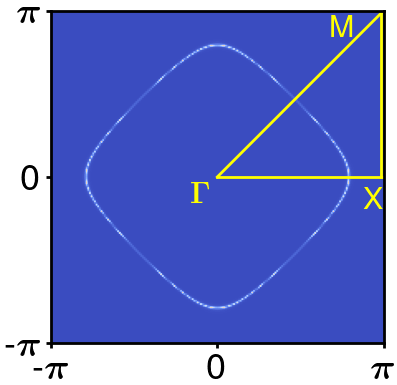}
\end{minipage}
\hspace{0.1mm}
\begin{minipage}{0.01\textwidth}
\includegraphics[height = 4.5cm,width=0.8cm]{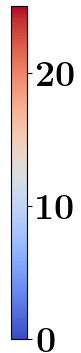}
\end{minipage}
\caption{Fermi Surfaces obtained from the spectral function (imaginary part of Eq. (\ref{Gandkappa})) at $\omega=0$, associated to the bands depicted in (\ref{Fig2:bandsNN}). The parameters are $t=1, n=0.8, U=8t$ and $T=0$. top left: COM1, top right: COM2, bottom left: Roth, bottom right: Non interacting (tight binding)}
\label{fig3:FSNN}
\end{figure}

The parameter $e$ contains correlations between neighboring composite operators and will mainly re-normalize the chemical potential because it always appear in M in front of a $\delta_{ij}$.  The $p$ parameter contains charge-charge, spin-spin and pair-pair correlations and will affect the bandwidth.

We now assume translational invariance and lattice inversion and rotation ($C_4$) symmetries in order to take n, p and e as constants. 
Under these assumptions the coefficients of the $\textbf{M}$ matrix become

\begin{equation}
\hspace{-0.1cm}\begin{aligned}
m_{ij}^{11}=& \hspace{-0.05cm}  -\mu \left(1-\frac{n}{2} \right)\delta_{ij} - \delta_{ij} 4 t e - t \alpha_{ij}^1 (1-n + p) \\
m_{ij}^{12}=&  \delta_{ij}4 t e - t \alpha_{ij}^1 \left(\frac{n}{2}-p \right) \\
m_{ij}^{22}=& -(\mu - U) \frac{n}{2} \delta_{ij} - \delta_{ij} 4 t e  - t \alpha_{ij}^1 p
\end{aligned}
\end{equation}

And the $\textbf{E}$ matrix defined by Eq. (\ref{EMIS:E}) is given by
\begin{equation}
\textbf{E}_{ij}=
\begin{pmatrix}
\frac{2}{2-n} m_{ij}^{11} & \frac{2}{n} m_{ij}^{12} \\
\frac{2}{2-n} m_{ij}^{12} & \frac{2}{n} m_{ij}^{22}
\end{pmatrix}
\end{equation}

This matrix is diagonalizable and Eq. (\ref{CfctE}) allows to express the correlation function $\textbf{C}_{ij\sigma}=\langle \psi_{i\sigma} \psi_{j\sigma}^\dagger \rangle$ as a function of its eigenvalues. Thus in order to close the system and be able to solve it self-consistently we need to express the parameters in the $\textbf{M}$ matrix as a function of the correlation functions. A diagram of the self-consistent loop is given in Fig. \ref{Fig1:selfcon} for clarity.

Since $e$ and $n$ are one-body parameters, therefore they can directly be expressed as
\begin{equation}
\begin{aligned}
n=&2(1 - C^{11}_{0} - 2 C^{12}_{0} - C^{22}_{0} ) \\
e=& C^{11}_{ij} - C^{22}_{ij}
\end{aligned}
\end{equation}

In this equation $\textbf{C}_{0} = \textbf{C}_{ii}$ are constants by translational invariance. and $\textbf{C}=\textbf{C}_{ij}=\textbf{C}_{i-j}$ is treated as e and p using lattice inversion and rotation ($C_4$) symmetry, so it is also a constant but different from $\textbf{C}_0$. 

Expressing $p$ as a function of the correlations functions is not so direct since $p$ is composed of two-bodies operators while $e$, $n$ and $C$ are one-body operators. A full detailed computation of p using Roth's decoupling scheme can be found in appendix D.

Finally, the expressions for the parameters appearing in $\textbf{M}$ and $\textbf{I}$ as a function of the $C^{nm}_{ij}=\langle \psi^n_i (\psi_j^m)^\dagger \rangle$ are the following

\begin{equation}
\begin{cases}
e &=C^{11} - C^{22} \\
n&=2(1-C^{11}_0 - 2 C_0^{12} - C_0^{22} ) \\
p&= \frac{n^2}{4} - \frac{\rho_1}{1-\phi^2} - \frac{\rho_1}{1-\phi} - \frac{\rho_3}{1+\phi}
\end{cases}
\label{selfcon}
\end{equation}

With 
\begin{equation}
\begin{cases}
\phi &= - \frac{2}{2-n}(C^{11}_0 + C^{12}_0) + \frac{2}{n}(C^{12}_0+C^{22}_0) \\
\rho_1 &= \frac{2}{2-n}(C^{11}+C^{12})^2 + \frac{2}{n}(C^{22}+C^{12})^2 \\
\rho_3&=\frac{4}{n(2-n)}(C^{11}+C^{12})(C^{22} + C^{12})
\end{cases}
\end{equation}
\\
With the self consistent equations Eq. (\ref{selfcon}) on the parameters n, e and p we just closed the system. In Fig. \ref{Fig1:selfcon}, we represented the self-consistency pattern. Starting from initial guess for e, n and p, we compute the $\textbf{M}$ and $\textbf{I}$ matrices. We can then obtain $\textbf{E}$ and diagonalize it. Then, using Eq. (\ref{CfctE}), we can express the correlation functions from its eigenvalues. Finally, using the self consistent Eq. (\ref{selfcon}), we compute again e, p and n. We stop when $f(x)-x<\delta$ where $x=(e,p,\mu)$ and f are given in Eq. (\ref{selfcon}).  We chose $\delta=10^{-8}$. Once the system converges, we use the parameters $(e,p,\mu)$ to compute the electronic Green's function 

\begin{equation}
\begin{aligned}
G_{k}(\omega)=&S_{k}^{11}(\omega)+S_{k}^{12}(\omega)+S_{k}^{21}(\omega)+S_{k}^{22}(\omega) \\
=& \sum \limits_{l=1}^2 \frac{(\kappa^l)^{11}_{k}+(\kappa^l)^{12}_{k}+(\kappa^l)^{21}_{k}+(\kappa^l)^{22}_{k}}{\omega-\epsilon^l_{k}+i 0^+}
\end{aligned}
\label{Gandkappa}
\end{equation}

\begin{figure*}[ht]
\begin{center}
\includegraphics[width=1.\linewidth,height=9cm]{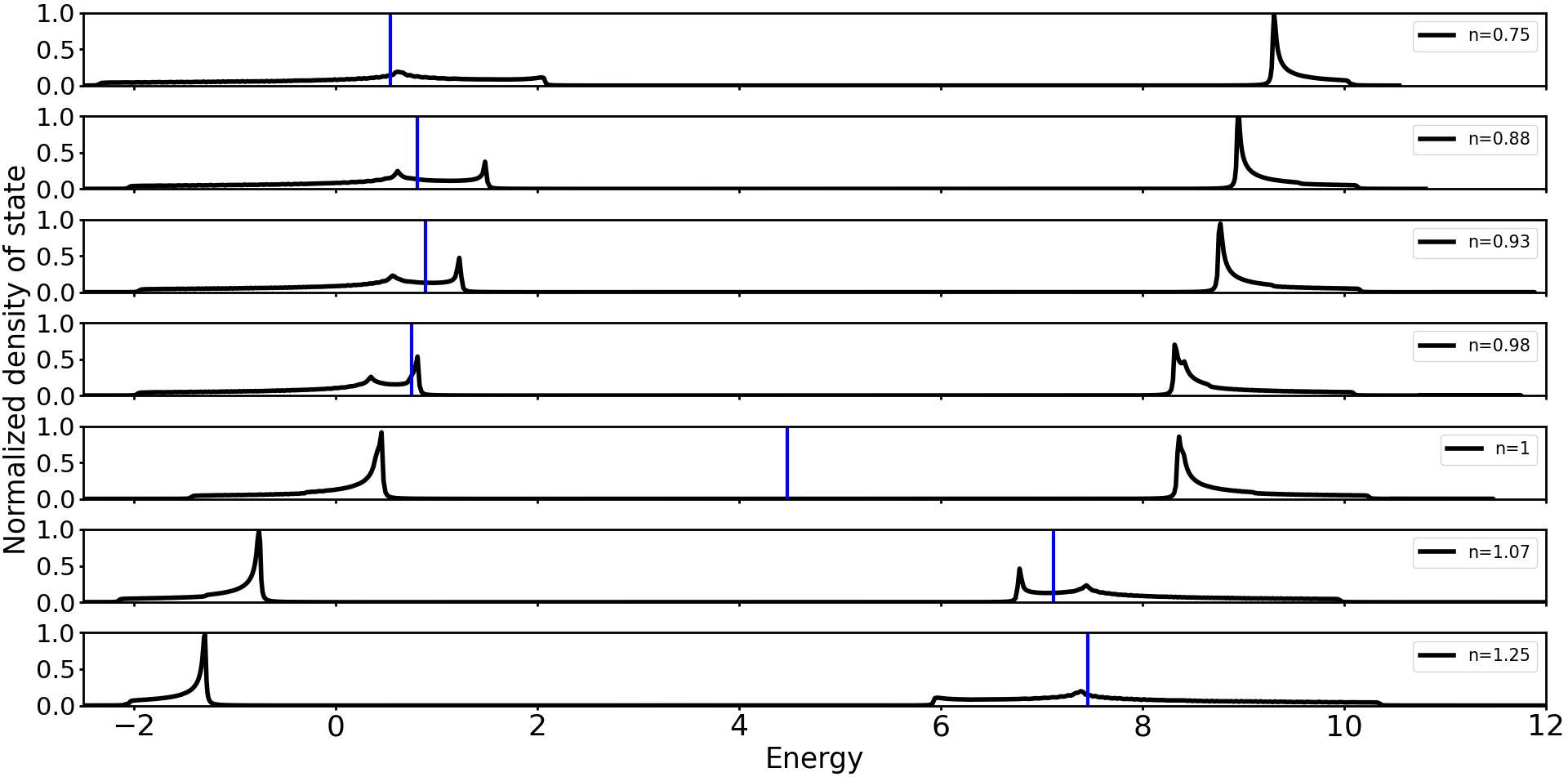}
\end{center}
\caption{Density of states in energy with parameters $t=1, U=8t$ and $T=0$, using Roth minimization with nearest neighbors. From top to bottom the filling is respectively: n=0.75, 0.88, 0.93, 0.98, 1 (half filling), 1.07 and 1.25. The blue line on each panel corresponds to the chemical potential.}
\label{Fig4:DOS}
\end{figure*}

Where the $\boldsymbol{\kappa}$ act as spectral weights and are defined in Eq. (\ref{kappa}) and $\epsilon^{1}$, $\epsilon^{2}$ are the eigenvalues of the $\textbf{E}$ matrix, and can be built from the parameters $(e,p,\mu)$.

\section{\label{sec:II} Nearest-neighbors study}

This self-consistent scheme has a drawback. With the numerical solution of the self consistent equations we obtain a non vanishing $C^{12}_0$. However, analytically this correlation function is zero because of the Pauli principle.

\begin{equation}
\begin{aligned}
C_0^{12}&=\langle \xi_{i \sigma} \eta^\dagger_{i \sigma }\rangle \\
&= \langle c_{i \sigma} c^\dagger_{i\sigma} n_{i\bar{\sigma}} + c_{i \sigma} c^\dagger_{i \bar{\sigma}} c^\dagger_{i\sigma}c_{i \bar{\sigma}} n_{i \bar{\sigma}} \rangle \\
&= \langle -  c_{i \sigma} (c^\dagger_{i \bar{\sigma}})^2  c^\dagger_{i \sigma}   (c_{i \bar{\sigma}})^2  \rangle \\
&=0 
\end{aligned}
\label{PauliC012}
\end{equation}

This numerical violation of the Pauli principle gets smaller as U becomes larger than t \cite{avella_hubbard_1998}.

It is possible to solve a different set of self-consistent equations by imposing $C^{12}_0=0$ \cite{avella_composite_2012} as the third self consistent equation instead of $p$ in Eq. (\ref{selfcon}). From now on, we will call the minimization with $p$ the ``Roth minimization" since it uses the decoupling formalized by L. Roth and the minimization with $C^{12}_0$ the ``Pauli minimization".

In the following section we consider both minimizations and discuss the bands and Fermi surfaces obtained for each solution we found. We compare to the non interacting tight-binding model, and study the particle-hole symmetry, the Luttinger theorem as well as superconductivity.

\subsection{Comparison of Pauli and Roth minimization}
By varying the initial conditions and using a minimizer, we isolate two distinct solutions with the Pauli minimization and one unique solution with the Roth minimization. Following notations from Ref \cite{avella_hubbard_1998}, we call these three solutions COM1, COM2, and Roth solutions. In Fig. \ref{Fig2:bandsNN}, we plot the bands along high symmetry points for these solutions. 

These bands correspond to the eigenvalues of the $\mathbf{E}$ matrix and act as the poles of the electronic Green's function (cf Eq. (\ref{Gandkappa})). The solutions have two bands associated with the two eigenvalues of the energy matrix, split by the interaction strength $U$. The COM2 and Roth solutions exhibit Mott insulator physics at half-filling as the chemical potential resides between the two bands. In contrast, contrary to the conventional understanding, the COM1 solution represents a metallic phase at half-filling for $t=1$, $U=8t$. Consequently, COM1 cannot be deemed a physically viable solution for the Hubbard model in strong coupling regimes.

The COM2 solution is very different from the non interacting case (Fig. \ref{Fig2:bandsNN}b), but always presents two holes pockets, leading to two Fermi Surfaces in Fig. \ref{fig3:FSNN}. This is unexpected, as this has never been observed by ARPES experiments for strongly correlated materials such as cuprates where this treatment of the Hubbard model is relevant. However, in Ref. \citep{avella_composite_2012} the basis has been extended to take into account dynamical corrections of the self-energy (cf appendix A for details). The lifetime of the second hole pocket is then computed and happens to be small, which can explain why it is not observed experimentally. Finally, the Roth solution (Fig. \ref{Fig2:bandsNN}c) has a very different shape from the non-interactive solution. It also has the advantages of presenting only one hole pocket and a maximum at $(\pi, \pi)$. 

Note we plotted the bands we obtained at half filling for the Roth solution in appendix E, on Fig. \ref{Fig14}. Close to half filling (around 3\% hole doping), the Roth solution exhibits a second small hole pocket at $(\pi, \pi)$ (In appendix E we plot the bands and Fermi Surface of Roth decoupling at 2\% hole doping in Fig. \ref{Fig13}). This second hole pocket around $(\pi, \pi)$ may be the consequence of our paramagnetic assumption $\langle n_{i \up} \rangle = \langle n_{i \down} \rangle = \frac{n}{2}$. It appears around half filling where we know the antiferromagnetic phase dominates \cite{bergeron_breakdown_2012}. The wave-vector $(\pi, \pi)$ is also associated with antiferromagnetism, so this second hole pocket might be an instability of the system because we neglected it.

For comparison, in the Hubbard I approximation (the first and simplest approximation developed by Hubbard in Ref. \citep{Hubbard_electron_1963}) a simple factorization procedure of the 2-bodies Green functions is used. According to Ref. \cite{avella_hubbard_1998}, Hubbard I approximation in the composite operator framework would be equivalent to set 
\begin{equation}
\begin{aligned}
e&=0 \\
p&=\frac{n^2}{4}
\end{aligned}
\end{equation}

It is known that the Hubbard I approximation does exhibit a Mott-insulator transition as long as $U\neq 0$ (cf Ref. \cite{Hubbard_electron_1963}). However this approximation only treat the currents partially compared to the Composite operator Method. Therefore the composite operator method is a more refined approximation than
Hubbard I.

\begin{figure}[h]
\begin{center}
\begin{subfigure}{0.49\linewidth}
\includegraphics[width=4cm,height=3cm]{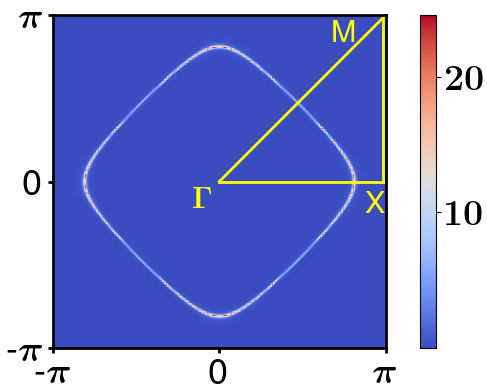}
\end{subfigure}
\begin{subfigure}{0.49\linewidth}
\includegraphics[width=4cm,height=3cm]{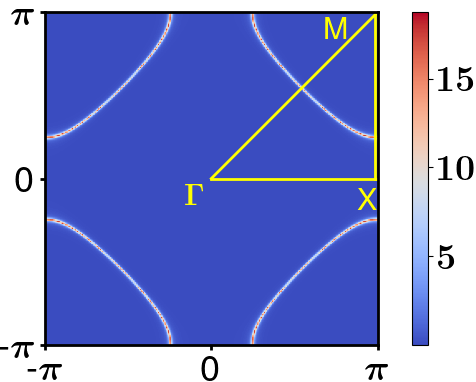}
\end{subfigure}
\caption{Fermi Surface obtained with Roth solution, at n=0.7 and n=0.9.The chemical potential or the Fermi energy coincides with the Van Hove singularity for some specific electron density (here n=0.8). Below the singularity the Fermi surface is centered around $(0,0)$, and above the singularity it is centered around $(\pi,\pi)$}
\label{fig_van_hove_fs}
\end{center}
\end{figure}

It is also instructive to consider the density of states. In Fig. \ref{Fig4:DOS}, we plot it as a function of energy for several doping with the Roth solution. Only Roth solution is considered because its Fermi surface is the closest to the non-interacting one, but similar behaviors are expected for COM1 and COM2. The density of states has been computed from the spectral function using the following formula
\begin{equation}
\mathcal{D}(\omega)=\frac{1}{N_k^2} \sum \limits_{k_x, k_y=1}^{N_k} \left(-\frac{1}{2\pi} \right) Im(G(\textbf{k},\omega))
\label{DOS}
\end{equation}
Where $N_k$ denotes the number of considered points for sampling $k_x$ and $k_y$. At half filling we do not have any states at the Fermi energy, since the model leads to an insulator for this doping. Around the gap two peaks can be distinguished. For every doping except half-filling, a third peak is observable and corresponds to the Van Hove singularity. It is associated to a flattening of the bands, meaning there are a lot of states associated to this energy. In term of Fermi surface for a square lattice the Van Hove singularity corresponds to the doping below which the Fermi Surface is centered on $(0,0)$ and above which it is centered on $(\pi, \pi)$, as shown in Fig. \ref{fig_van_hove_fs}. On Fig. \ref{Fig2:bandsNN}, this flatness of the bands can be found for every solutions around $(\pi,0)$. The Fermi energy is exactly at the Van Hove singularity for the Roth solution at n=0.8 or n=1.2. On Fig. \ref{Fig2:bandsNN}, the Roth solution is plotted exactly at this doping and we can check the flat band around $(\pi,0)$ lies exactly at the Fermi energy (corresponding to the dotted black line). 

Regarding the Mott transition, we see at half filling no quasiparticle peak is observed around the Fermi energy. Instead, the density of states is closer to what was observed with determinantal quantum Monte Carlo simulation (DQMC) in Ref. \cite{osborne_fermi-surface_2020}. DQMC is a stochastic algorithm which allows under some limitations to perform direct studies of complex condensed matter problems. As predicted by DQMC in Ref \cite{osborne_fermi-surface_2020}, at high doping we have only one peak corresponding to a Van Hove singularity (Fermi liquid behavior), and when approaching the Mott transition a transfer of spectral weight occurs, changing the density of states, without creating a quasiparticle peak at the Fermi energy at half filling. Therefore in this regime where $U \gg t$ the density of states of the lower and upper Hubbard bands are the only contribution.

\begin{figure*}[ht]
\begin{subfigure}{\linewidth}
\begin{center}
\begin{subfigure}{0.49\linewidth}
\includegraphics[width=8.9cm, height=5cm]{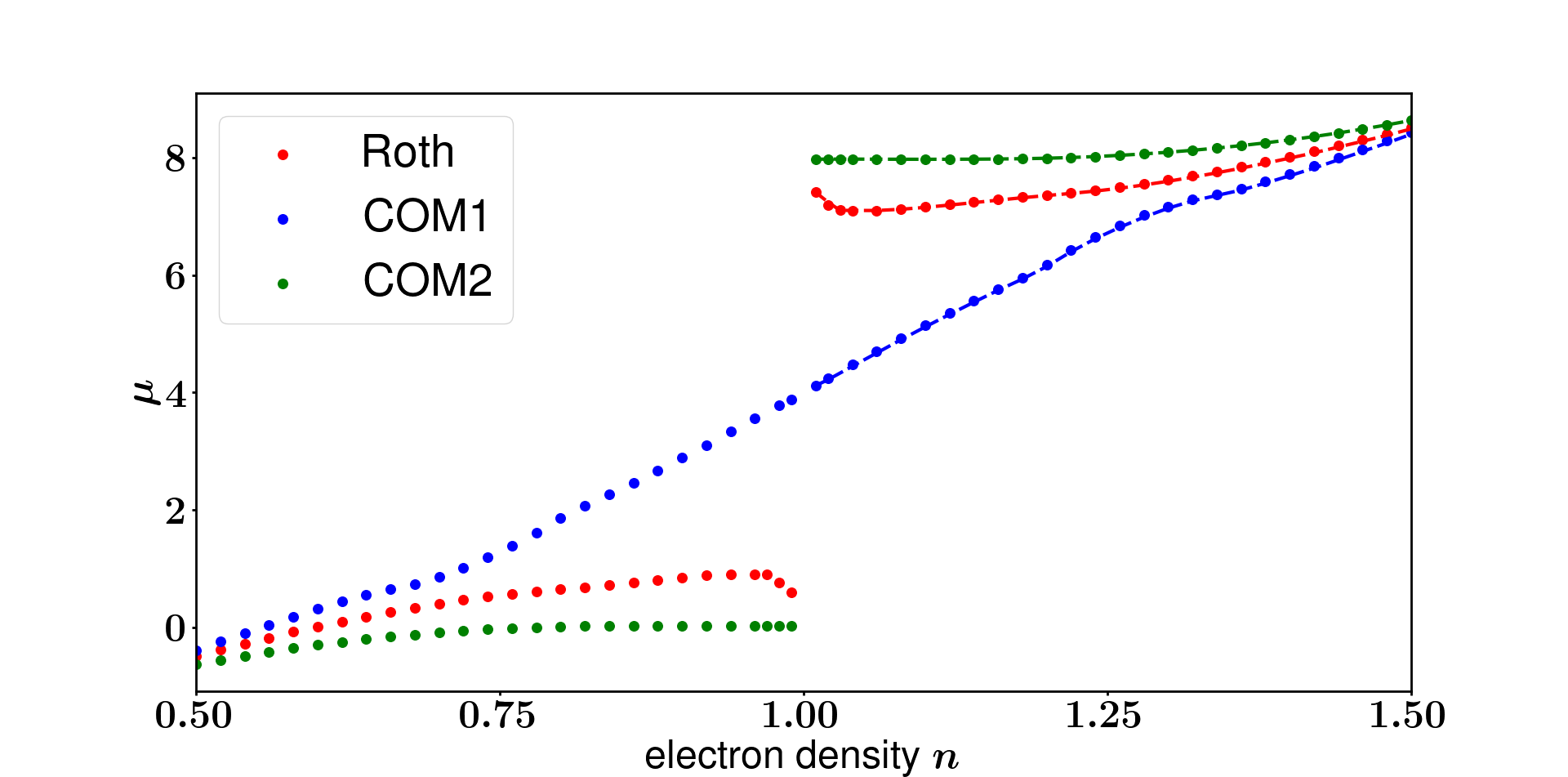}
\caption{Chemical potential}
\end{subfigure}
\begin{subfigure}{0.49\linewidth}
\includegraphics[width=8.9cm, height=5cm]{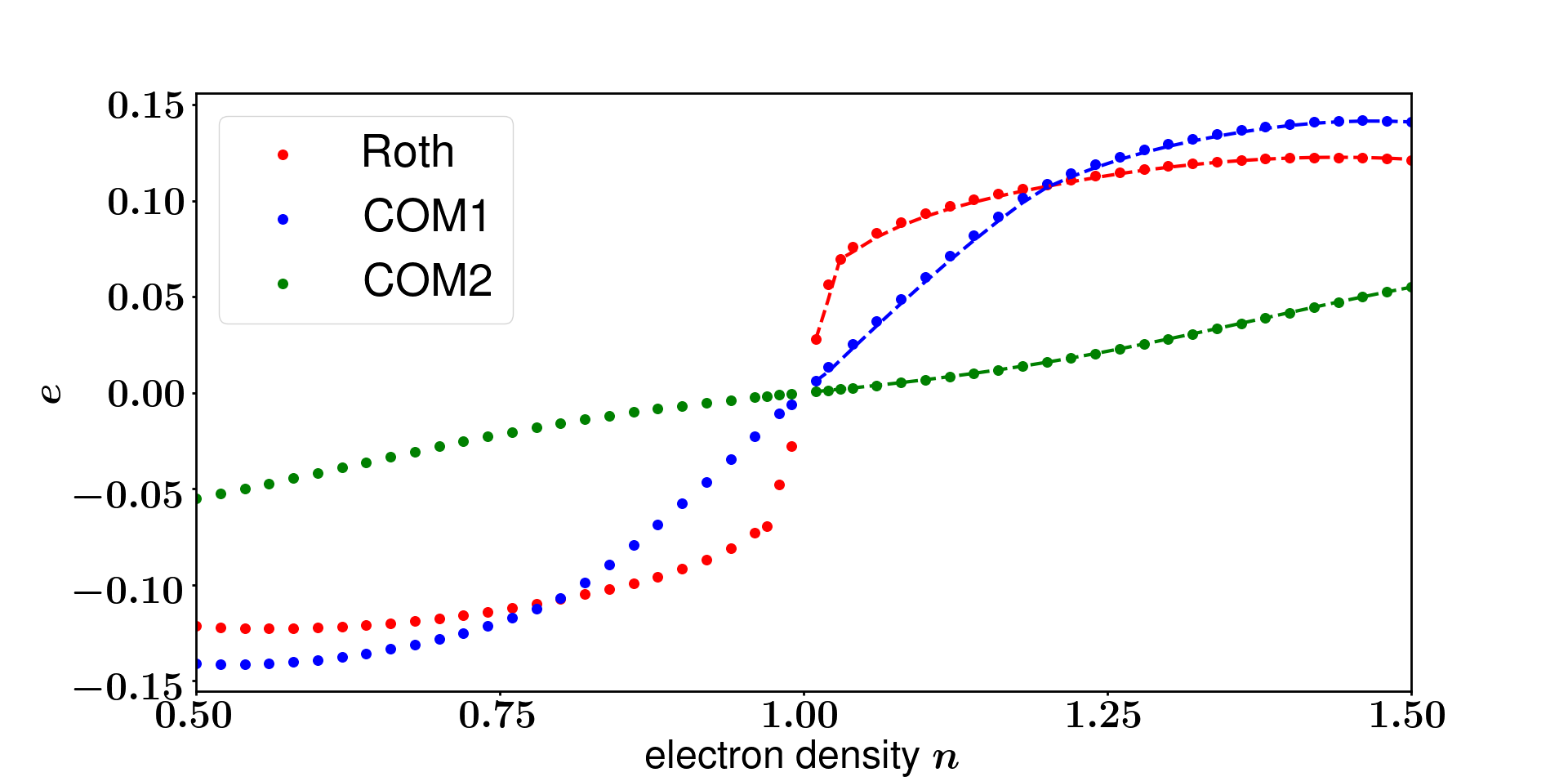}
\caption{Parameter e}
\end{subfigure}
\\
\begin{subfigure}{0.49\linewidth}
\centering
\includegraphics[width=8.9cm, height=5cm]{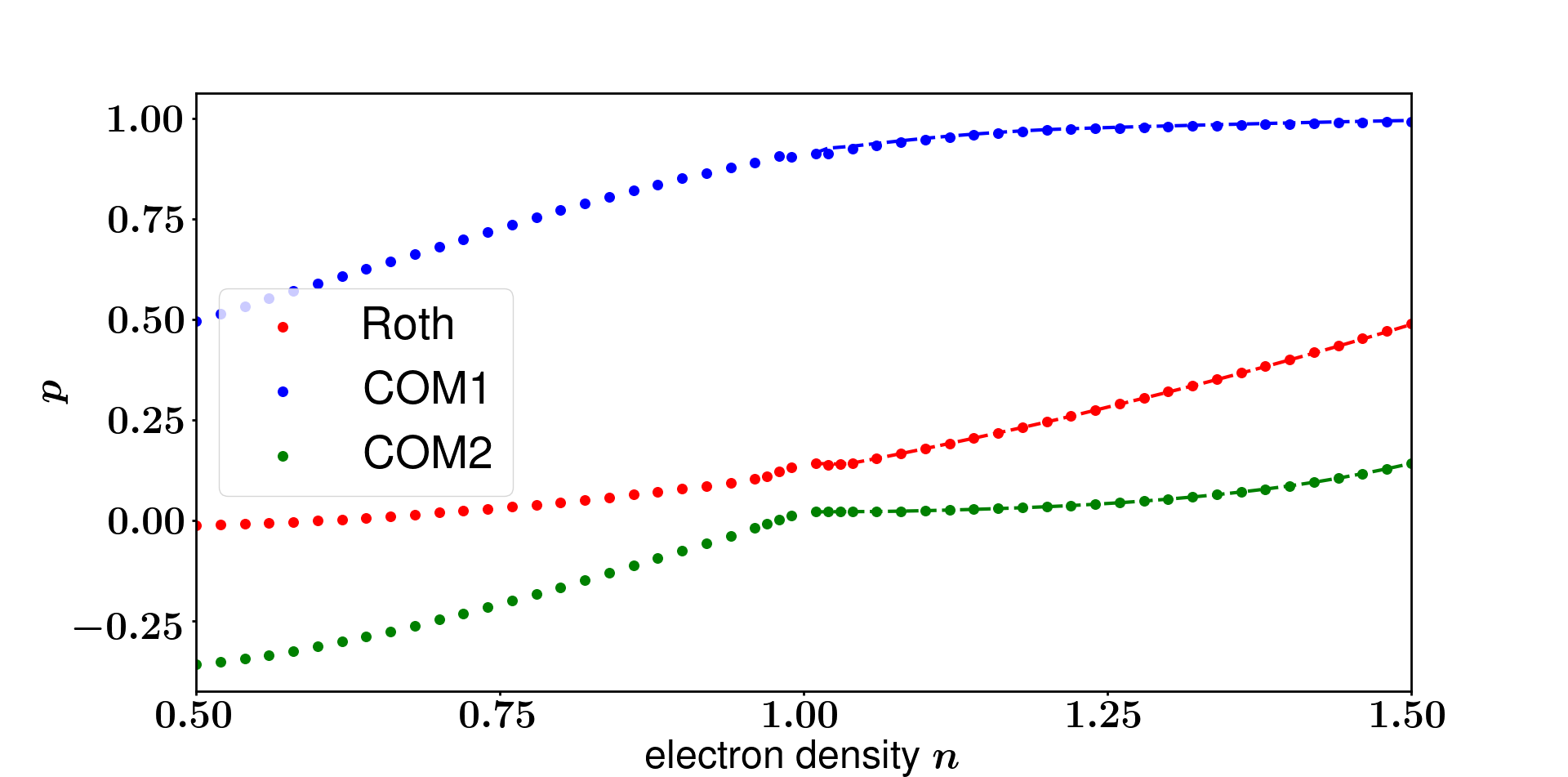} 
\caption{Parameter p}
\end{subfigure}
\begin{subfigure}{0.49\linewidth}
\includegraphics[width=8.9cm, height=5cm]{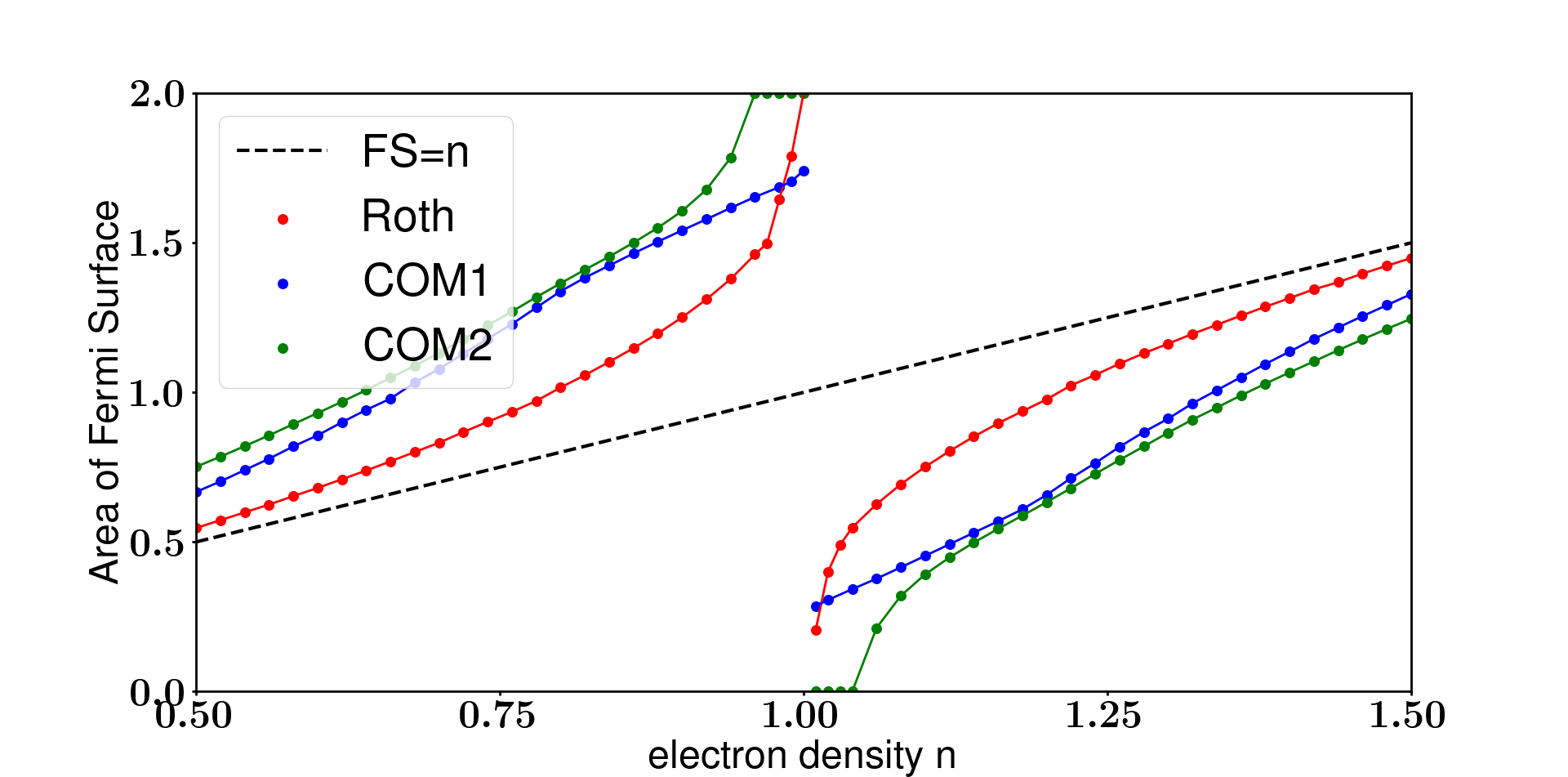} 
\caption{Luttinger Violation}
\end{subfigure}
\end{center}
\end{subfigure}
\caption{ (a), (b) and (c): parameters as a function of doping for each solution. The dash lines in the electron dope ($n>$1) region are the particle-hole symmetric (\ref{ph_param}) of their equivalent in the hole doped region. Bottom right: area of Fermi surface as a function of doping. It is computed using Eq. (\ref{SpecFuncExpr}). We observe a violation of Luttinger theorem for all solutions. All of these figures are obtained at $T=0$K ($\beta \approx 10^5 t$) and $U=8t$.}
\label{Fig5:phLuttNN}
\end{figure*}

\begin{figure}[h]
\begin{center}
\includegraphics[width=9cm,height=6cm]{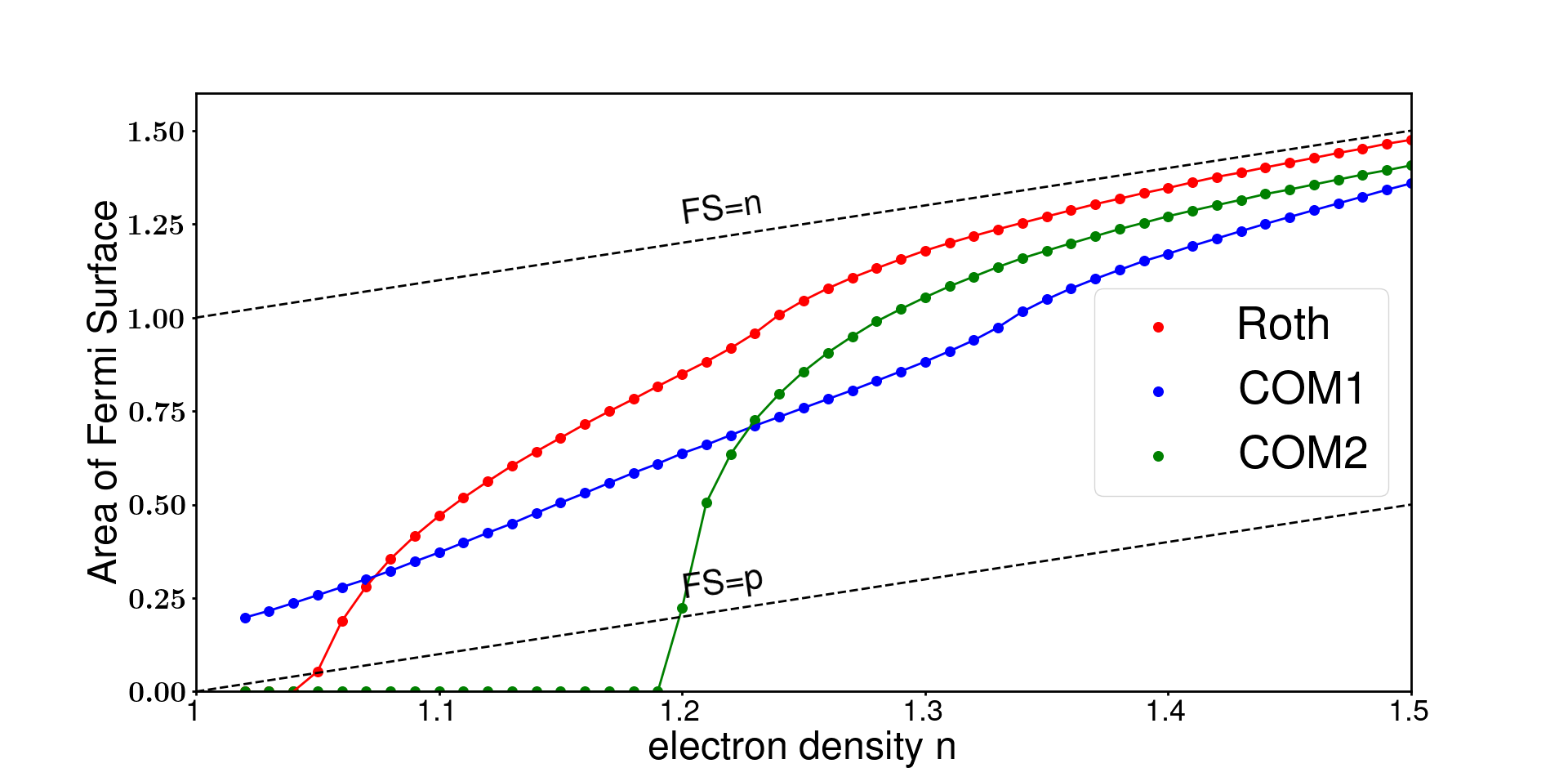}
\caption{Luttinger violation as a function of electron density. The parameters are with $\beta=2t$ and U=10t. The solutions are proportional to the electron density only asymptotically (the curves and the black dotted line ``FS=n" associated to Luttinger theorem are not parallel). The system is in a ``Luttinger breaking phase". These behaviors are similar to what was obtained in Ref. \cite{osborne_fermi-surface_2020} from Determinantal Monte Carlo.}
\label{Fig:LuttTrivedi}
\end{center}
\end{figure}

\subsection{Particle-hole symmetry and Luttinger theorem}

\subsubsection{Particle-hole symmetry}

It is well known that the Hubbard model with nearest neighbour hoppings only is particle-hole symmetric. This symmetry exchanges particles and holes with the transformation
\begin{equation}
c_{i \sigma} \rightarrow (-1)^i c_{i \sigma}^\dagger \ \ \ \ \ \ \ \ \ \ \ \ \ \ \ c_{i \sigma}^\dagger \rightarrow  (-1)^i c_{i \sigma}
\end{equation}

We could also have taken another convention for this transformation without any $(-1)^i$, as long as we change t to -t to keep the Hamiltonian invariant.
Applied to the composite operators one can show that it becomes
\begin{equation}
\eta_{i \sigma} \rightarrow (-1)^i \xi_{i \sigma}^\dagger \ \ \ \ \xi_{i \sigma} \rightarrow (-1)^i \eta_{i \sigma}^\dagger
\label{compos_ph_sym}
\end{equation}


Let us study the behaviour of the parameters $\mu$, p and e under the particle-hole symmetry. With the paramagnetic assumption, the particle-hole transformation is rewritten 
\begin{equation}
\begin{aligned}
\langle c_{i \sigma}^\dagger c_{i \sigma} \rangle 
&\rightarrow \langle c_{i \sigma} c_{i \sigma}^\dagger \rangle \\
\Leftrightarrow \frac{n}{2} &\rightarrow 1- \frac{n}{2}
\end{aligned}
\end{equation}
Hence applying the particle-hole transformation on electronic filling gives $n \rightarrow 2-n$. The transformation changes p and e (Eq. (\ref{paramdef})) as follow (details are given in appendix G)
\begin{equation}
\begin{aligned}
e(2-n)=& -e(n) \\
p(2-n)=& p(n) + (1-n) \\
\mu(2-n) = & U - \mu(n)
\end{aligned}
\label{ph_param}
\end{equation}

The relation of the chemical potential can be obtained by using the fact that the Hubbard Hamiltonian must stay invariant under this symmetry for nearest-neighbour hoppings. 
Finally, applying this transformation on the composite bands $\epsilon^1$ and $\epsilon^2$ leads to 
\begin{equation}
\begin{aligned}
\epsilon^1_k(2-n) =& -\epsilon^2_{k+(\pi,\pi)}(n) \\
\epsilon^2_k(2-n)=& -\epsilon^1_{k+(\pi, \pi)}(n)
\end{aligned}
\label{bandsph}
\end{equation}

In Fig. \ref{Fig5:phLuttNN}a, \ref{Fig5:phLuttNN}b and \ref{Fig5:phLuttNN}c we plot e, p and $\mu$ as a function of doping for the three solutions we studied (COM1, COM2 and Roth). The dashed-lines on the electron-doped region is the value the parameter must have to satisfy the particle-hole relations Eq. (\ref{ph_param}). We see that particle-hole symmetry is respected for every parameters for the three solutions. The chemical potential presents a jump of the order of U at half filling for the COM2 and Roth solutions. This is because the lower Hubbard band is filled and the upper Hubbard band starts to be occupied at half filling. COM1 does not exhibit this feature.

We obtain a different result from Ref. \cite{avella_hubbard_1998}. It is possible to have a particle-hole symmetric solution which violates Pauli principle. This is indeed the case of the Roth solution which has a non vanishing $C^{12}_0$ despite the fact it is zero analytically. The solution is particle-hole symmetric as long as $C^{12}_0$ is not put to zero by hand in the self-consistent equations it must appear both in the equation of n and in the $\phi$ term in the equation of p.

\subsubsection{Luttinger theorem}

We now turn our attention to the Luttinger theorem. This theorem states that the volume enclosed by the Fermi surface is proportional to the electron density \cite{luttinger_fermi_1960}. The regime of validity of Luttinger theorem is still a very debated topic \cite{heath_necessary_2020} \cite{dzyaloshinskii_consequences_2003} \cite{skolimowski_luttingers_2022}. To compute the volume enclosed by the Fermi surface, we need to remember the relation between the composite and electronic Green's function Eq. (\ref{Gandkappa}) The $\boldsymbol{\kappa}$ act as spectral weights of the electronic Green's function.
The Fermi surface is given by the imaginary part of this electronic Green's function at the Fermi energy

\begin{equation}
\begin{aligned}
A_{k}(\omega=0)=&-\frac{1}{\pi} Im(G_{k}(\omega = 0))\\=& \sum \limits_l [ (\kappa^l)^{11}_{k}+(\kappa^l)^{12}_{k}+(\kappa^l)^{21}_{k}+(\kappa^l)^{22}_{k}] \delta(\epsilon_{k}^l)
\end{aligned}
\end{equation}

\begin{figure*}[ht]
\begin{center}
\begin{subfigure}{0.48\linewidth}
\centering
\includegraphics[width=7cm,height=6cm]{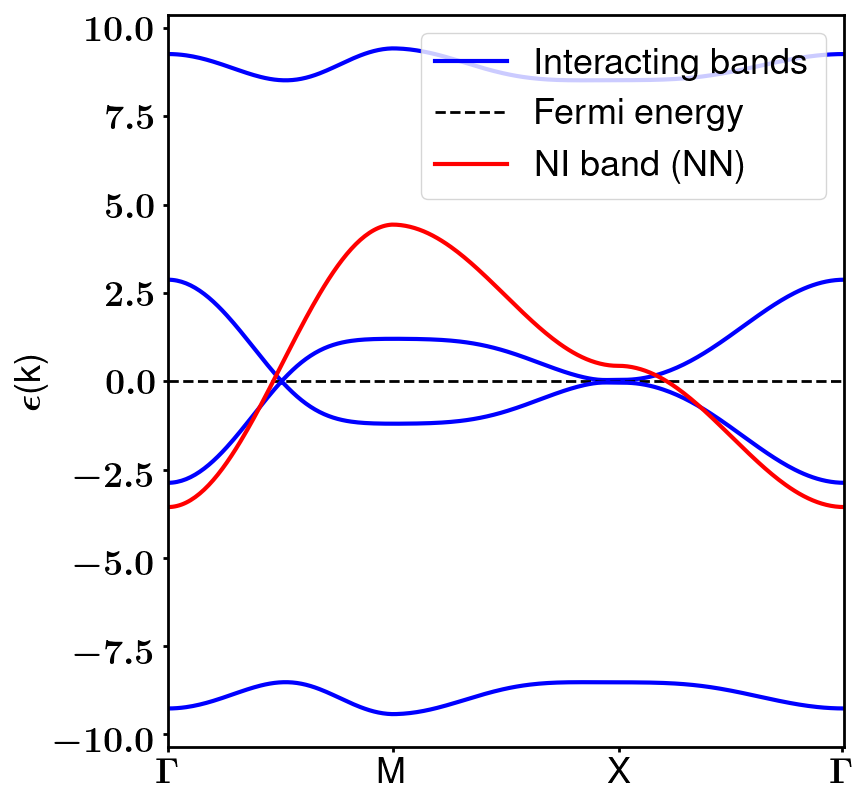}
\begin{picture}(0,0)
\put(-120,10){\includegraphics[height=2.3cm,width=3cm]{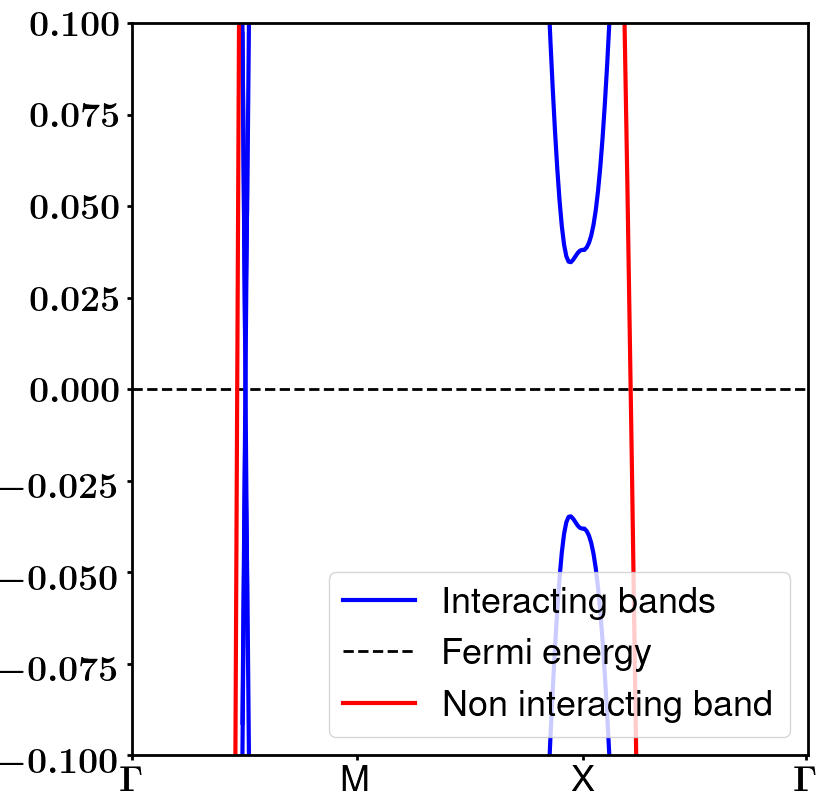}}
\end{picture}
\caption{Bands for the Roth solution}
\end{subfigure}
\begin{subfigure}{0.48\linewidth}
\centering
\includegraphics[width=7cm,height=6cm]{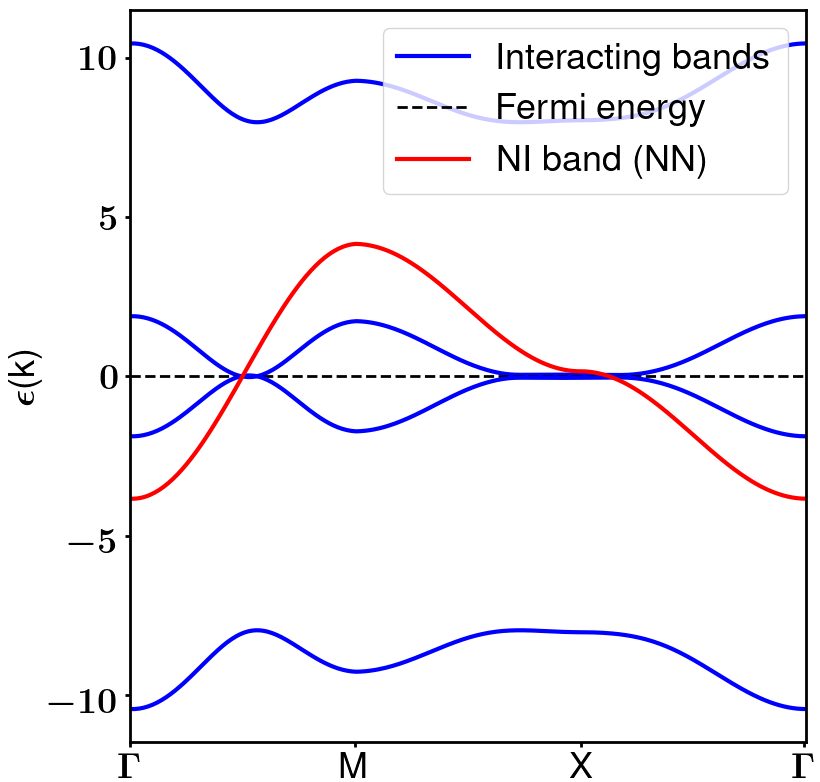}
\begin{picture}(0,0)
\put(-120,10){\includegraphics[height=2.3cm,width=3cm]{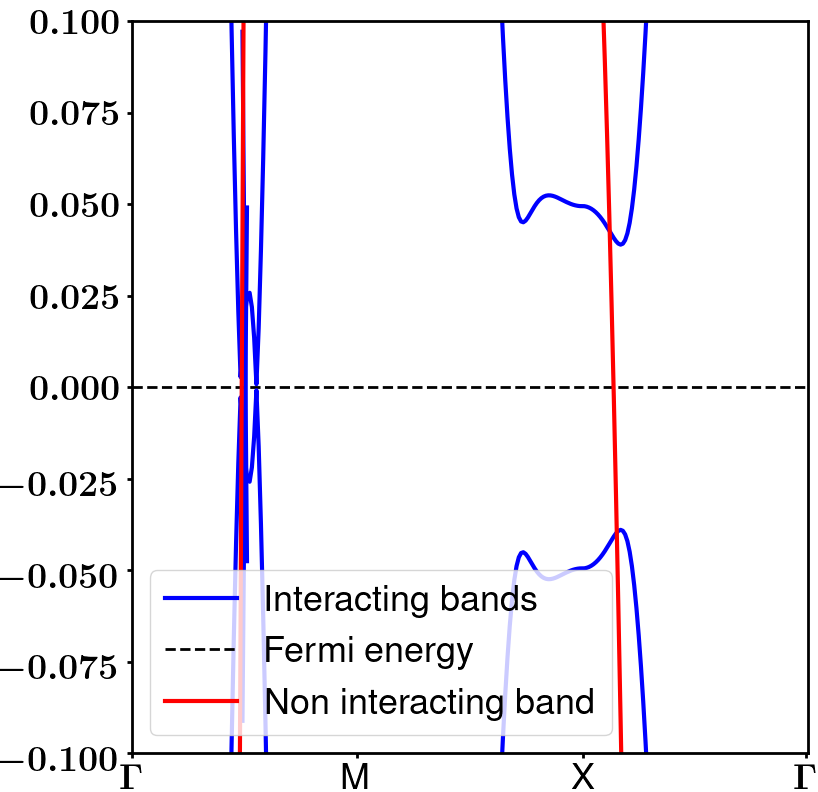}}
\end{picture}
\caption{Bands for the COM2 solution}
\end{subfigure} \\
\begin{subfigure}{0.48\linewidth}
\centering
\includegraphics[width=4cm,height=3cm]{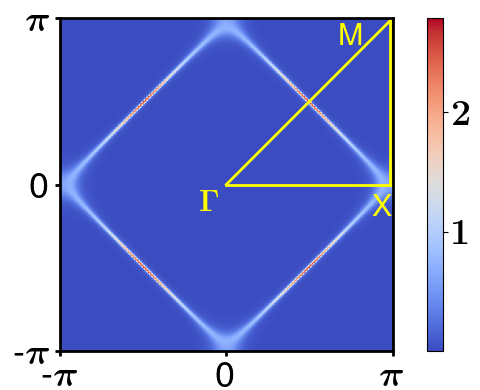}
\caption{Fermi Surface for the Roth solution}
\end{subfigure}
\begin{subfigure}{0.48\linewidth}
\centering
\includegraphics[width=4cm,height=3cm]{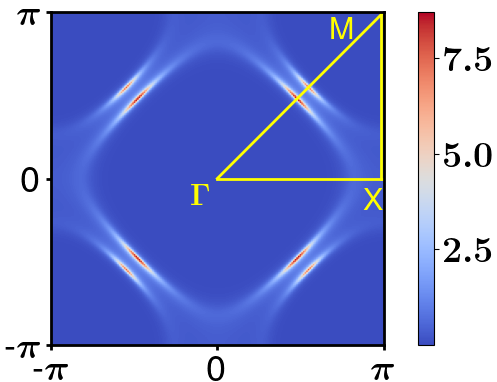}
\caption{Fermi Surface for the COM2 solution}
\end{subfigure}
\end{center}
\caption{Bands and Fermi Surface with superconductivity at $t=1$, $U=8t$ and $T=0$. The plots on the left correspond to the Roth solution at $n=0.8$, whereas the plot on the right correspond to COM2 at $n=0.9$. These are the respective doping at which superconductivity is maximum. We observe a doubling of the bands associated to the doubling of the basis. The insets on the top plots correspond to a zoom around zero energy of the bands: we see a gap opening at \textbf{k}=$(\pi,0)$. There is no gap opening between \textbf{k}=$(0,0)$ and \textbf{k}=$(\pi,\pi)$ because of d-wave symmetry. Both Fermi surfaces on the bottom plots have a loss of spectral weight around \textbf{k}=$(\pi,0)$.}
\label{fig6:SCvanHoeve}
\end{figure*}

Thus, the Fermi contour is the set of points associated to a vanishing $\epsilon_{ij}^l$.
Therefore a way to compute the enclosed area $A_{FS}$ is simply by considering the following equation

\begin{equation}
\begin{aligned}
A_{FS}^l=&\frac{1}{N^2}  \sum \limits_{k_x, k_y=1 }^N  \theta_H(\epsilon^l_k)
\end{aligned}
\label{SpecFuncExpr}
\end{equation}

Note in this last equation we do not have the sum over the two eigenvalues. Indeed, one must only consider the bands that are not completely filled or empty, so l has to been chosen accordingly. For instance, in the hole-doped regime the upper band is going to be empty, so l must correspond to the eigenvalue of the lower upper band. Fig. \ref{Fig5:phLuttNN}d reveals that the Luttinger theorem is violated. This violation is analogous to what was observed by Ref. \cite{osborne_fermi-surface_2020} using determinantal quantum Monte-Carlo simulations. In Fig. \ref{Fig:LuttTrivedi}, we plot Luttinger violation observed with the same parameters as in Ref. \cite{osborne_fermi-surface_2020} (U=10t and $\beta$=2). We obtain analogous results: while none of the solutions we considered are precisely similar to what was observed with determinantal quantum Monte-Carlo, we have the same overall behavior. Our curves are not parallel to the black dotted line representing the Luttinger theorem (where Fermi surface area equals electron density). The system is in a ``Luttinger breaking phase". Ref. \cite{osborne_fermi-surface_2020} claimed this phase is a consequence of a topological order because of the proximity to the Mott transition. Contrary to their results, Luttinger theorem with composite operators seems broken at every doping and verified only asymptotically at maximum and minimum doping.


\subsection{Superconductivity and Van Hove singularity}

\subsubsection{Method}

\begin{figure*}[ht]
\begin{center}
\begin{subfigure}{0.49\linewidth}
\includegraphics[width=8.9cm, height=6cm]{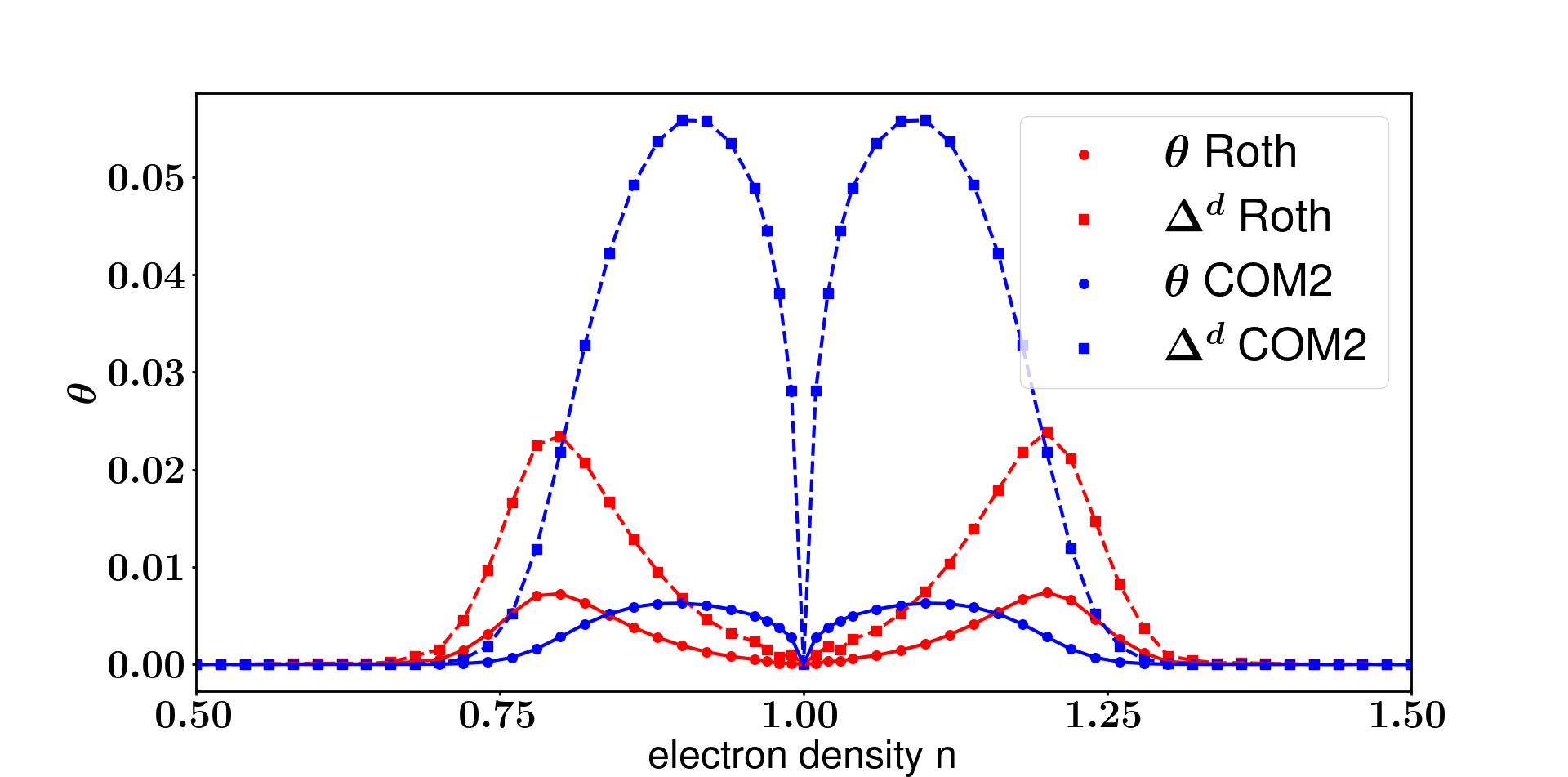}
\caption{$\theta$ and $\Delta^d$ parameters as a function of electron density $n$}
\end{subfigure}
\begin{subfigure}{0.49\linewidth}
\includegraphics[width=8.9cm, height=6cm]{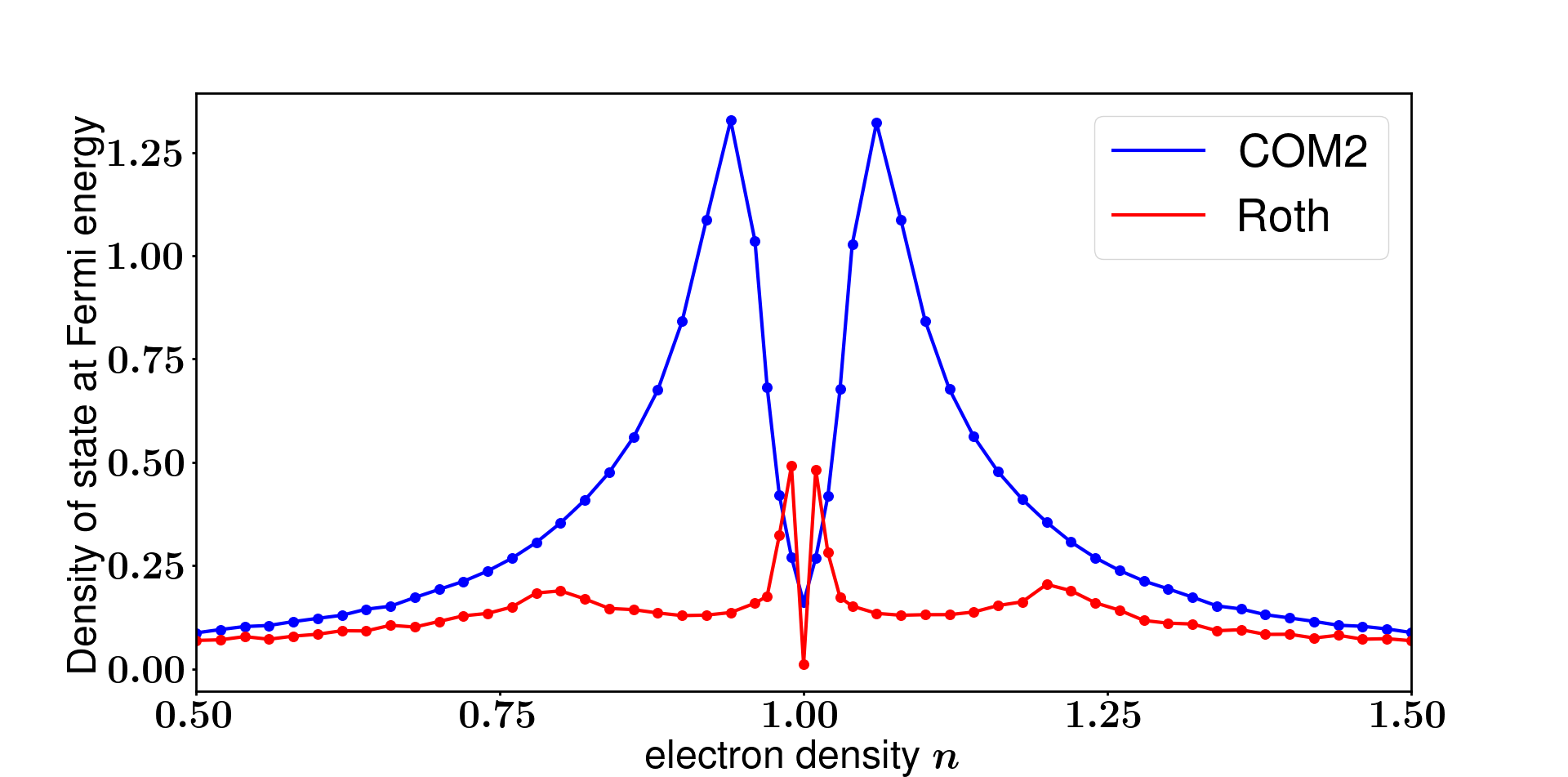}
\caption{Density of states at the Fermi Energy}
\end{subfigure} 
\end{center}
\caption{(a): Anomalous superconducting mean field parameter $\theta$ as a function of doping for Roth minimization and Pauli minimization (COM2 solution). $\theta$ satisfy particle-hole symmetry in both cases. The dashed lines corresponds to the d-wave superconducting order parameter $\Delta^d$. (b): Density of states at the Fermi energy with no superconductivity. We see a clear correlation between enhancement of the density of states and superconductivity. $\theta$ and $\Delta^d$ are maximum at the Van Hove singularity. It lies at n=0.8 for Roth and around n=0.9 for COM2, as shown in Fig. \ref{fig6:SCvanHoeve}.}
\label{fig7:theta(n)}
\end{figure*}

Superconductivity can be studied by extending the initial basis. The new spinor $\psi$ to consider is
\begin{equation}
\psi = \begin{pmatrix}
\xi_{i \sigma}\\
\eta_{i \sigma}\\
\xi_{i \bar{\sigma}}^\dagger\\
\eta_{i\bar{\sigma}}^\dagger
\end{pmatrix}
\end{equation}

The method described before remains the same except for the $\textbf{I}$ and $\textbf{M}$ matrix which are now $4 \times 4$ matrices. Since we are interested in cuprates, we consider only the case of d-wave superconductivity. Therefore, $\langle c_{i\up} c_{i\down} \rangle =0$. The expression of the M and I matrices can be found in appendix C. It is important to note that a new superconducting parameter, $\gamma_{ij} \theta_{ij}$ will now appear in the M matrix.

$\gamma_{ij}$ is a d-wave coefficient such that $\gamma_{i,i\pm \delta_y}=-\gamma_{i\pm\delta_x,i}=1$ and $\delta_x$/$\delta_y$ is the lattice constant along x/y axis. The parameter $\theta$ is given by
\begin{equation}
\theta_{ij}=\langle c_{i\sigma} c_{i\bar{\sigma}} n_{j \sigma} \rangle
\end{equation}

$\theta_{ij}$ can be expressed in several symmetry channels. For the sake of giving an intuition of this, we apply Wick theorem on $\theta_{ij}$ (this cannot be done since Wick theorem is only valid for weak correlations but it will give an insight of the physics)

\begin{equation}
\begin{aligned}
\theta_{ij}=& \langle c_{i \sigma} c_{i\bar{\sigma}} \rangle n_{j \sigma} - \langle c_{i \sigma} c^\dagger_{j \sigma} \rangle c_{i \bar{\sigma}} c_{j \sigma} \\ &+ \langle c_{i \sigma} c_{j \sigma} \rangle c_{i\bar{\sigma}} c^\dagger_{j \sigma} - \langle c_{i \bar{\sigma}} c_{j \sigma} \rangle c_{i \sigma} c^\dagger_{j \sigma} + ...
\end{aligned}
\end{equation}

There is more to $\theta_{ij}$ than just superconductivity. Namely, it has a charge and a spin sector too. But, because of the d-wave form factor we impose, we will focus only on superconductivity. $\theta_{ij}$ can therefore been seen as an anomalous d-wave superconductivity mean field parameter. Since we consider singlet pairing, we have
\begin{equation}
\langle c_{i \sigma} c_{i \bar{\sigma}} n_{j \sigma} \rangle = \langle c^\dagger_{i \bar{\sigma}} c^\dagger_{i \sigma} n_{j \sigma} \rangle
\end{equation}

We can still apply translationnal invariance to treat p, n and e as a constant. We can do the same for $\theta$, but because of d-wave symmetry $\theta$ is not the same along the x and y axis ($\theta_x=-\theta_y$). We will therefore only average on one axis in order to have a non zero $\theta$. The self-consistent equations remain the same for n and e since they are only one body correlations. However extending the basis changes the self-consistent equations of $p_{ij}$ and $\theta_{ij}$. We obtain (cf appendix F for details)

\begin{equation}
\begin{cases}
p&=\frac{n^2}{4}-\frac{\rho_1+\phi \rho_2}{1-\phi^2}-\frac{\rho_1-\rho_2}{1-\phi}-\frac{\rho_3}{1+\phi}\\
\theta &=  \frac{\zeta}{1+\phi}
\end{cases}
\end{equation}

With
\begin{equation}
\begin{cases}
\phi &= -\frac{2}{2-n}(C^{11}_0 + C^{12}_0) + \frac{2}{n}(C^{12}_0+C^{22}_0) \\
\rho_1 &= \frac{2}{2-n}(C^{11}+C^{12})^2 + \frac{2}{n}(C^{22}+C^{12})^2 \\
\rho_2&=\frac{2}{2-n}(C^{13}+C^{14})^2 + \frac{2}{n}(C^{23}+C^{24})^2 \\
\rho_3&=\frac{4}{n(2-n)}(C^{11}+C^{12})(C^{22} + C^{12}) \\
\zeta&=\frac{2}{2-n}(C^{11}+C^{12})(C^{13}+C^{14})\\ &+ \frac{2}{n}(C^{12}+C^{22})(C^{23}+C^{24}) 
\end{cases} 
\end{equation}

Let us note that this decoupling is not unique. Several choices can be made. These choices give similar results but tend to overestimate or underestimate some quantities, depending on which regime we consider \cite{Bednorz1988}.
Except for the larger $\textbf{M}$ and $\textbf{I}$ matrices and these changes in the self-consistent equation, everything else remains the same. The expression of $\boldsymbol{\kappa}$ from Eq. (\ref{kappa}) remains unchanged, but will involve $4 \times 4$ matrices.

\subsubsection{Results}

On Fig. \ref{fig6:SCvanHoeve} we plot the bands for the Roth and COM2 solutions. There is a doubling of the bands due to the particle-hole symmetry of the basis: we have four distincts eigenvalues $\epsilon^l$ of the E matrix verifying the property $\epsilon^1=-\epsilon^3$ and $\epsilon^2=-\epsilon^4$. Beside this doubling, the bands are almost unmodified compared to what we have without superconductivity. Only one difference can be seen: a gap opening at $(\pi,0)$. We performed a zoom around zero energy in order to see the gap better on the insets. The presence of the gap also appears on the Fermi surface: there is less weight near the ($\pi,0$) compared to what we had in Fig. \ref{Fig2:bandsNN} without superconductivity. 

On Fig. \ref{fig7:theta(n)}a, we plot the parameter $\theta$ as a function of the electron density n for the COM2 and the Roth solutions. The dashed-line corresponds to the usual d-wave superconducting order parameter $\Delta^d_{ij} = \langle c_{i\sigma} c_{j \bar{\sigma}} \rangle$. We can recover it directly from the correlation functions involving nearest-neighbors $C_{ij}^{nm}=\langle \psi^n_i (\psi^m_j)^\dagger \rangle$ using the following equation
\begin{equation}
\Delta_{ij}^d=C_{ij}^{13}+C_{ij}^{14}+C_{ij}^{23}+C_{ij}^{24}
\end{equation}

The maximum of $\theta$ and $\Delta^d$ are at the same electron density. For the Roth solution this corresponds to n=0.8, while for the COM2 solution it is around n=0.9. We already showed that n=0.8 corresponds to the Van Hove singularity for the Roth solution in the discussion of Fig. \ref{Fig4:DOS} and \ref{fig_van_hove_fs}. This is in agreement with other studies \cite{calegari_superconductivity_2005} \cite{beenen_superconductivity_1995}. We claim the same phenomenon occurs for the COM2 solution. In Fig. \ref{fig6:SCvanHoeve} we plotted the bands and Fermi surfaces for the COM2 solution at n=0.9 where superconductivity is maximum. Beside the gap opening, the band for COM2 exhibits some flatness at $(\pi,0)$ (it is at least flatter than the Roth solution) and its Fermi Surface is almost diamond-like. We justified this claim by plotting the density of states at the Fermi energy as a function of electron density on Fig. \ref{fig7:theta(n)}b. Let us note we did not considered superconductivity to compute this density of states (in order to see the Van Hove singularity): this is why we do not have any superconducting gap. The density of states was computed using Eq. (\ref{DOS}) at $\omega=0$. The maximum of the density of states can be seen at n=0.9 for the COM2 solution. For the Roth solution a maximum can be found near half-filling but this maximum is associated with the extremum of the band and does not improve superconductivity because it is too close to half-filling. Another maximum is found at n=0.8 and corresponds to the flatness at $(\pi,0)$ we already associated this with the Van Hove singularity. This is in agreement with our physical intuition: a high density of states means there are a lot of available electrons available to form Cooper pairs.

We checked that the $\theta$ parameter also verifies the following particle-hole symmetry
\begin{equation}
\theta \rightarrow \theta^*
\end{equation}
In order to satisfy particle-hole symmetry, there must be another maximum of $\theta$, therefore another Van Hove singularity in the electron doped regime. On Fig. \ref{fig7:theta(n)}b, we indeed see another peak both for superconductivity and the density of states in the $n > 1$ area. They correspond to the particle-hole symmetric of the peaks in the hole doped region.

In this method, the gap opening observed on the bands in Fig. {\ref{fig6:SCvanHoeve}} is of the order of $\Delta^d$, as it is expected. The value of $\theta$ affects both superconductivity and the density n, since $\theta$ involves both quantities.   
 
\begin{figure}[h]
\centering
\includegraphics[width=0.7\linewidth, height=5cm]{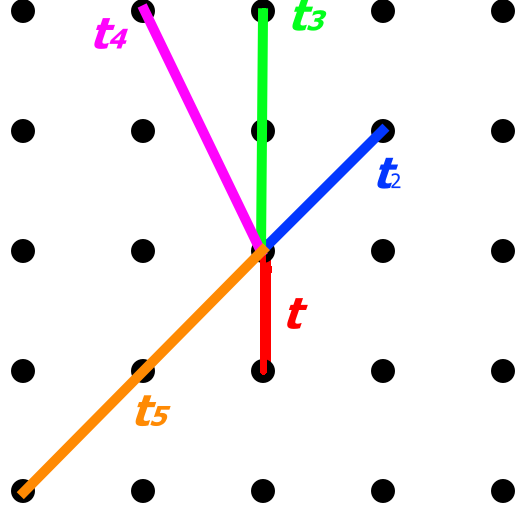}
\caption{Square lattice with the considered hopping. We will consider long ranged hoppings up to $t_5$. The associated values of these hoppings can be found in Table. \ref{Tablebands}}
\label{Fig:tb}
\end{figure} 
 
\begin{figure*}[ht]
\begin{center}
\begin{subfigure}{0.48\linewidth}
\centering
\includegraphics[width=7cm, height=6cm]{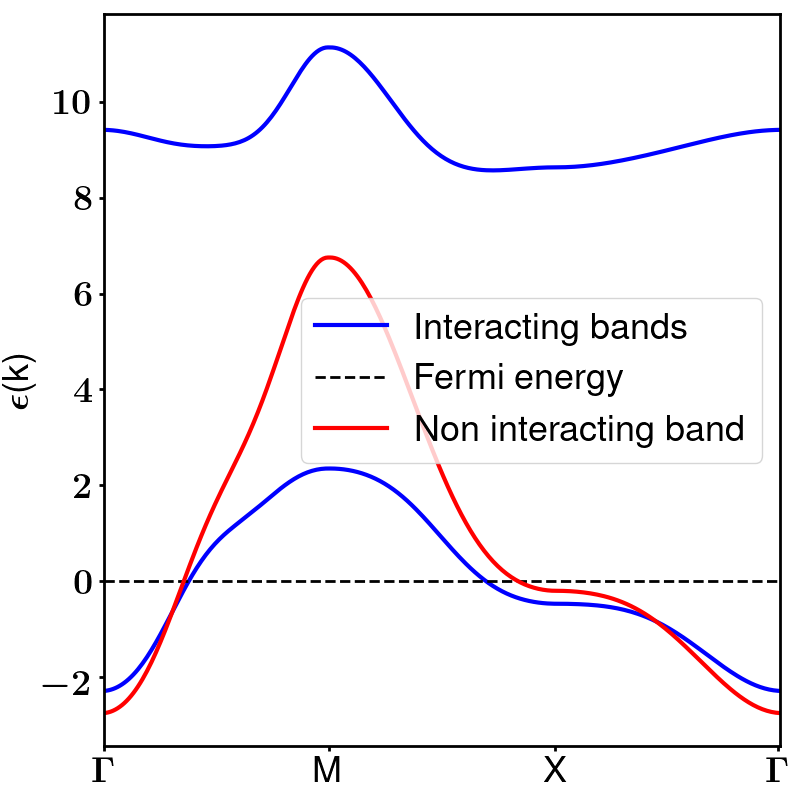}
\caption{\centering tb1: ARPES fit Bi2212}
\end{subfigure}
\begin{subfigure}{0.48\linewidth}
\centering
\includegraphics[width=7cm, height=6cm]{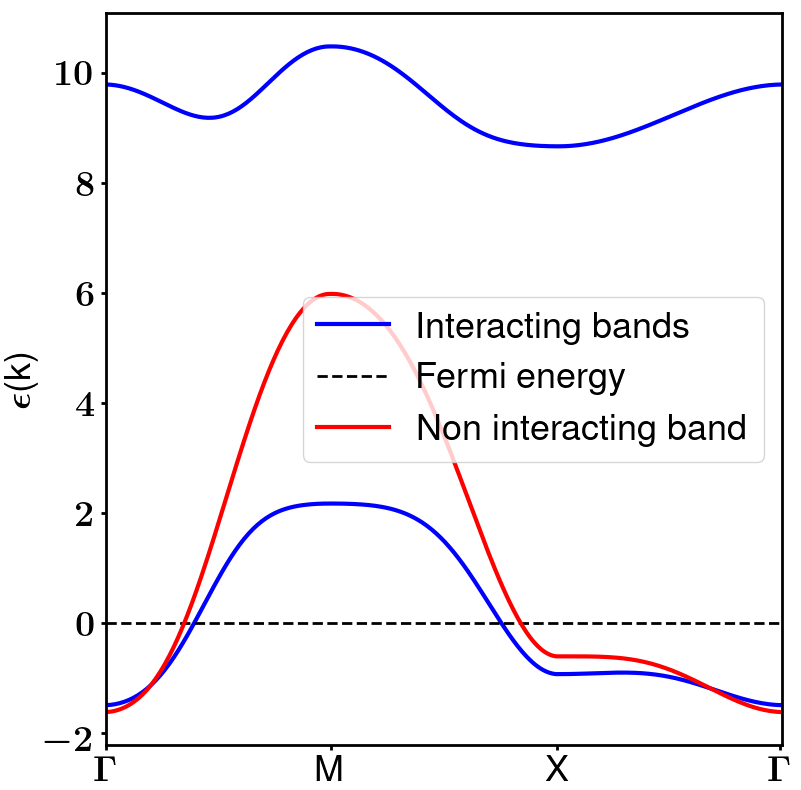}
\caption{\centering tb2: fit of bonding surface of Bi2212}
\end{subfigure}
\\
\begin{subfigure}{0.48\linewidth}
\centering
\includegraphics[width=7cm, height=6cm]{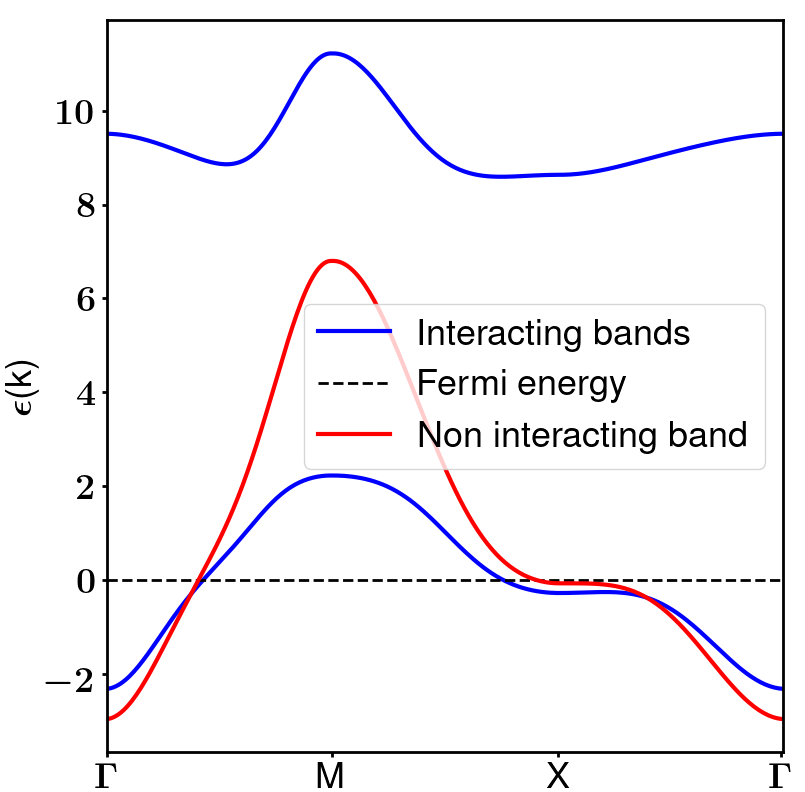} 
\caption{\centering tb 3: Modified tb2 to get better incommensurability}
\end{subfigure}
\begin{subfigure}{0.48\linewidth}
\centering
\includegraphics[width=7cm, height=6cm]{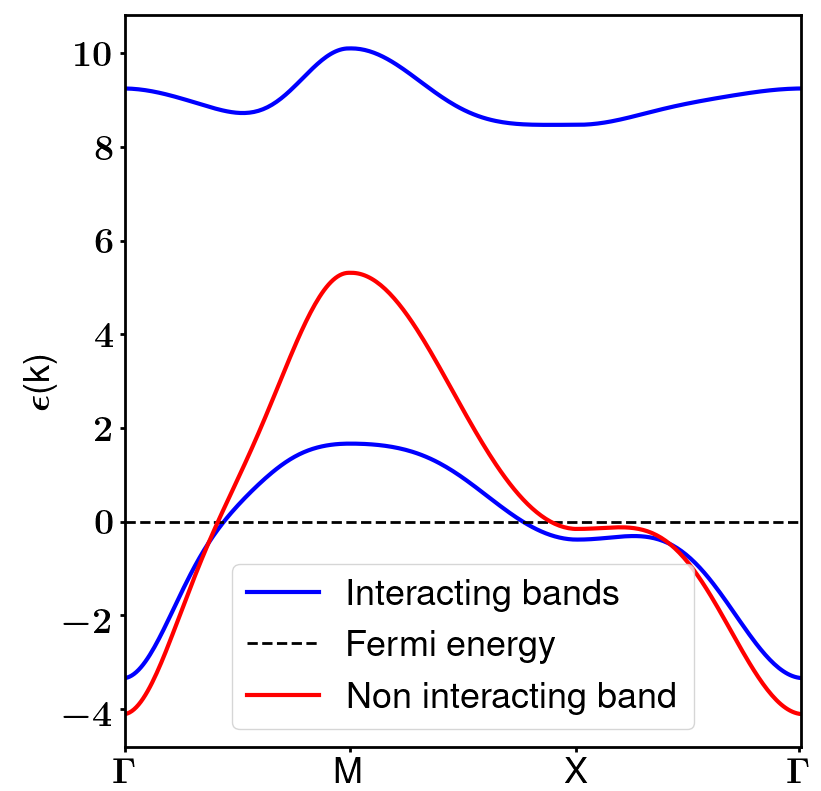} 
\caption{\centering tb 4: underdoped LSCO}
\end{subfigure}
\caption{Bands renormalization using Roth minimization with hoppings up to order 5. The tight bindings parameters are taken from Ref \cite{norman_linear_2007} and are given in table I. The red line is the non-interactive, tight binding dispersion. We are at 20\% hole doping.
\label{Fig8:BandsAll}}
\end{center}
\end{figure*}

\section{\label{sec:III}Further hoppings study}

We now want to consider the effect of higher hopping terms in order to get closer to more realistic materials. We want to see if the results we had with nearest-neighbours hoppings on the bands, the validity of the Luttinger theorem and superconductivity are modified by further hoppings. Including at least next nearest neighbors in the model is enough to break the particle-hole symmetry. We will only consider the Roth solution in this section, since COM2 and COM1 solutions have been studied with next-nearest neighbors in Ref. \citep{avella_t_t'_2001}, and Roth presents Fermi surface closer to what is observed by ARPES for cuprates.

\subsection{Effect of further nearest neighbours and bands analysis}

From now on we will consider four different sets of tight-binding parameters all corresponding to a square lattice as in Fig. \ref{Fig:tb}.

These tight-binding parameters correspond to cuprates Bi2212 and LSCO, which are strongly correlated. Their values, taken from Ref. \cite{norman_linear_2007},are given in Table \ref{Tablebands}. They are such that the energy for a tight binding model of a square lattice is given by 
\begin{equation}
\begin{aligned}
\epsilon_{tb}(k)&= 2t(cos(k_x)+cos(k_y))\\&+4 t_2 cos(k_x) cos(k_y) \\&+ 2 t_3(cos(2k_x)+cos(2k_y)) \\&+ 4 t_4 (cos(k_x)cos(2k_x)+cos(2k_x)cos(k_y)) \\&+ 4t_5 cos(2k_x)cos(2k_y)
\end{aligned}
\label{EqTB}
\end{equation}

In the following we will normalize every plots so we have t=1 (we will divide every tight binding parameter by t in absolute value).

\begin{table}[h]
\begin{tabular}{c|c|c|c|c|c}
\multicolumn{1}{c|}{ } & \multicolumn{1}{c|}{t} & \multicolumn{1}{c|}{$t_2$} & \multicolumn{1}{c|}{$t_3$} &
\multicolumn{1}{c|}{$t_4$} &
\multicolumn{1}{c}{$t_5$} \\
\hline
tb1 & -0.2956 & 0.0818  & -0.0260  &-0.0280 & 0.0255 \\
\hline
tb2 &-0.3399 & 0.1184 & -0.0397   & 0.0086 & 0.0006\\
\hline
tb3 & -0.2941 & 0.0731 & 0.0048   & -0.0325 & 0.0035\\
\hline
tb4 & -0.3912 & 0.0370 & -0.0294  &-0.0350 & -0.0087
\end{tabular}
\captionof{table}{Values of the 4 tight bindings we are going to consider. tb1 is based on an ARPES fit of Bi2212. tb2 corresponds to the bonding surface of Bi2212, tb3 is a modified version of tb2 to get a flatter band and tb4 corresponds to underdoped LSCO.}
\label{Tablebands}
\end{table}

In this section we include hoppings up to $t_5$ and study the bands and Fermi surface behavior. With additional hopping terms, only the expression of the matrix $\textbf{M}$ changes.
In addition to the  $\alpha_{il}^1$ parameter appearing in Eq. (\ref{Mcoefs}) higher hopping terms will appear in the $\textbf{M}$ matrix. It becomes

\begin{equation}
\begin{aligned}
m_{ij}^{11}=& \hspace{-0.05cm}  -\mu (1-\frac{n_i}{2})\delta_{ij} - \sum \limits_{k=1}^N t \left[ \alpha_{ij}^k(1-\frac{n_i+n_j}{2} + p_{ij}) +  \delta_{ij} \sum \limits_{l}  \alpha^k_{il} e_{il} \right]  \\
m_{ij}^{12}=&  \sum \limits_{k=1}^N t\left[   \alpha_{ij}^k (\frac{n_j}{2}-p_{ij}) - \delta_{ij} \sum \limits_l \alpha_{il}^k e_{il}\right]  \\
m_{ij}^{22}=& -(\mu - U) \frac{n_i}{2} \delta_{ij} + \sum \limits_{k=1}^N t \left[  \alpha_{ij}^k p_{ij} - \delta_{ij} \sum \limits_l \alpha_{il}^k e_{il} \right]
\end{aligned}
\label{McoefsNNN}
\end{equation}

where 
\begin{equation}
\begin{cases}
\alpha^N_{il}&=1 \ \ \ \ \text{if i and l are  }\overbrace{\text{next-...}}^{N-1 \text{ times}}\text{-nearest neighbour} \\
\alpha^N_{il}&=0 \ \ \ \ \text{Otherwise}
\end{cases}
\end{equation}

Each new hopping considered adds a term in the tight-binding Hamiltonian which is then added in the $\textbf{M}$ matrix. 
The parameters p and e depend on i-j, so we should make a distinction between $e^1_{ij}$ with i and j nearest neighbors (NN), $e^2_{ij}$ with i and j next nearest neighbors (NNN)... $e^n_{ij}$ and $p^n_{ij}$ will be associated with their corresponding hopping as in Fig. \ref{Fig:tb}. Translationnal invariance still allows us to treat $e^1$, $e^2$,...,$e^5$, $p^1$, $p^2$,...,$p^5$ as constants. Correlation functions $\textbf{C}_{ij}=\langle \psi_i; \psi^\dagger_j \rangle$ are at different sites too, so we will also have to make a distinction in the self-consistent equations between $\textbf{C}^1$ for NN, $\textbf{C}^2$ for NNN and so on.

\begin{figure}[h]
\begin{minipage}{0.2\textwidth}
\includegraphics[scale=0.41]{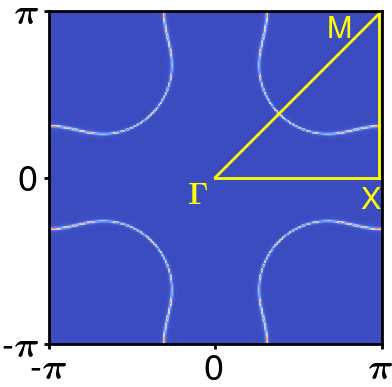}
\includegraphics[scale=0.41]{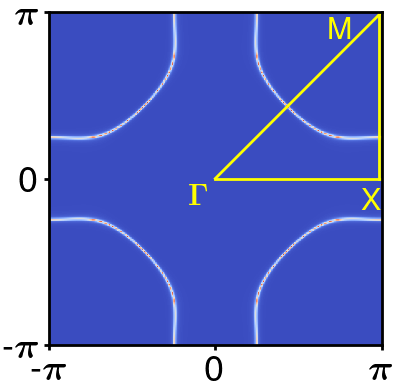}
\end{minipage}
\hspace{0.1mm}
\begin{minipage}{0.2\textwidth}
\includegraphics[scale=0.41]{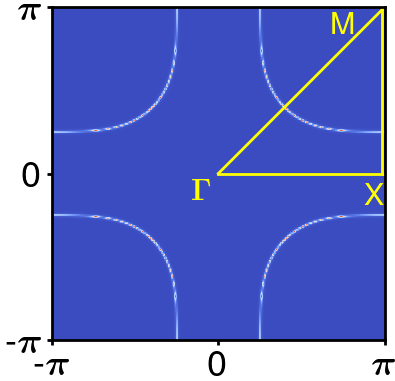}
\includegraphics[scale=0.41]{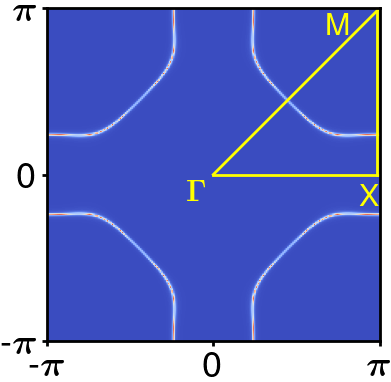}
\end{minipage}
\hspace{0.1mm}
\begin{minipage}{0.01\textwidth}
\includegraphics[height = 4.5cm,width=0.8cm]{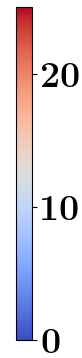}
\end{minipage}

\caption{Fermi Surfaces renormalized by the composite operator methods using Roth minimization with the parameters of Ref \cite{norman_linear_2007}. We are at 20\% hole doping. Top left: tb1, top right: tb2, bottom left: tb3, bottom right: tb4.}
\label{Fig9:FSAll}
\end{figure}

\begin{figure}[h]
\begin{minipage}{0.2\textwidth}
\includegraphics[scale=0.41]{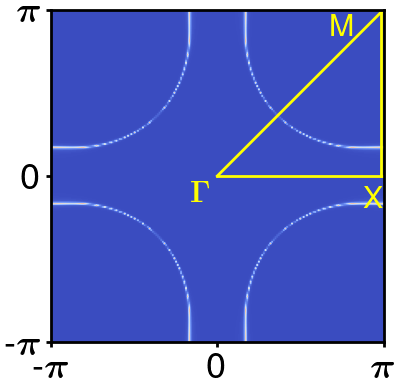}
\includegraphics[scale=0.41]{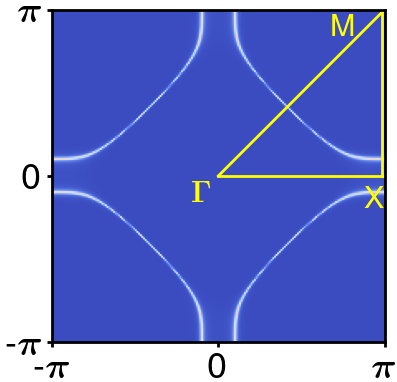}
\end{minipage}
\hspace{0.1mm}
\begin{minipage}{0.2\textwidth}
\includegraphics[scale=0.41]{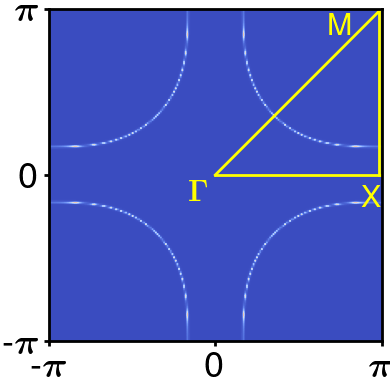}
\includegraphics[scale=0.41]{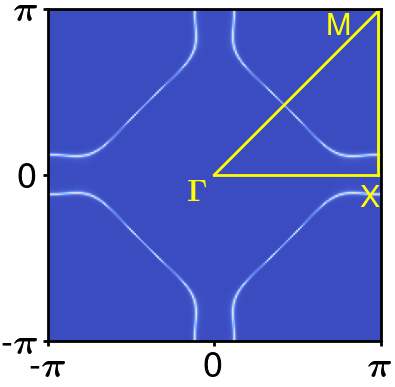}
\end{minipage}
\hspace{0.1mm}
\begin{minipage}{0.01\textwidth}
\includegraphics[height = 4.5cm,width=0.8cm]{cbar_FS.png}
\end{minipage}

\caption{Non interacting (NI) Fermi Surfaces for the parameters of Ref \cite{norman_linear_2007} at 20\% hole doping. Top left: tb1, top right: tb2, bottom left: tb3, bottom right: tb4.}
\label{Fig10:FSAllNI}
\end{figure}

Fig. \ref{Fig8:BandsAll} presents the bands we obtain for Roth solutions for the four sets of tight-binding parameters in Table. (\ref{Tablebands}). In Fig. \ref{Fig9:FSAll} and \ref{Fig10:FSAllNI} we plotted respectively the Fermi surfaces obtained with the method and the Fermi surfaces of the non-interactive tight-binding dispersions (corresponding to Eq. (\ref{EqTB})).


\begin{figure*}[ht]
\begin{subfigure}{\linewidth}
\begin{center}
\begin{subfigure}{0.49\linewidth}
\includegraphics[width=8.9cm, height=6cm]{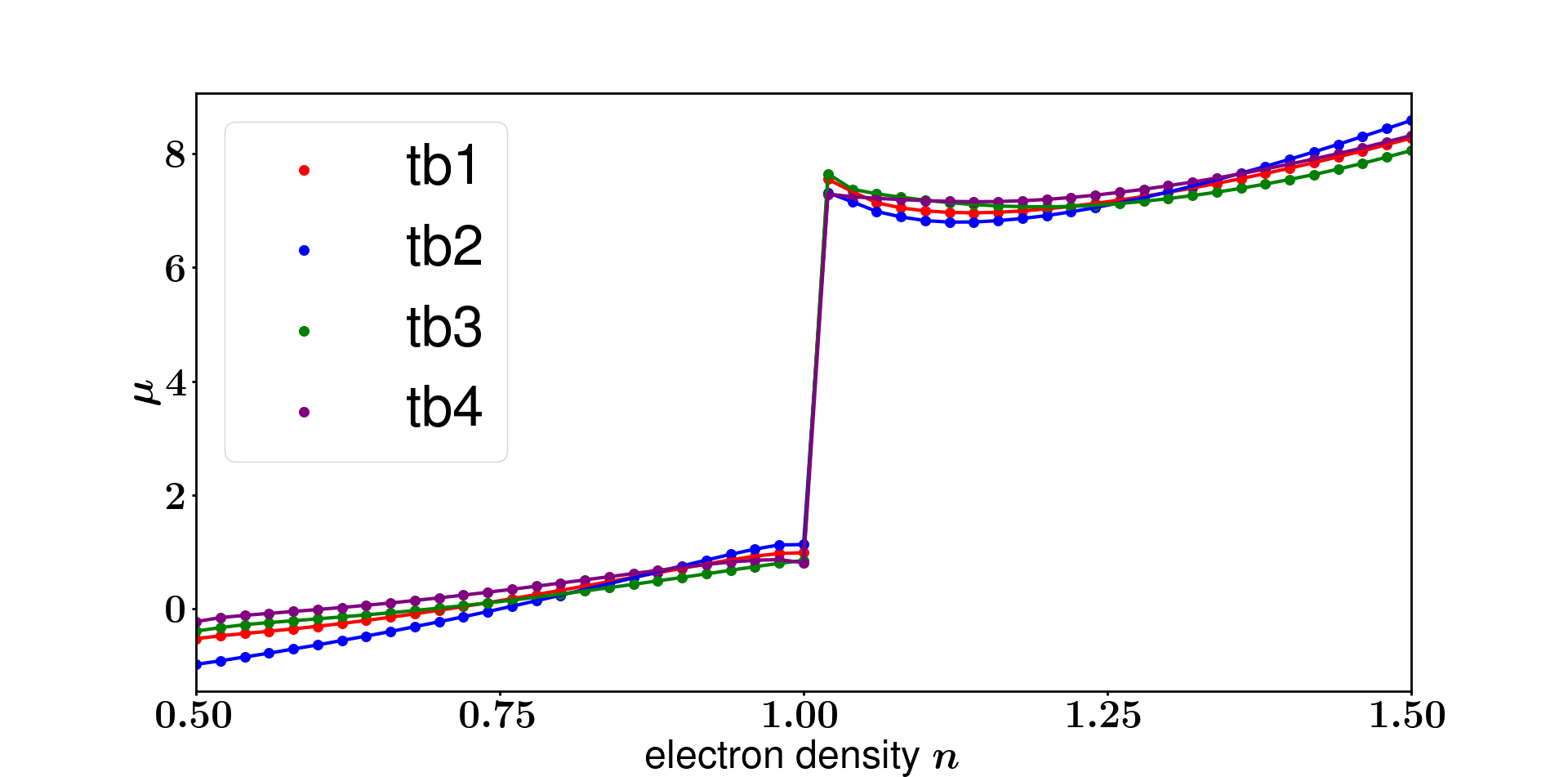}
\caption{Chemical potential}
\end{subfigure}
\begin{subfigure}{0.49\linewidth}
\includegraphics[width=8.9cm, height=6cm]{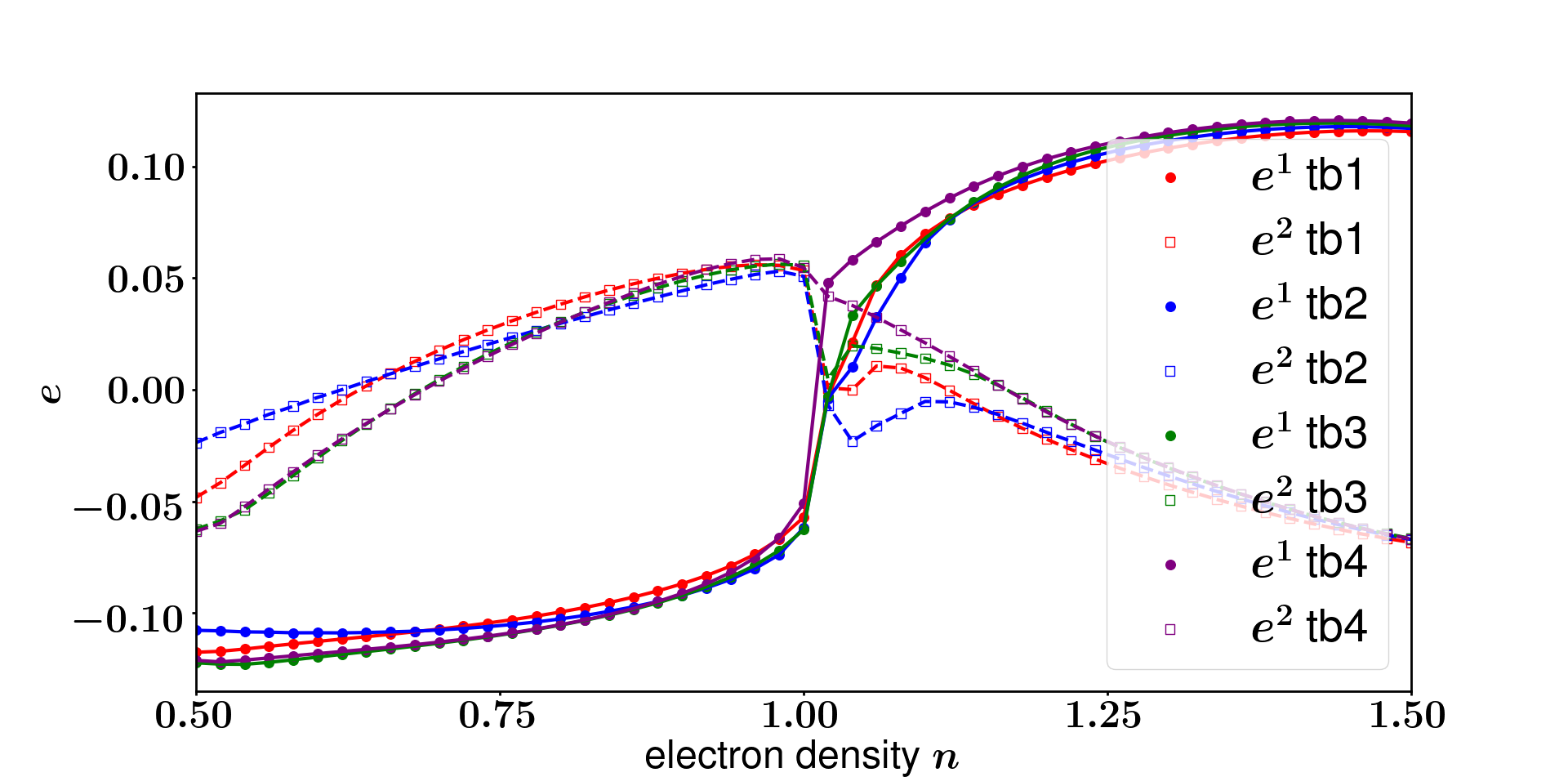}
\caption{Parameters e}
\end{subfigure}
\\
\begin{subfigure}{0.49\linewidth}
\centering
\includegraphics[width=8.9cm, height=6cm]{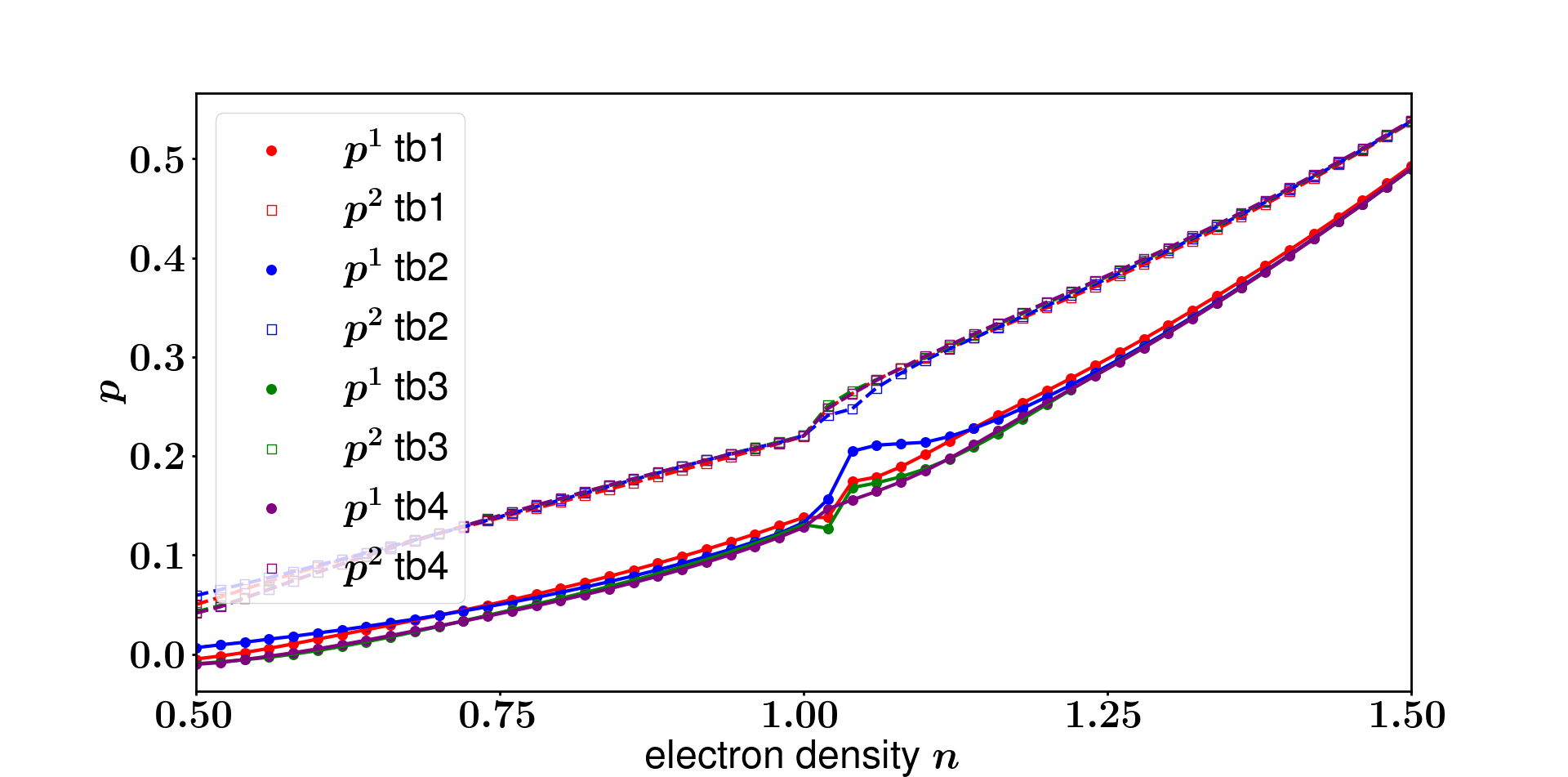} 
\caption{Parameters p}
\end{subfigure}
\begin{subfigure}{0.49\linewidth}
\includegraphics[width=8.9cm, height=6cm]{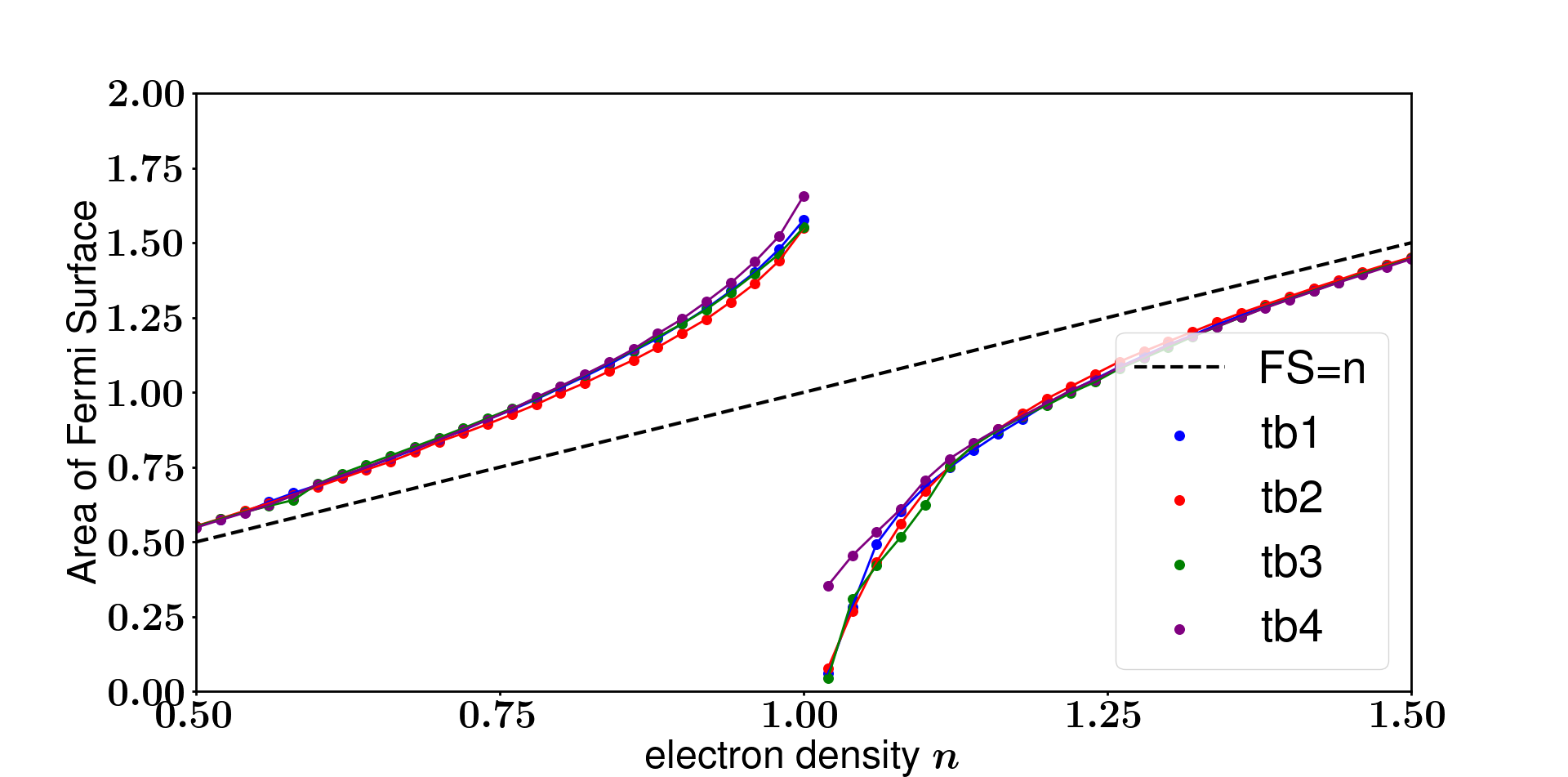} 
\caption{Luttinger Violation}
\end{subfigure}
\end{center}
\end{subfigure}
\caption{(a), (b) and (c) :Parameters as a function of doping for each tight binding with next nearest neighbors. The circles corresponds to p1 and e1 (NN) while the squares corresponds to p2 and e2  (NNN). (d) : Luttinger violation for each tight binding.}
\label{Fig11:phLuttAll}
\end{figure*}

In the hole doped regime, the Fermi surface we obtained from the Roth solution has the same general shape as the non interacting Fermi Surface. The composite operators method produces Fermi surfaces that appear to be smaller/larger than the tight-binding ones. This is in agreement with the violation of the Luttinger theorem observed with nearest-neighbour hopping, and indicates that it is still violated with further hoppings. We checked the opposite situation is accordingly observed in electron doped area: the Fermi Surface obtained with the method is at a lower doping than the non interacting one. In the following we study the particle-hole symmetry and the Luttinger theorem violation with next-nearest neighbours.
\begin{figure*}[ht]
\begin{center}
\begin{subfigure}{0.49\linewidth}
\includegraphics[width=8.9cm, height=6cm]{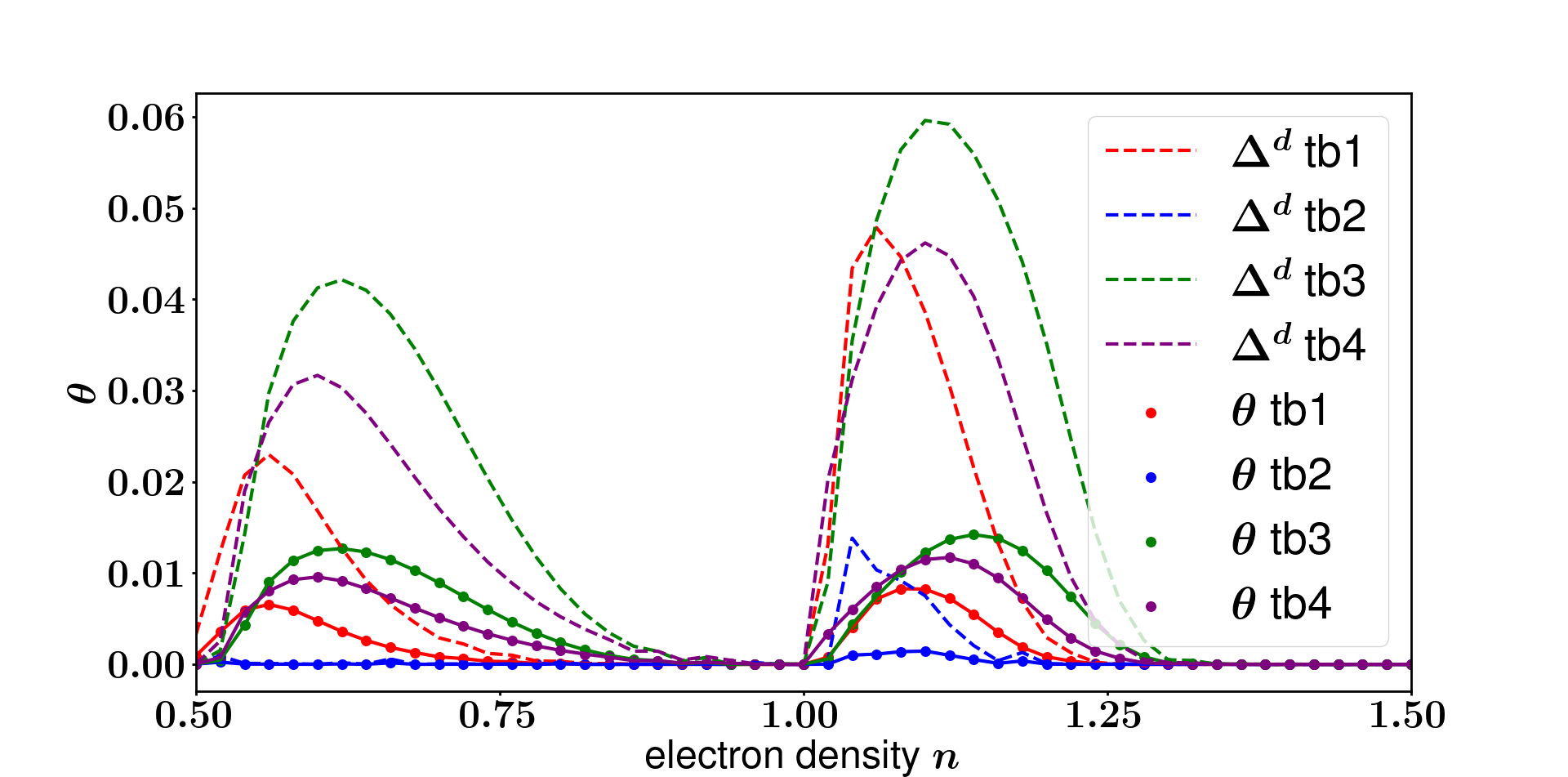}
\caption{$\theta$ and $\Delta^d$}
\end{subfigure}
\begin{subfigure}{0.49\linewidth}
\includegraphics[width=8.9cm, height=6cm]{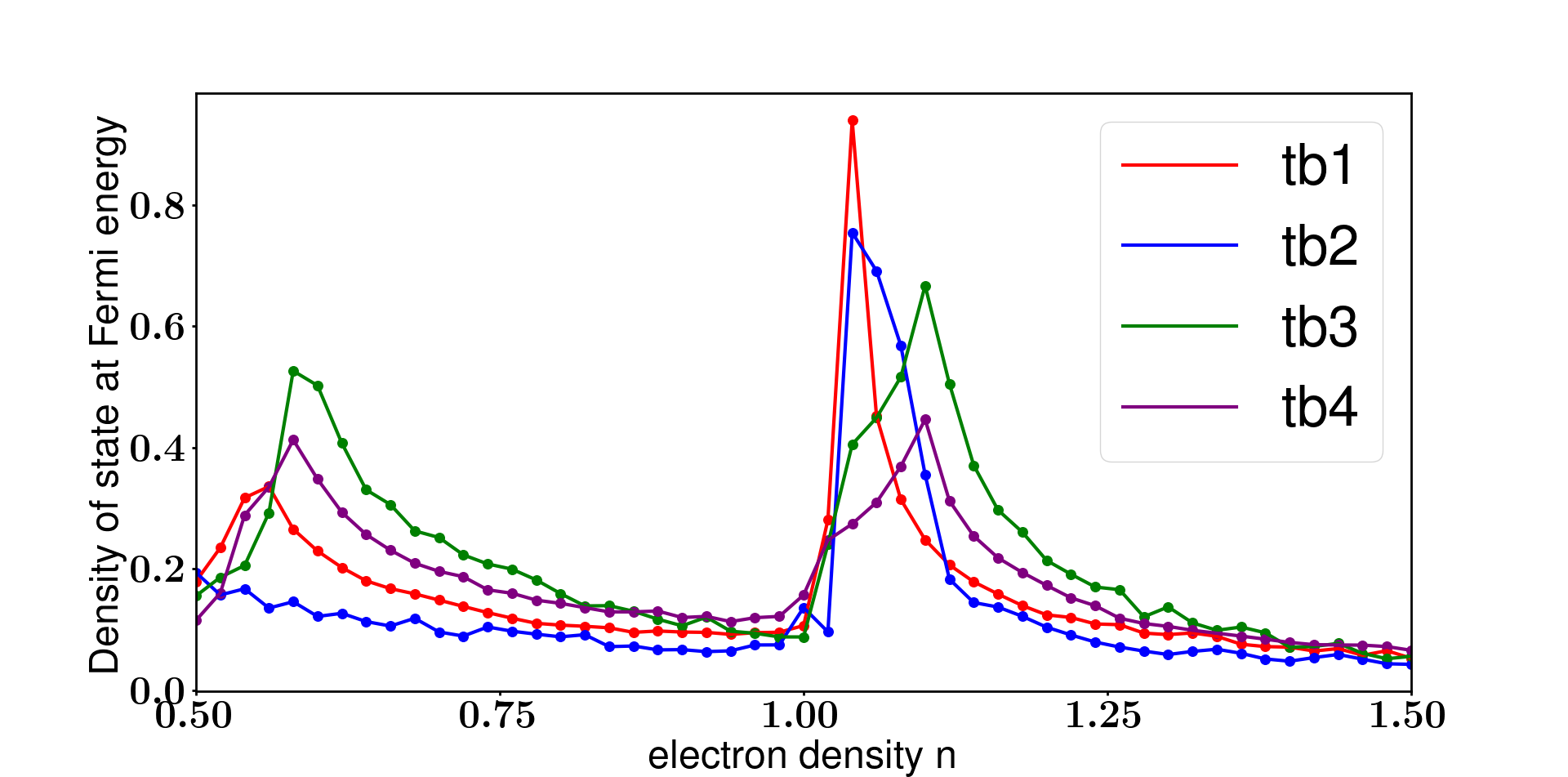}
\caption{Density of states at the Fermi energy}
\end{subfigure} \\
\begin{subfigure}{0.49\linewidth}
\includegraphics[width=8.cm, height=6cm]{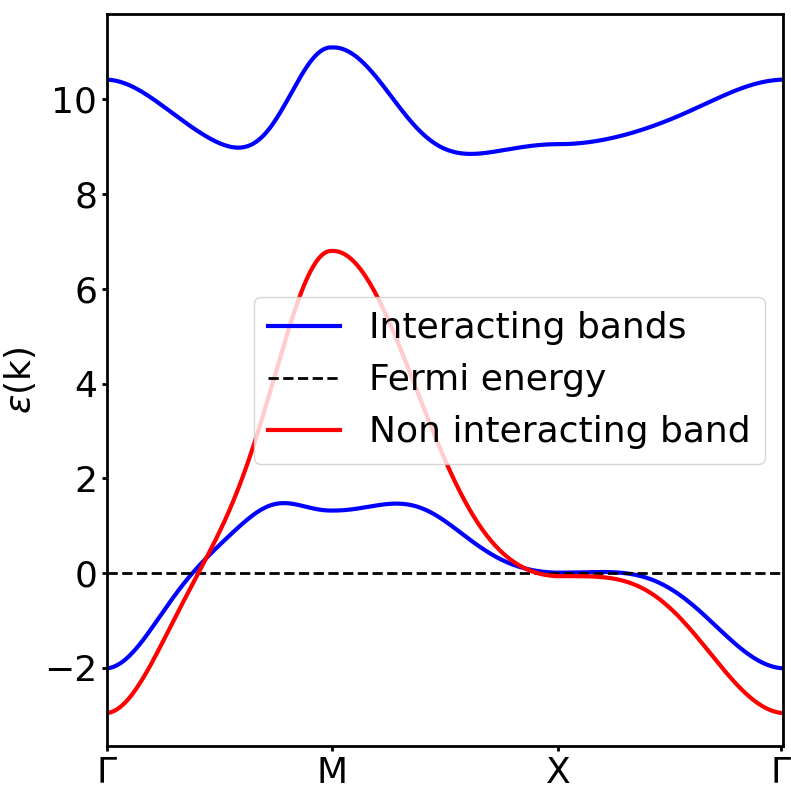}
\caption{Bands for tb3 at n=0.6 with no superconductivity}
\end{subfigure}
\begin{subfigure}{0.49\linewidth}
\includegraphics[width=8.cm, height=6cm]{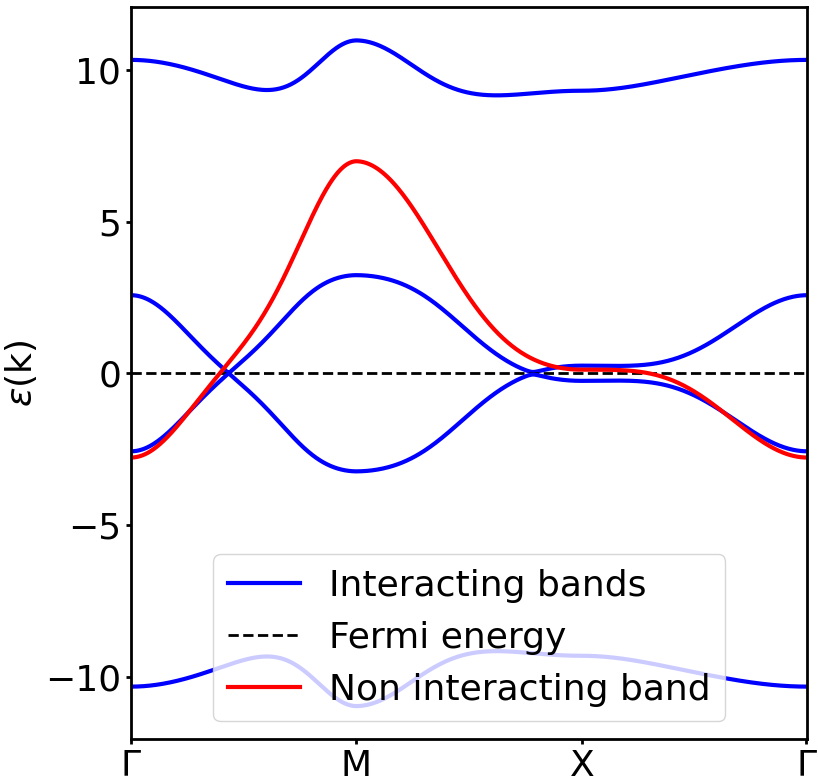}
\begin{picture}(0,0)
\put(-75,90){\includegraphics[height=3cm,width=3cm]{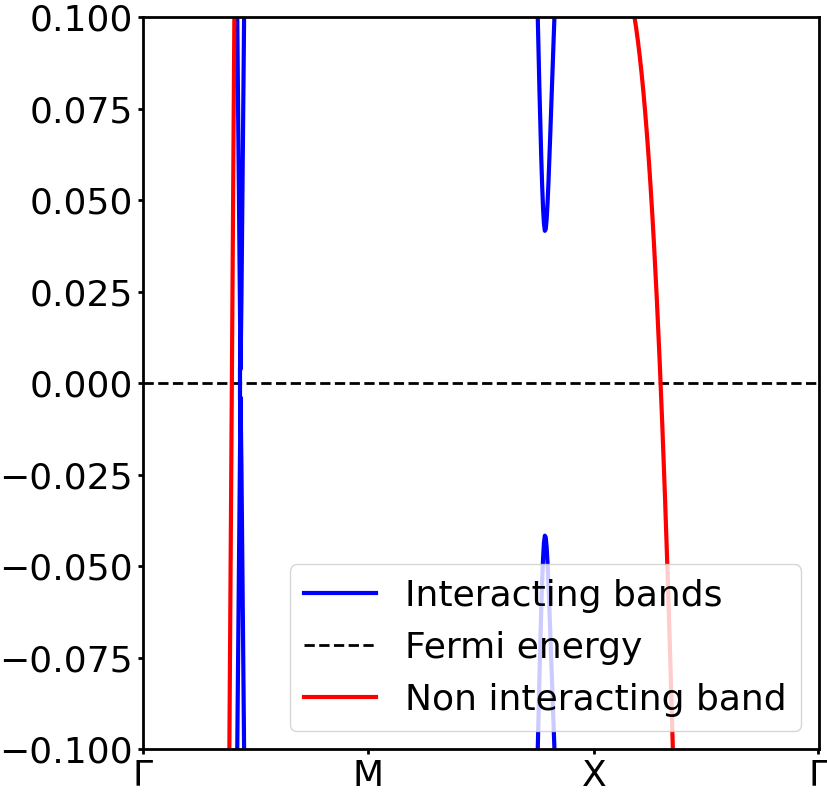}}
\end{picture}
\caption{Bands for tb3 at n=0.6 with superconductivity}
\end{subfigure}
\end{center}
\caption{(a): Anomalous superconducting order parameter $\theta$ as a function of doping for the four sets of tight bindings parameters with Roth minimization. The dashed line corresponds to the usual d-wave superconducting order parameter we rebuilt from the correlation function. We are at U=8t. (b): Density of states (with no superconductivity in order to see the peaks) at the Fermi energy. The 2 peaks corresponds to the 2 Van Hove singularity in hole and electron doping respectively. To illustrate this the bottom plots (c) and (d) are the bands at the Van Hove singularity with tight binding parameters tb3 at n=0.6 respectively with: (c) no superconductivity, where we clearly see the flat band associated to Van Hove singularity (d) with superconductivity, where we see on the zoom in the inset the gap of the order of $\Delta^d$.}
\label{fig12:theta(n)All}
\end{figure*}

\subsection{Particle-hole symmetry and Luttinger theorem with further neighbours}

For the sake of simplicity we restrict ourselves to next-nearest neighbors in this part. In order to keep the full Hamiltonian invariant under particle-hole symmetry the following relation is needed
\begin{equation}
t_2 \rightarrow -t_2
\label{t2}
\end{equation}

This is because the hopping term of the Hamiltonian $\sum \limits_{ij} t_{ij} c_{i \sigma}^\dagger c_{j\sigma}$ transforms into $\sum \limits_{ij} t_{ij} (-1)^{i+j} c_{i \sigma}c^\dagger_{j\sigma}$. For i and j next-nearest neighbour we have $(-1)^{i+j}=1$. We obtain $-\sum \limits_{\langle \langle i,j \rangle \rangle} t_2 c^\dagger_{j \sigma} c_{i \sigma}$, which has an additional minus sign, breaking the symmetry unless $t_2$ changes its sign with the transformation.

Hence, we know that at least next-nearest neighbors will break particle-hole symmetry (because $t_2$ is an external parameter that will always keep the same sign, therefore we do not verify Eq. (\ref{t2})). The particle-hole transformation works the same for n and $\mu$
\begin{equation}
\begin{aligned}
n &\rightarrow 2-n \\
\mu &\rightarrow U-\mu 
\end{aligned}
\end{equation}

The main difference with the nearest-neighbors case comes from the necessity to differentiate the e and p parameters depending on the considered hopping. We already know the transformation for $p^1$ and $e^1$. The only difference for $p^2$ and $e^2$ is $(-1)^{i+j}=1$. $p^2$ will not be affected by the $(-1)^{i+j}$ coming from the transformation because it involves only two bodies operators. Thus only $e^2$ has an additional minus sign under the transformation, and we obtain the following relations
\begin{equation}
\begin{aligned}
p^1 &\rightarrow 1-n+p^1 \\
p^2 &\rightarrow 1-n+p^2 \\
e^1 &\rightarrow -e^1 \\
e^2 &\rightarrow e^2
\end{aligned}
\label{phNNN}
\end{equation}

In Figs. \ref{Fig11:phLuttAll}a, \ref{Fig11:phLuttAll}b and \ref{Fig11:phLuttAll}c we plot the parameters $\mu$, e and p as a function of doping for the four considered tight-binding parameters of Table. (\ref{Tablebands}), only considering t and $t_2$. Parameters $e^2$ and $p^2$ indeed break the particle-hole symmetry from Eq. (\ref{phNNN}). The chemical potential, as well as $e^1$ and $p^1$ behave as in the nearest neighbor case. We also checked that Eq. (\ref{phNNN}) is respected if we impose $t_2 \rightarrow - t_2$ when $n \rightarrow 2-n$.
\\
\\
Finally we can again study the Luttinger theorem. In Fig. \ref{Fig11:phLuttAll}d, the area of the Fermi surface is plotted as a function of electron density. Interestingly we observe an analogous behavior as in the nearest-neighbor case. The Luttinger violation does not seem to be modified by next-nearest neighbors hoppings and is barely affected when we modify tight-binding parameters. As already mentioned before, the Luttinger theorem is strongly violated around half filling and is recovered further away. This confirms why in the previous section the Fermi surfaces seemed to be at a higher doping than the electron density we considered when we are in the hole doped regime (and conversely, at a lower doping in the electron doped regime).

\subsection{Superconductivity, Van Hove singularity and density of states}

Superconductivity can also be included in the model with more hoppings. In this section we include hoppings up until $t_5$. We assume again a d-wave symmetry and only consider nearest-neighbor pairing. As before, the $\textbf{M}$ and $\textbf{I}$ matrices become 4x4 and have the same symmetries as the nearest neighbours case. The $\textbf{I}$ matrix is independent of t and t' and is thus the same as before. The main difference is that the coefficients $m_{ij}^{11}$, $m_{ij}^{12}$ and $m_{ij}^{22}$ are now given by Eq. (\ref{McoefsNNN}). $m_{ij}^{13}$ stays identical to its expression in the nearest neighbor case since we only consider superconductivity for nearest neighbors. Therefore $m_{ij}^{13}$ is proportionnal to the $\theta$ parameter, defined the same way as before ($\theta_{ij}=\langle c_{i \up} c_{i \down} n_{j \sigma} \rangle$).

In Fig. \ref{fig12:theta(n)All}a, the $\theta$ and $\Delta^d$ parameters are plotted as a function of electron density for the four tight-binding parameters with the Roth solution. Since we have included further neighbors, $\theta$ is not particle-hole symmetric anymore. In Fig. \ref{fig12:theta(n)All}b, we plotted the density of states at the Fermi energy without superconductivity using Eq. (\ref{DOS}) at $\omega=0$ (in order to see the peaks with no superconducting gap). In Fig. \ref{fig12:theta(n)All}c and \ref{fig12:theta(n)All}d, we plotted without and with superconductivity the bands at electron densities corresponding to the maximum of the density of states for tb3 when the system is hole doped ($n \approx 0.6$). The bands with no superconductivity on \ref{fig12:theta(n)All}c are flats at the Fermi energy. This proves that the maximum of Fig. \ref{fig12:theta(n)All}b correspond to the Van Hove singularity. On \ref{fig12:theta(n)All}d we see in the inset the gap at $(\pi,0)$ is again of the order of $2\Delta^d$. The maximum of the $\theta$ and $\Delta^d$ parameters are thus at the same electron density as the Van Hove singularity. Hence, the situation is the same as in the nearest-neighbor case. In Table \ref{Tablesing} we give the electron densities associated to the Van Hove singularities for the four sets of tight-binding parameters. Let us note there is no proportionality between the peak in the density of states and $\Delta^d$: it is different for every set of tight-binding parameters. This is seemingly a consequence of the electronic correlations which are treated differently for each tight-binding parameters as a consequence of the main approximation of the method Eq. (\ref{EMIS:curr}).

\begin{table}[h]
\begin{tabular}{c|c|c|c|c}
\multicolumn{1}{c|}{ } & \multicolumn{1}{c|}{tb1} & \multicolumn{1}{c|}{tb2} & \multicolumn{1}{c|}{tb3} & \multicolumn{1}{c}{tb4} \\
\hline
Van Hove (electron) & 1.04 & 1.05  & 1.10  & 1.11 \\
\hline
Van Hove (hole) & 0.57 & 0.45 & 0.6   & 0.58 \\

\end{tabular}
\captionof{table}{Electron density at which there is a Van Hove singularity at the chemical potential. At this values a flat band lies at the Fermi energy and the associated Fermi surface is diamond like. These singularities occur both in electron and in hole.}
\label{Tablesing}
\end{table}


The most striking feature is that superconductivity in the method seems to be induced by the proximity of the Van Hove singularity. This was already the case with nearest neighbors but this property seems unaffected by further hopping terms. For cuprates there exist a consensus that antiferromagnetism is the interaction necessary to explain the pairing mechanism \cite{taillefer_scattering_2010}. This has also the advantage to explain why experimentally superconductivity is observed around 15\% hole doping. Since this method predicts superconductivity only close to the Van Hove singularity, it is non zero at some doping values which does not correspond to what is expected. This flaw is maybe a consequence of the main hypothesis of the method (detailed in Appendix A): it is maybe necessary to consider dynamical corrections to the self-energy in order to observe a different behavior for superconductivity.


\section{Conclusion}

Considering only nearest neighbors first, three solutions have been isolated and studied: COM1, COM2 and Roth solutions. The Roth solution is the unique solution obtained if the self-consistency is performed on a parameter we called p and which enforced charge-charge, spin-spin and pair-pair correlations. This solution violates the Pauli principle because the correlation function $C^{12}_0=\langle \xi_{i \up} \eta_{i\up}^\dagger \rangle$ obtained after convergence is numerically non zero, even though it should be analytically because of the Pauli principle. COM1 and COM2 are two independant solutions obtained by performing a self-consistency imposing the correlation function $C^{12}_0$ instead of the parameter p to enforce the Pauli principle. 

We have performed a systematic comparison of the bands and Fermi surfaces for these three solutions. The three solutions present two Hubbard bands associated to the two eigenvalues of the method. COM1 is not renormalized by the interactions. COM2 has two hole pockets whatever the doping. Roth only has one hole pocket but a second small one appears close to half filling at $(\pi, \pi)$, possibly because antiferromagnetism has been neglected. Both Roth and COM2 solutions are insulators at half filling because the chemical potential lies in between the two Hubbard bands. The density of states for the Roth solution does not present the usual quasiparticle peak at $\omega=0$ at half-filling and is in agreement with Determinantal quantum Monte-Carlo simulations.

The particle-hole symmetry is respected with nearest-neighbour hoppings: we checked that all solutions verify particle-hole relations, including the Roth solution. This result is different from what was predicted by ref. \cite{avella_hubbard_1998}: a solution of this method can violate the Pauli principle and still be particle-hole symmetric. 
All solutions break the Luttinger close to half filling. This results is also in agreement with the determinantal quantum Monte-Carlo simulation \cite{osborne_fermi-surface_2020}. The Luttinger theorem is only proven for weakly interacting systems. We think the Roth solution is the most physical. Although there is a violation of the Pauli principle, this solution exhibits Fermi surfaces typical of strongly correlated materials such as cuprates, contrary to COM1 and COM2. The COM2 solution presents a second hole pocket for every value of the electron density which is not observed in ARPES experiments for Bi2212 and LSCO.

We extended the method to study d-wave superconductivity. Four composite operators are necessary to perform this study. The doubling of the basis leads to four eigenvalues, which are particle-hole symmetric because of the symmetry of the basis. We observed a gap opening at $(\pi, 0)$ for the COM2 and Roth solutions, of the order of the usual d-wave superconducting order parameter $\Delta_{ij}^d= \langle c_{i\up} c_{j \down} \rangle$. We only observe non zero superconductivity close to the Van Hove singularity. While this has already been predicted for the Roth solution \cite{calegari_superconductivity_2005}, we showed that the COM2 solution behaves the same. We observed two Van Hove singularities for the COM2 and Roth solutions: one in the hole-doped and one in the electron-doped regime. Both singularities are a consequence of a flat band at $(\pi,0)$. Lastly we noticed $\Delta^d=\langle c_{i\up} c_{j \down} \rangle$ is three or four times bigger than $\theta$. $\theta$ has no proportionnality to $\Delta^d$ because it includes both a superconducting and a charge channel when we try to decouple it.

We studied the effect of longer ranged tight-binding parameters fitted from ARPES experiments on cuprates Bi2212 and LSCO from ref.\citep{norman_linear_2007}, with the Roth minimization. If hole doped, the Fermi surfaces obtained with the method are similar to the non interacting Fermi surface, but at a higher doping. The opposite situation happens if the system is doped in electron: the Fermi surfaces appear similar but at a lower doping than the non interacting one. This is in agreement with the violation of the Luttinger theorem we still observe around the Mott transition. Adding further hoppings does not seem to modify this property. We checked that adding further hoppings breaks particle-hole symmetry. Finally, superconductivity also behaves mostly the same with further neighbors compared to the nearest neighbor case. The main difference is that Van Hove singularities as well as superconductivity peaks are moved at different electron density compared to the nearest-neighbors case. While this result is not at all in agreement with what was observed experimentally \citep{norman_electronic_2003}, it may be an indication that the main approximation done in this scheme needs to be refined. The dynamical corrections to the self-energy need to be considered in order to maybe have a behavior for superconductivity which is more in agreement with experimental observations.

\bibliographystyle{ieeetr}
\bibliography{biblio}

\section*{Appendix A : physical meaning of the approximation on the current}

In order to be able to compute the $\textbf{M}$ matrix with the full Hubbard Hamiltonian in the strongly coupled regime, we neglected all terms in the current (Eq. (\ref{currents})) that are not along $\psi$. Though afterwards we computed explicitly the currents, the terms orthogonal to $\psi$ are neglected through the self consistency because we use the relation $\textbf{M}=\textbf{E} \textbf{I}$, which is a consequence of our hypothesis.\\
\\
To better understand the consequences of such approximation we need to consider the self energy. To do so, we follow the step of Ref. \cite{avella_anomalous_2008}. Let's therefore consider the full expression of the current 
\begin{equation}
J_i= \sum \limits_l \textbf{E}_{il} \psi_l + \delta \phi_i
\label{fullcurr}
\end{equation}

Our approximation, $\langle \{ \delta \phi_i, \psi_j^\dagger \} \rangle=0$ allows us to write $\textbf{E}_{k}=\textbf{M}_{k} \cdot \textbf{I}^{-1}_{k}$. From this, we know the 0-th order Green's function is defined for $\delta \phi=0$ by $\textbf{S}^0=\frac{\textbf{I}}{\omega-\textbf{E}}$. 
\\
\\
Without this approximation, the full equations of motion for the composite Green's function is
\begin{equation}
\frac{d}{d\tau} \textbf{S}_{ij}(\tau)= \delta(\tau) \langle \{ \psi_i(\tau); \psi^\dagger_j \} \rangle + \theta_H(\tau) \langle \{ J_i(\tau); \psi_j^\dagger \} \rangle
\end{equation}

In Fourier space, and with Eq. (\ref{fullcurr}), this becomes
\begin{equation}
\textbf{S}_{k}(\omega)= \textbf{S}^0_{k}(\omega) + \textbf{I}^{-1} \textbf{S}^0_{k}(\omega) \langle \{ \delta \phi_k(\omega); \psi^\dagger(\omega) \} \rangle
\end{equation}

Now we know that
\begin{equation}
\frac{d}{d\tau} \psi_i^\dagger = \sum \limits_l \textbf{E}_{il} \psi_l^\dagger + \delta \phi_i
\end{equation}

Therefore in Fourier space
\begin{equation}
\psi^\dagger_k(\omega)= (\omega - \textbf{E}_{k}+i0^+)^{-1} \delta \phi_k = \textbf{S}^0_{k}(\omega) \textbf{I}^{-1} \delta \phi_k
\end{equation}

Hence we can inject this in the equations of motion on $\textbf{S}_{k}$ to get
\begin{equation}
\textbf{S}_{k}(\omega)=\textbf{S}^0_{k}(\omega) + \textbf{S}^0_{k}(\omega) \textbf{I}^{-1}  \langle \{ \delta \phi_k; \delta \phi^\dagger \} \rangle \textbf{I}^{-1}\textbf{S}^0_{k}(\omega)
\end{equation}

We can introduce the scattering matrix $\textbf{T}$
\begin{equation}
\textbf{T}=\textbf{I}^{-1} \langle \{ \delta \phi; \delta \phi^\dagger \} \rangle \textbf{I}^{-1}
\end{equation}

We therefore obtain the familiar form
\begin{equation}
\textbf{S}=\textbf{S}^0 + \textbf{S}^0 \textbf{T} \textbf{S}^0
\end{equation}

We can introduce the self-energy through the relation $\textbf{T} \textbf{S}^0= \textbf{I}^{-1} \boldsymbol{\Sigma} \textbf{S}$, and obtain in reciprocal space the Dyson equation

\begin{equation}
\textbf{S}(k,\omega)=\frac{\textbf{I}}{\omega - \textbf{E}(k) - \boldsymbol{\Sigma}(k, \omega) +i0^+}
\end{equation}

From this equation we clearly see the consequences of our approximation. Neglecting $\delta \phi$ in the current, which are all the contributions orthogonal to $\psi$, is equivalent to neglecting $\boldsymbol{\Sigma}(k, \omega)$, therefore working with a static self energy. This approximation neglects all dynamical dependencies of the self energy.

\section*{Appendix B : Spectral representation and residue theorem of the composite Green's function}

We start from the following expression for the composite Green's function
\begin{equation}
\textbf{S}_{k}(\omega) = \frac{\textbf{I}}{\omega-\textbf{E}_{k}+i0^+}
\end{equation}

$\textbf{E}$ is diagonalizable, its eigenvalues are called $\epsilon^1$ and $\epsilon^2$, and we can rewrite this as
\begin{equation}
\begin{aligned}
\textbf{S}_{ij}(\omega) =& \frac{\textbf{I} \ Com(\omega - \textbf{E}_{k}+i0^+)^T}{det(\omega-\textbf{E}_{k}+i0^+)} \\
=&\frac{\textbf{I} \ Com(\omega - \textbf{E}_{k}+i0^+ )^T}{(\omega - \epsilon^1_{k}+i0^+) (\omega - \epsilon^2_{k}+i0^+)}
\end{aligned}
\end{equation}

Where Com($\textbf{A}$) is the cofactors matrix of $\textbf{A}$. We now want to apply residue theorem. 


We consider the function $\frac{\textbf{S}_{ij}(z)}{z-\omega+i\eta}$ with $\eta$ small. Since we know any Green's functions, including S, is analytical on the upper half-circle of the complex plane we integrate over this contour $C_u$. The poles of S are real and we have another pole at $z=\omega-i\eta$, which is on the lower half-circle of the complex plane. Therefore no poles lies in $C_u$ and we have

\begin{equation}
\oint \limits_{C_u} \frac{\textbf{S}_{k}(z)}{z-\omega - i \eta} dz = 0
\end{equation}

We now use the previous expression of $\textbf{S}$ to get

\begin{equation}
\oint \limits_{C_u} \frac{\textbf{I} \ Com(z-\textbf{E}_{k})^T}{det(z-\textbf{E}_k)(z-\omega - i \eta)} dz = 0
\end{equation}

Replacing the integrals by the sum of all the residues, we obtain

\begin{equation}
\sum \limits_{z_0 \in \mathcal{P}} Res\left( \frac{\textbf{I} \  Com(z-\textbf{E}_{k})^T}{(z - \epsilon^1_{k})(z - \epsilon^2_{k}) (z-\omega - i \eta)} , z \rightarrow z_0 \right) =0
\end{equation}

In this equation, $\mathcal{P}$ denotes the poles of $\frac{\textbf{S}_{k}(z)}{z-\omega+i \eta}$. It has three poles
\begin{equation}
\begin{aligned}
z_1&=\omega + i \eta \\
z_2&= \epsilon^1_{k} \\
z_3&= \epsilon^2_{k}
\end{aligned}
\end{equation}

These poles are all non degenerated, meaning the residue can easily be computed using
\begin{equation}
Res(f(z), z \rightarrow z_0 ) = \lim \limits_{z \rightarrow z_0} (z-z_0) f(z)
\end{equation}

Doing so leads to
\begin{equation}
\begin{aligned}
0=& \frac{\textbf{I} \ Com(\omega - \textbf{E}_{k} + i\eta)^T}{(\omega - \epsilon^1_{k} + i \eta)(\omega - \epsilon^2_{k} + i \eta)} \\+& \frac{\textbf{I} \ Com(\epsilon^1_{k} - \textbf{E}_{k} )^T}{(\epsilon^1_{k} - \epsilon^2_{k})(\epsilon^1_{k} - \omega - i \eta)} \\+& \frac{\textbf{I} \ Com(\epsilon^2_{k} - \textbf{E}_{k})^T}{(\epsilon^2_{k} - \epsilon^1_{ikj})(\epsilon^2_{k} - \omega - i \eta)}
\end{aligned}
\end{equation}

We recognize the first term can be rewritten as 
\begin{equation}
\frac{\textbf{I} \ Com(\omega - \textbf{E}_{k} + i\eta)^T}{det(\omega-\textbf{E}_{k} + i \eta)} = \textbf{I} (\omega - \textbf{E}_{k} + i \eta)^{-1} = \textbf{S}_{k}(\omega + i\eta)
\end{equation}

Finally we rearrange the equation (and evaluate $\textbf{S}$ in $\omega - i \eta$ to get retarded composite Green's functions) to recover the desired decomposition (replacing $\eta$ by $0^+$)
\begin{equation}
\begin{aligned}
\textbf{S}_{k}(\omega) =& \frac{\textbf{I}\ Com(\epsilon^2_{k} - \textbf{E}_{k})^T}{(\epsilon^1_{k}-\epsilon^2_{k})(\omega - \epsilon^2_{k} + i 0^+)} \\ -&\frac{\textbf{I} \ Com(\epsilon^1_{k} - \textbf{E}_{k})^T}{(\epsilon^1_{k}-\epsilon^2_{k})(\omega - \epsilon^1_{k} + i 0^+)}
\end{aligned}
\end{equation}

We pose 
\begin{equation}
\boldsymbol{\kappa}^a_{k}=(-1)^{a+1} \frac{\textbf{I}_{ij} \ Com(\epsilon^a_{k}-\textbf{E}_{k})^T}{(\epsilon^1_{k}-\epsilon^2_{k})}
\end{equation}

With $a\in \{1,2 \}$ to obtain the form given in Eq. (\ref{Sresidue}).\\
\\

\section*{Appendix C : Computations of M and I matrices}

In this section we first derive from the composite operator algebra the currents, then we use them to obtain the $\textbf{M}$ and $\textbf{I}$ matrices. Let us start by writing few of the most useful commutation relation that we used for the computations
\begin{equation}
\begin{aligned}
\{ \eta_{i \sigma}; \eta_{j \sigma'}^\dagger \} =& \delta_{ij}( \delta_{\sigma \sigma'} n_{i \bar{\sigma}} - \delta_{\sigma \bar{\sigma^{\prime}}} c^\dagger_{i \sigma} c_{i \bar{\sigma}}) \\
\{ \xi_{i \sigma}; \xi_{j \sigma^{\prime}}^\dagger \} =& \delta_{ij}( \delta_{\sigma \sigma^{\prime}} (1-n_{i \bar{\sigma}}) + \delta_{\sigma \bar{\sigma^{\prime}}} c^\dagger_{i \sigma} c_{i \bar{\sigma}}) \\
\{ \xi_{i \sigma}; \eta_{j \sigma^{\prime}}^\dagger \} =&0 \\
\{ c_{i \sigma}; \xi_{j \sigma^{\prime}}^\dagger \} =& \delta_{ij} ( \delta_{ \sigma \sigma^{\prime}} (1-n_{j \sigma} ) + \delta_{\sigma \bar{\sigma^{\prime}}} c^\dagger_{i \sigma^{\prime}} c_{i \bar{\sigma^{\prime}}} )\\
\{ c_{i \sigma}; \eta_{j \sigma^{\prime}}^\dagger \} =& \delta_{ij} (\delta_{\sigma \sigma^{\prime}} n_{i \bar{\sigma}} - \delta_{\sigma \bar{\sigma^{\prime}}} c^\dagger_{i \sigma^{\prime}} c_{i \bar{\sigma^{\prime}}}) \\
\{ c_{i \sigma}; \xi_{j \sigma^{\prime}} \} =& \delta_{ij} \delta_{\sigma \bar{\sigma^{\prime}}} c_{i \sigma^{\prime}} c_{ i \bar{\sigma^{\prime}}} \\
\{ c_{i \sigma}; \eta_{j \sigma^{\prime}} \} =& -\delta_{ij} \delta_{\sigma \bar{\sigma^{\prime}}} c_{i \sigma^{\prime}} c_{i \bar{\sigma^{\prime}}}
\end{aligned}
\end{equation}

From this we can explicitly compute the commutators of the composite operators $\psi$ with the hamiltonian to get the following currents
\begin{equation}
\begin{aligned}
j_i^1=&  -\mu \xi_{i \sigma} - \sum \limits_l t_{il} \left(c_{l\sigma} - n_{i \bar{\sigma}} c_{l \sigma} + S_i^- c_{l \bar{\sigma}} - \Delta_{i} c^\dagger_{l \bar{\sigma}} \right) \\
j_i^2=&  -( \mu - U) \eta_{i \sigma} + \sum \limits_l t_{il} \left( -n_{i \bar{\sigma} }c_{l \sigma } + S_i^- c_{l \bar{\sigma} } - \Delta_{i} c^\dagger_{l \bar{\sigma}} \right)
\end{aligned}
\end{equation}

The $\textbf{I}=\langle \{ \psi; \psi^\dagger \} \rangle$ and $\textbf{M}=\langle \{ j; \psi^\dagger \} \rangle$ matrices can be explicitly computed from these expressions. Note that
\begin{equation}
\begin{aligned}
\textbf{I}_{ij}^{12}=&\textbf{I}_{ij}^{21} \\
\textbf{M}_{ij}^{12}=&\textbf{M}_{ij}^{12}
\end{aligned}
\end{equation}

In the extended $\psi_i$ basis with superconductivity, the M and I matrices are 4x4 matrices and take the following form:
In this framework, the $\textbf{M}$ and $\textbf{I}$ matrices are given by

\begin{equation}
\textbf{I}_{i}=\begin{pmatrix}
1-\frac{n_i}{2} & 0 & 0 & 0 \\
0 & \frac{n_i}{2} & 0 & 0 \\
0 & 0 & 1-\frac{n_i}{2} & 0 \\
0 & 0 & 0 & \frac{n_i}{2}
\end{pmatrix} \\
\textbf{M}_{ij}=\begin{pmatrix}
m_{ij}^{11} & m_{ij}^{12} & m_{ij}^{13} & -m_{ij}^{13} \\
m_{ij}^{12} & m_{ij}^{22} & -m_{ij}^{13} & m_{ij}^{13} \\
m_{ij}^{13} & -m_{ij}^{13} & -m_{ij}^{11} & -m_{ij}^{12} \\
-m_{ij}^{13} & m_{ij}^{13} & -m_{ij}^{12} & -m_{ij}^{22}
\end{pmatrix}
\end{equation}

The expression of $m_{ij}^{11}$, $m_{ij}^{12}$ and $m_{ij}^{22}$ are the same as before (cf Eq. (\ref{Mcoefs})). The off diagonal coefficient is given by
\begin{equation}
m_{ij}^{13}=-t \gamma_{ij} \theta_{ij}
\label{m13}
\end{equation}

\section*{Appendix D : Roth decoupling and computation of p}

In this appendix we derive the self-consistent equation of 
\begin{equation}
p(i-j)= \langle n_{i\up} n_{j\up} \rangle + \langle S_i^+ S_j^- \rangle - \langle \Delta_{i} \Delta_{j}^* \rangle 
\end{equation}

\subsection*{Pair-pair term}
Following the step of L. Roth \cite{roth_electron_1969}, we express p as a function of correlation functions by mean of equations of motion. 

First, notice we can write
\begin{equation}
\langle \Delta_i \Delta_j^* \rangle = \langle \xi_{i \up} c_{i \down} \Delta_j^* \rangle + \langle \eta_{i \up} c_{i \down} \Delta_j^* \rangle
\end{equation}

Note we illustrate this decoupling with $\xi_{i\up}$ and $\eta_{i \up}$ but the idea is exactly the same with $\psi_{i \down}$. We introduce the following Green's functions
\begin{equation}
\begin{cases}
F_{ijl}(\tau)&= \langle \langle \xi_{i \up}(\tau); c_{j \down} \Delta_{l}^* \rangle \rangle \\
G_{ijl}(\tau)&=\langle \langle \eta_{i \up}(\tau); c_{j \down} \Delta_{l}^* \rangle \rangle 
\end{cases}
\end{equation}

We then consider the equations of motion for these Green's functions

\begin{equation}
\begin{aligned}
\partial_\tau \begin{pmatrix} F_{ijl}(\tau) \\ G_{ijl}(\tau) \end{pmatrix}  &=  \theta_H(\tau) \begin{pmatrix} \langle \{ \partial_t \xi_{i \up}(\tau) ; c_{j \down} \Delta_{l}^* \} \rangle \\ \langle \{ \partial_\tau \eta_{i \up}(\tau); c_{j \down} \Delta_{l}^* \} \rangle \end{pmatrix}  \\& + \delta(\tau) \begin{pmatrix} f^1_{ijl} \\ f^2_{ijl} \end{pmatrix} 
\end{aligned}
\end{equation}

where $f^n_{ijl}= \langle \{ \psi_i^n; c_{j\down} \Delta_l^* \} \rangle$. We now use Eq. (\ref{EMIS:curr}) to obtain

\begin{equation}
\partial_\tau \begin{pmatrix} F_{ijl}(\tau) \\ G_{ijl}(\tau) \end{pmatrix}  =  \sum \limits_k E_{ik} \begin{pmatrix} F_{kjl}(\tau) \\ G_{kjl}(\tau) \end{pmatrix} + \delta(\tau) \begin{pmatrix} f^1_{ijl} \\ f^2_{ijl} \end{pmatrix}
\end{equation}

We then time and space Fourier transform associating the Fourier variable $k_1$ to $r_i-r_l$ and $k_2$ to $r_i-r_j$. The equation becomes

\begin{equation}
\begin{pmatrix}
F_{k_1 k_2} (\omega) \\ G_{k_1 k_2}(\omega)
\end{pmatrix} = (\omega \textbf{Id}_2 - E_{k_1+k_2})^{-1} \begin{pmatrix}
f_{k_1 k_2}^1 \\ f_{k_1 k_2}^2
\end{pmatrix}
\end{equation}

Finally we use Eq. (\ref{EMIS:S}) to obtain

\begin{equation}
\begin{pmatrix} F_{k_1 k_2}(\omega) \\ G_{k_1 k_2}(\omega) \end{pmatrix}  = S_{k_1+k_2}(\omega) I^{-1} \begin{pmatrix} f^1_{k_1 k_2} \\ f^2_{k_1 k_2} \end{pmatrix} 
\end{equation}

Finally, we can extract $\langle \Delta_i \Delta_j^* \rangle$ by summing F and G, integrating over $\omega$ and taking the imaginary part to use Eq. (\ref{CfctS}) in order to replace the composite Green's functions by correlation functions. We get \\

\begin{equation}
\begin{aligned}
TF[\langle c_{i \sigma} c_{j \bar{\sigma}} \Delta_{l}^* \rangle](k_1,k_2) &=\frac{2}{2-n} \sum \limits_k (C^{11}_{k_1+k_2}+C^{12}_{k_1+k_2})f^1_{k_1 k_2} \\&+ \frac{2}{n} \sum \limits_k (C^{12}_{k_1+k_2} + C^{22}_{k_1+k_2}) f^2_{k_1 k_2}
\end{aligned}
\label{Deltadeltaf}
\end{equation}

We compute $f_{kjl}^n=\langle \{ \psi^n_k; c_{j \down} \Delta_l^* \} \rangle$ using the algebraic relations given in appendix C leads to

\begin{equation}
\begin{aligned}
f^1_{ijl}=& \delta_{ij} \langle \Delta_{i} \Delta_{l}^* \rangle + \delta_{il} (C^{21}_{ij} + C^{22}_{ij}) \\
f^2_{ijl}=& - \delta_{ij} \langle \Delta_{i}\Delta_{l}^* \rangle + \delta_{il} (C^{11}_{ij} + C^{12}_{ij})
\end{aligned}
\end{equation}
Performing a Fourier transform of $f^1_{ijl}$ and $f^2_{ijl}$ and setting i=j by integrating on $k_2$, then finally inverse Fourier transform on $k_1$ leads to:

\begin{equation}
    \langle \Delta_{i} \Delta_{l}^* \rangle =\frac{4}{n(2-n)} \frac{(C^{11}_{il}+C^{12}_{il})(C^{22}_{il}+C^{21}_{il})}{1-\frac{2}{2-n}(C^{11}_0 + C^{12}_0) + \frac{2}{n} (C^{21}_0 + C^{22}_0)}
\end{equation}

Which is the form in the main text. We pose
\begin{equation}
\phi=-\frac{2}{2-n}(C^{11}_0 + C^{12}_0) + \frac{2}{n} (C^{21}_0 + C^{22}_0)
\end{equation}

Replacing $C^{11}_0$ and $C^{22}_0$ by their definitions allows us to express these correlations function explicitly as a function of n. We don't explicit $C^{12}_0$ however, else it will be zero while it is not numerically: $C^{12}_0$ should stay in the numerical minimization process to obtain our results. Doing so leads to the following expression for $\phi$

\begin{equation}
    \phi = \frac{n^2 - 4 (\frac{n}{2}-\langle n_{i \up} n_{i \down} \rangle - C^{12}_0)}{n(2-n)}
\end{equation}

With our notations the pair-pair term becomes
\begin{equation}
    \langle \Delta_{i} \Delta_{l}^* \rangle = \frac{\rho_3}{1+\phi}
\end{equation}

\subsection{Spin-Spin term}

The spin-spin term is defined as $ \langle S_i^+ S_l^- \rangle = \langle c^\dagger_{i \up} c_{i \down} c^\dagger_{l \down} c_{l \up} \rangle$. In order to have our basis element as the first term, we rewrite it as \begin{equation}\langle S_i^+ S_l^- \rangle = - \langle c^\dagger_{i \up} c_{i \down} c_{l \up} c^\dagger_{l \down} \rangle = - \langle c_{l \up} c^\dagger_{l \down}  c^\dagger_{i \up} c_{i \down} \rangle 
\end{equation}

We therefore introduce the following Green's functions (setting $\tau'=0$)
\begin{equation}
\begin{cases}
F_{ijl}(\tau) =& \langle \langle \xi_{i \up}(\tau); c^\dagger_{j \down} S_{l}^+ \rangle \rangle \\
G_{ijl}(\tau) =& \langle \langle \eta_{i \up}(\tau); c^\dagger_{j\down} S_{l}^+ \rangle \rangle
\end{cases}
\end{equation}

The next steps are the same as with the pair-pair term. The only difference lies in the definition of $f^n$ in the resulting equations of motion. For the spin-spin term it is defined as $f^n_{ijl}= \langle\{ \psi^n_i; c^\dagger_{j \down} S_{l}^+ \}\rangle$. Hence we arrive at the following equation

\begin{equation}
\begin{aligned}
    TF(\langle c_{i \up} c^\dagger_{j \down} S_{l}^+ \rangle)(k_1, k_2) &=\frac{2}{2-n} (C^{11}_{k_1+k_2}+C^{12}_{k_1+k_2})f_{k_1 k_2}^1 \\&+ \frac{2}{n} (C^{12}_{k_1+k_2} + C^{22}_{k_1+k_2}) f^2_{k_1 k_2}
    \end{aligned}
\end{equation}

A computation of the $f^n$ leads to

\begin{equation}
    \begin{aligned}
    f^1_{ijl}&= \delta_{ij} \langle S_i^- S_l^+ \rangle - \delta_{kl}(C_{ij}^{11}+C_{ij}^{12}) \\
    f^2_{ijl}&= - \delta_{ij} \langle S_i^- S_l^+ \rangle - \delta_{kl} (C_{ij}^{12} + C_{ij}^{22} )
    \end{aligned}
\end{equation}

Therefore by Fourier transform the expression of f, then by integrating over $k_2$ to set i=j and by inverse Fourier transform on $k_1$, we obtain
%

\begin{equation}
\langle S_i^- S_l^+ \rangle= - \frac{\frac{2}{2-n}(C^{11}_{il}+C^{12}_{il})^2 + \frac{2}{n}(C^{12}_{il}+C_{22}^{il})^2 }{1+\frac{2}{2-n}(C^{11}_0+C^{12}_0)-\frac{2}{n}(C^{12}_0+C^{22}_0)}
\end{equation}

Which become with our notations 

\begin{equation}
\langle S_i^- S_l^+ \rangle = \frac{\rho_1}{1-\phi}
\end{equation}

\subsection{Charge-Charge term}

As we did for the $\langle S_i^- S_l^+ \rangle$ term we need to commute the charge charge term so the first element can be decomposed using our spinor. We then rewrite
\begin{equation}
\begin{aligned}
\langle c^\dagger_{i \up} c_{i \up} n_{l \up} \rangle = \frac{n}{2} - \langle c_{i \up} c^\dagger_{i \up} n_{l \up} \rangle
\end{aligned}
\end{equation}

We introduce the following Green's functions

\begin{equation}
\begin{cases}
F_{ijl}(\tau)=& \langle \langle \xi_{i \up}(\tau) ; c^\dagger_{j \up} n_{l \up} \rangle \rangle \\
G_{ijl}(\tau)=& \langle \langle \eta_{i \up}(\tau); c^\dagger_{j \up} n_{l \up} \rangle \rangle 
\end{cases}
\end{equation}

Once again the general form of the equation for $\langle c_{i \sigma} c^\dagger_{i \sigma} n_{l \sigma} \rangle$ will be the same as for the other 2 terms. However the definition of the involved $f^n_{ijl}$ will not be the same. We compute $f^n_{ijl}=\langle \{ \psi^n_i; c^\dagger_{j\up} n_{l \up} \} \rangle$ 

\begin{equation}
\begin{aligned}
f^1_{ijl}&=\delta_{ij}(\frac{n}{2}-\langle n_{i\down} n_{l \up} \rangle) +  \delta_{il}(C^{11}_{ij} + C^{12}_{ij}) \\
f^2_{ijl}&=\delta_{ij} \langle n_{i\down} n_{l\up} \rangle + \delta_{il}(C^{12}_{ij}+C^{22}_{ij})
\end{aligned}
\end{equation}

We hence obtain
\begin{equation}
\begin{aligned}
\langle n_{i \up} n_{l \up} \rangle = &\frac{n}{2}- \rho_1-\phi \langle n_{i \up} n_{l \down} \rangle + \frac{n}{2-n}(C^{11}_0+C^{12}_0)
\end{aligned}
\end{equation}

We don't know how to express $\langle n_{i\down} n_{l\up} \rangle $ as a function of the correlations functions. So we need to redo this decoupling on this term. This time we use
\begin{equation}
\langle n_{i \down} n_{l \up} \rangle = \frac{n}{2}-\langle  c_{l\up}c^\dagger_{l \up} n_{i \down} \rangle
\end{equation}
We therefore introduce
\begin{equation}
\begin{cases}
F_{ijl}(\tau)=& \langle \langle \xi_{i \up}(\tau) ; c^\dagger_{j \up} n_{l \down} \rangle \rangle \\
G_{ijl}(\tau)=& \langle \langle \eta_{i \up}(\tau); c^\dagger_{j, \up} n_{l \down} \rangle \rangle 
\end{cases}
\end{equation}
The $f^n_{ijl}=\langle \psi^n_i, c^\dagger_{j \up} n_{l \down} \rangle$ are given by

\begin{equation}
\begin{aligned}
f^1_{ijl}&=\delta_{ij}(\frac{n}{2}-\langle n_{i\down} n_{l\down} \rangle ) \\
f^2_{ijl}&=\delta_{ij} \langle n_{i\down} n_{l \down} \rangle &
\end{aligned}
\end{equation}

Using the paramagnetic assumption we have $\langle n_{i \up} n_{l \up} \rangle = \langle n_{i \down} n_{l\down} \rangle$, leading to

\begin{equation}
\begin{aligned}
\langle n_{i\up} n_{l\down} \rangle = & \frac{n}{2} - \phi \langle n_{i \up} n_{l \up} \rangle+ \frac{n}{2-n}(C^{11}_0+C^{12}_0) 
\end{aligned}
\end{equation}

\begin{figure*}[ht]
\begin{subfigure}{0.47\linewidth}
\centering
\includegraphics[width=7.9cm,height=6cm]{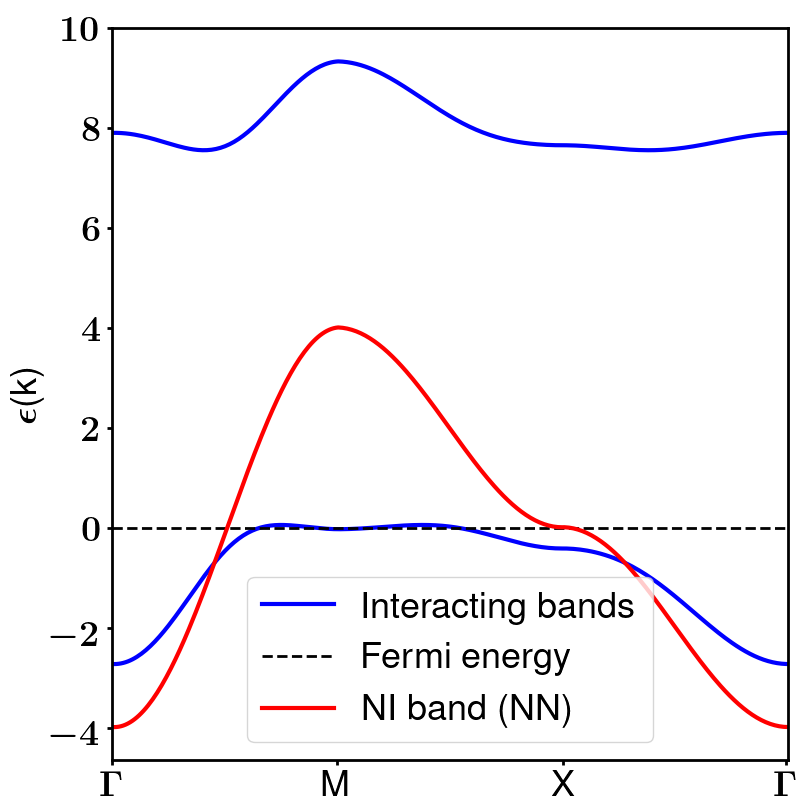}
\end{subfigure}
\begin{subfigure}{0.49\linewidth}
\centering
\includegraphics[width=5cm]{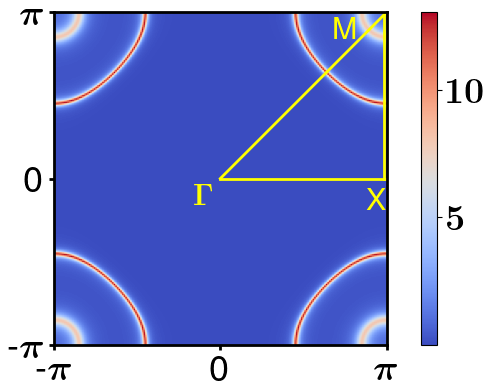}
\end{subfigure}
\\
\begin{subfigure}{0.49\linewidth}
\centering
\includegraphics[width=8cm,height=6cm]{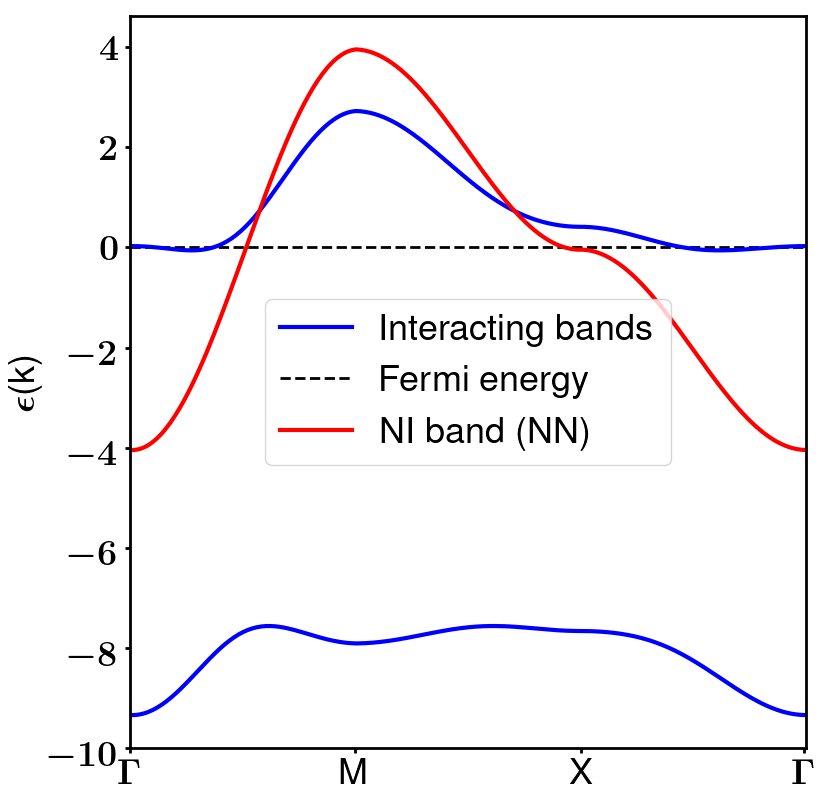}
\end{subfigure}
\begin{subfigure}{0.49\linewidth}
\centering
\includegraphics[width=5cm]{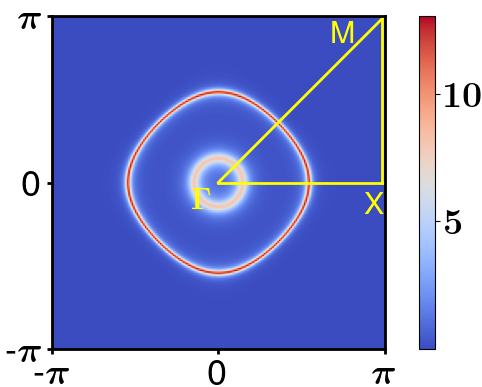}
\end{subfigure}
\caption{Fermi Surfaces and bands obtained by Roth minimization in the neighboorhood of half filling. Top row: 2\% hole doped. Bottom row: 2\% electron doped.}
\label{Fig13}
\end{figure*} 
If we inject this in the equation of $\langle n_{i\up} n_{l\up} \rangle$, we obtain
\begin{equation}
\begin{aligned}
\langle n_{i\up} n_{l \up} \rangle &=- \rho_1 + \phi^2 \langle n_{i\up} n_{l \up} \rangle + \frac{n}{2-n}(C^{11}_0+C^{12}_0)(1-\phi)
\end{aligned}
\end{equation}

The last term can be simplified. An explicit computations of the $C_0$ leads to $C^{11}_0+C_0^{12}= 1 - n + \langle n_{i \up} n_{i \down} \rangle$, allowing us to show the last term is in fact just equal to $\frac{n^2}{4}$. We therefore obtain

\begin{equation}
\langle n_{i \up} n_{l \down} \rangle = \frac{n^2}{4}-\frac{\rho_1}{1-\phi^2}
\end{equation}
Which is the self consistent equation we have. Combining the three terms, since $p=\langle n_{i \up} n_{l \up} \rangle + \langle S_i^+ S_l^- \rangle - \langle \Delta_i \Delta_l^* \rangle$, we obtain the following self consistent equation
\begin{equation}
p= \frac{n^2}{4}-\frac{\rho_1}{1-\phi^2}-\frac{\rho_1}{1-\phi}-\frac{\rho_3}{1+\phi}
\end{equation}

\section*{Appendix E: Bands at and around half-filling}

Here we plot the bands for nearest-neighboors at half filling for the 3 minimization. We see the Fermi energy in between the two bands therefore we have an insulator. Note we get the same result with next-nearest neighboors.
We plot the bands for Roth solution at 2\% hole doping with its Fermi Surface: we see the formation of a small hole pocket at $(\pi, \pi)$ when approaching half filling. This can be interpreted as a magnetic instability \cite{bergeron_breakdown_2012}, our assumption of paramagnetism becomes invalid.

\begin{figure}[H]
\begin{center}
\includegraphics[width=7cm, height=6cm]{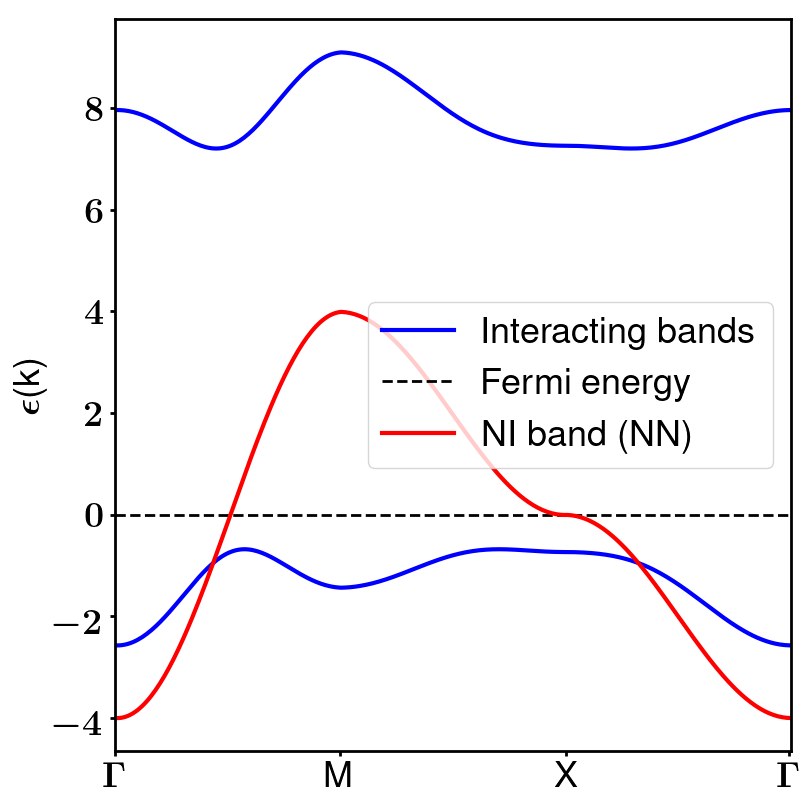}
\caption{Band at half filling with t=1, U=8t. Note that both COM2 and Roth are alike.}
\label{Fig14}
\end{center}
\end{figure}

\section*{Appendix F: Roth decoupling with superconductivity}

In this appendix we show how Roth decoupling changes when we include superconductivity with 4x4 basis. We will redo one of the terms of p as an example, and perform the decoupling for the $\theta$ parameter. Let us start by considering $\langle S_i^- S_l^+ \rangle$ in p for example. The superconducting basis is of size 4 and define by
\begin{equation}
\psi = \begin{pmatrix}
\xi_{i \up} \\
\eta_{i \up} \\
\xi^\dagger_{i\down} \\
\eta^\dagger_{i\down}
\end{pmatrix}
\end{equation} 

To include the full basis we need to introduce 4 Green's functions now

\begin{equation}
\begin{aligned}
A_{ijl}=&\langle \langle \xi_{i \up} ; c_{j \down}^\dagger S_{l}^+ \rangle \rangle \\
B_{ijl}=&\langle \langle \eta_{i \up} ; c_{j \down}^\dagger S_{l}^+ \rangle \rangle \\
F_{ijl}=&\langle \langle \xi_{i \down}^\dagger ; c_{j \down}^\dagger S_{l}^+ \rangle \rangle \\
G_{ijl}=&\langle \langle \eta_{i \down}^\dagger ; c_{j \down}^\dagger S_{l}^+ \rangle \rangle 
\end{aligned}
\end{equation}

Now the equations of motion has to be defined for the 4 Green's functions. Hence
\begin{equation}
\begin{aligned}
\partial_\tau \begin{pmatrix}
A_{ijl} \\ B_{ijl} \\ F_{ijl} \\ G_{ijl}
\end{pmatrix} (\tau) &= \sum \limits_k E_{ik} \begin{pmatrix}
A_{ijl} \\ B_{ijl} \\ F_{ijl} \\ G_{ijl}
\end{pmatrix}(\tau) +\delta(\tau) \begin{pmatrix}
f^1_{ijl} \\ f^2_{ijl} \\ f^3_{ijl} \\ f^4_{ijl}
\end{pmatrix}
\end{aligned}
\end{equation}

With $f^n_{ijl} = \langle \{ \psi^n_i; c_{j \down}^\dagger S_l^+ \} \rangle$. 
As before we can use $(\omega - E_{k})=S_{k}(\omega)I^{-1}$ after a space Fourier transform and integrate over $\omega$ after we took the imaginary part to replace the composite Green's function matrix by a correlation function matrix. Thus
\begin{equation}
\begin{aligned}
-TF[\langle c_{j \down}^\dagger c_{i \up} S_l^+ \rangle](k_1, k_2) =& \frac{2}{2-n}[ (C_{k_1+k_2}^{11}+C_{k_1+k_2}^{12})f^1_{k_1 k_2} + (C_{k_1+k_2}^{12}+C_{k_1+k_2}^{22})f^2_{k_1 k_2}\\&+(C^{13}_{k_1+k_2}+C^{23}_{k_1+k_2})f^3_{k_1 k_2}+(C^{14}_{k_1+k_2}+C^{24}_{k_1+k_2})f^4_{k_1 k_2} ]
\end{aligned}
\end{equation}

In the last equation, the first line is the same as before, while the second line are additionnal terms appearing with the superconducting basis. A bit of algebra yields to
\begin{equation}
\begin{aligned}
f^3_{ijl}=& \delta_{il} (C^{13}_{ij}+C^{14}_{ij})   \\
f^4_{ijl}=& \delta_{il} (C^{23}_{ij}+C^{24}_{ij})
\end{aligned}
\end{equation}

Finally the spin-spin term in p becomes
\begin{equation}
\begin{aligned}
\langle S_i^- S_l^+ \rangle=& - \frac{\rho_1 + \rho_2}{1-\phi}
\end{aligned}
\end{equation}
With $\rho_2=\frac{2}{2-n}(C^{13}_{il}+C^{14}_{il})^2 + \frac{2}{n}(C^{23}_{il}+C^{24}_{il})^2$.
\\
\\

Now we move on to the Roth decoupling for $\theta$. As Beenen and Edwards \cite{beenen_superconductivity_1995} have mentioned, we have several ways of decoupling $ \theta$ depending on whether we consider $\langle c_{i \up} c_{i \down} n_{l \sigma} \rangle$ or for example $\langle c^\dagger_{i \down} c^\dagger_{i \up} n_{l \sigma} \rangle$. Depending on the decoupling scheme we will over estimate or underestimate the real value of $\theta$ but the behaviour will remain globally the same \cite{stanescu_d_2000}. Here we consider a decoupling starting from $\langle \langle c_{i \down}^\dagger, c_{i \up}^\dagger n_{l \sigma} \rangle \rangle$. We introduce the following Green's functions
\begin{equation}
\begin{pmatrix}
A_{ijl}=& \langle \langle \xi_{i \up}; c_{j \up}^\dagger n_{l \up} \rangle \rangle \\
B_{ijl}=& \langle \langle \eta_{i \up}; c_{j \up}^\dagger n_{l \up} \rangle \rangle \\
F_{ijl}=& \langle \langle \xi_{i \down}^\dagger; c_{j \up}^\dagger n_{l \up} \rangle \rangle \\
G_{ijl}=& \langle \langle \eta_{i \down}^\dagger; c_{j \up}^\dagger n_{l \up} \rangle \rangle \\
\end{pmatrix}
\end{equation}  

The decoupling is identical and by considering F+G we arrive to

\begin{equation}
\begin{aligned}
TF[\langle c^\dagger_{i \down} c^\dagger_{j \up} n_{l \sigma} \rangle](k_1,k_2) &= \frac{2}{2-n}[ (C_{k_1+k_2}^{13}+C_{k_1+k_2}^{14})f^1_{k_1 k_2}  \\&+ (C_{k_1+k_2}^{33}+C_{k_1+k_2}^{44})f^3_{k_1 k_2}] \\&+\frac{2}{n} [(C^{23}_{k_1+k_2}+C^{24}_{k_1+k_2})f^2_{k_1 k_2} \\&+(C^{34}_{k_1+k_2}+C^{44}_{k_1+k_2})f^4_{k_1 k_2} ]
\end{aligned}
\end{equation}

With $f^n_{ijl}=\langle \{ \psi^n_i;c^\dagger_{j \up} n_{l \sigma} \} \rangle$. Computing the f gives
\begin{equation}
\begin{aligned}
f^1_{ijl}=& \delta_{ij}(\frac{n}{2}-\langle n_{i \up} n_{l \up} \rangle) + \delta_{il} (C^{11}_{ij} + C^{12}_{ij}) \\
f^2_{ijl}=& \delta_{ij} \langle n_{i \up} n_{l \up} \rangle + \delta_{il} ( C^{12}_{ij} + C^{22}_{ij}) \\
f^3_{ijl}=& - \delta_{ij} \frac{\theta_{il}}{2} \\
f^4_{ijl}=& \delta_{ij} \frac{\theta_{il}}{2}
\end{aligned}
\end{equation}

We therefore obtain
\begin{equation}
\begin{aligned}
\frac{\theta}{2}=&\frac{ \langle n_{i\up} n_{l\up} \rangle [ \frac{2}{n}(C^{23}_0+C^{24}_0)-\frac{2}{2-n}(C^{13}_0+C^{14}_0)] + \zeta }{1+\phi} \\
&+ \frac{n}{2}\frac{C^{13}_0+C^{14}_0}{1+\phi}
\end{aligned}
\end{equation}

With $\zeta=\frac{2}{2-n}(C^{11}_{il}+C^{12}_{il})(C^{13}_{il}+C^{14}_{il})+ \frac{2}{n}(C^{12}_{il}+C^{22}_{il})(C^{23}_{il}+C^{24}_{il})$. Finally by noting that $C^{23}_0+C^{24}_0=0$ and $C^{13}_0+C^{14}_0=0$, we obtain the equation we used
\begin{equation}
\frac{\theta}{2}=\frac{  \zeta }{1+\phi}
\end{equation}

\section*{Appendix G: Effects of particle-hole transformation}

In this appendix we give some details on how we derived the particle-hole relations in Eq. (\ref{ph_param}). The relation for the chemical potential is obtained by using the fact that the Hubbard hamiltonian stays invariant under this transformation.

\begin{equation}
\begin{aligned}
H=& \sum \limits_{ij \sigma} t_{ij} c^\dagger_{i \sigma} c_{j \sigma} + U \sum \limits n_{i \up} n_{i \down} + \mu \sum \limits_{i \sigma} n_{i \sigma} \\
&\rightarrow \sum \limits_{ij\sigma} t_{ij} (-1)^{i+j} c_{i \sigma} c^\dagger_{j \sigma} + U \sum \limits_i (-1)^{4i} c_{i\up}c^\dagger_{i \up} c_{i\down}c^\dagger_{i\down} \\&- \mu \sum \limits_{i \sigma}(-1)^{2i} c_{i \sigma} c^\dagger_{i \sigma} \\
&= -\sum \limits_{ij\sigma} t_{ij} c_{i \sigma} c^\dagger_{j \sigma} + U \sum \limits_i \left( c_{i \down} c^\dagger_{i \down} - n_{i \up} c_{i \down} c^\dagger_{i \down} \right) + \mu \sum \limits_{i \sigma} n_{i \sigma} \\
&= \sum \limits_{ij\sigma} t_{ij} c^\dagger_{j \sigma} c_{i\sigma} + U \sum \limits_i \left( 1-n_{i \down} - n_{i \up} + n_{i \up} n_{i \down} \right) + \mu \sum \limits_{i \sigma} n_{i \sigma} \\
&= \sum \limits_{ij\sigma} t_{ij} c^\dagger_{i \sigma} c_{j \sigma} + U \sum \limits_i n_{i\up} n_{i \down} + (\mu-U) \sum \limits_{i \sigma} n_{i \sigma} +cste
\end{aligned}
\end{equation}

Thus, to keep the Hamiltonian invariant and therefore have under the particle-hole transformation $H \rightarrow H$, we need to impose
\begin{equation}
\mu(2-n) = - (\mu(n)-U)
\end{equation}

Which is the relation we get in the main text. For e and p we work directly with their definitions
\begin{equation}
\begin{aligned}
e=&\langle \xi_{i \sigma} \xi_{j \sigma}^\dagger \rangle - \langle \eta_{i \sigma} \eta^\dagger_{j \sigma} \rangle \\
p=& \langle n_{i \sigma} n_{j \sigma} \rangle + \langle  S_i^+ S_j^- \rangle - \langle \Delta_i \Delta_j^* \rangle
\end{aligned}
\end{equation}

Under the particle-hole transformation we have
\begin{equation}
\xi_{i \sigma} \rightarrow (-1)^i \eta_{i \sigma}^\dagger \ \ \ \ \ \ \ \ \ \ \ \ \eta_{i \sigma} \rightarrow (-1)^i \xi_{i\sigma}^\dagger
\end{equation}

Hence
\begin{equation}
\begin{aligned}
e(2-n)& \rightarrow (-1)^{i+j} ( \langle \eta_{i \sigma}^\dagger \eta_{j \sigma} \rangle - \langle \xi_{i\sigma}^\dagger \xi_{j \sigma} \rangle) \\
&=  \langle \eta_{j\sigma} \eta^\dagger_{i \sigma} \rangle  - \langle \xi_{j \sigma} \xi^\dagger_{i \sigma} \rangle \\
&=- (\langle \xi_{i\sigma} \xi^\dagger_{j \sigma} \rangle - \langle \eta_{i \sigma} \eta^\dagger_{j \sigma} \rangle) \\
&=-e(n)
\end{aligned}
\end{equation}

We didn't kept the terms with $\delta_{ij}$ because e and p always appear with a $t_{ij}$ prefactor and $t_{ij}=0$ if $i=j$. We used the fact that i and j are always nearest neighbours to get $(-1)^{i+j}=-1$. For $p$, we have

\begin{equation}
\begin{aligned}
p & \rightarrow (-1)^{2i+2j} \langle c_{i \sigma} c^\dagger_{i \sigma} c_{j\sigma} c^\dagger_{j \sigma} + c_{i \up} c^\dagger_{i \down} c_{j \down} c^\dagger_{j \up}\\& - c^\dagger_{i \up} c^\dagger_{i \down} c_{j \down} c_{j \up} \rangle \\
&=  \langle c_{j \sigma} c^\dagger_{j \sigma} - n_{i \sigma} c_{j \sigma} c^\dagger_{j \sigma} + S_i^- S_j^+ - \Delta_j^* \Delta_i \rangle \\
&=  \langle 1 - n_{j \sigma} - n_{i \sigma} + n_{i \sigma} n_{j \sigma} + S_j^+ S_i^- - \Delta_i \Delta_j^* \rangle \\
&= 1 - \frac{n}{2} - \frac{n}{2} + p(n) \\
&= (1-n)+p(n)
\end{aligned}
\end{equation}

Which is the other relation we had.

\clearpage

\end{document}